\pdfoutput = 1
\documentclass[12pt,a4paper]{article}

\usepackage[top=1in, bottom=1.25in, left=1in, right=1in]{geometry}

\usepackage{amsmath}
\usepackage{graphicx}
\usepackage[dvipsnames]{xcolor}
\usepackage[colorlinks=true, allcolors=Blue]{hyperref}

\makeatletter
\g@addto@macro\bfseries{\boldmath}
\makeatother

\usepackage{ifthen}
\newboolean{articletitles}
\setboolean{articletitles}{true}
\newboolean{inbibliography}
\setboolean{inbibliography}{false}
\usepackage{cite}
\usepackage{mciteplus}

\begin{document}
%
%
%
\begin{titlepage}
\vspace*{-0.7truecm}
\begin{flushright}
Nikhef-2022-012
\end{flushright}

\vspace{1.4truecm}

\begin{center}
{\Large{\textbf{New Physics in $B_q^0$--$\bar B_q^0$ Mixing: Present Challenges, Prospects, and Implications for $B_q^0\to\mu^+\mu^-$}}}
\end{center}

\vspace{0.8truecm}

\begin{center}
{\bf Kristof De Bruyn\,${}^{a,b}$, Robert Fleischer\,${}^{a,c}$, Eleftheria Malami\,${}^{a}$\\
and Philine van Vliet\,${}^{d}$}

\vspace{0.5truecm}

${}^a${\sl Nikhef, Science Park 105, 1098 XG Amsterdam, Netherlands}

${}^b${\sl Van Swinderen Institute for Particle Physics and Gravity, University of Groningen, 9747 Groningen, Netherlands}

${}^c${\sl  Faculty of Science, Vrije Universiteit Amsterdam,\\
1081 HV Amsterdam, Netherlands}

${}^d${\sl Deutsches Elektronen-Synchrotron DESY, Notkestr. 85, 22607 Hamburg, Germany}
\end{center}

\vfill

\begin{abstract}
The phenomenon of $B^0_q$--$\bar B^0_q$ mixing ($q=d,s$) provides a sensitive probe for physics beyond the Standard Model.
We explore the corresponding space for New Physics left through the current data, having a careful look at analyses of the Unitarity Triangle that are needed for the Standard Model predictions of the $B_q$ mixing parameters.
In particular, we explore the impact of tensions between inclusive and exclusive determinations of the CKM matrix elements $|V_{ub}|$ and $|V_{cb}|$.
Moreover, we focus on the angle $\gamma$ of the Unitarity Triangle, comparing measurements from $B\to DK$ and $B\to\pi\pi$, $\rho\pi$, $\rho\rho$ decays, where the latter are typically interpreted in terms of the angle $\alpha$.
We discuss various scenarios and present the corresponding state-of-the-art constraints on the New Physics parameters of $B_q^0$--$\bar B_q^0$ mixing.
We point out that these results have an interesting application in the analysis of rare $B_q^0\to \mu^+\mu^-$ decays, allowing us to minimise the impact of CKM parameters in the search for New Physics.
In view of the high-precision era, we make future projections.
Interestingly, we find that for the extraction of the New Physics parameters in the $B_d$ system the determination of the apex of the Unitarity Triangle results in a key limiting factor.
By contrast, the corresponding impact is negligible for the $B_s$ system, making it a promising candidate to reveal sources of New Physics.
\end{abstract}

\vfill

\noindent
August 2022

\end{titlepage}

\thispagestyle{empty}
~
\newpage

\setcounter{page}{1}

\section{Introduction}
Mixing between neutral $B_q^0$ and $\bar B_q^0$ mesons, where $q$ can be either a down or a strange quark, offers interesting opportunities to test the Standard Model (SM) paradigm at energy scales well beyond the centre-of-mass energy of the Large Hadron Collider (LHC).
The mixing process is an example of a flavour-changing neutral current and proceeds via box diagrams, which are loop-suppressed.
This makes it an ideal place to search for the effects of New Physics (NP) contributions.
The parameter space for NP in $B_q$-meson mixing can be determined in a model-independent way \cite{Ball:2006xx}, and has been analysed in the past \cite{UTfit:2005lis,UTfit:2007eik,Lenz:2010gu,Lenz:2012az,Charles:2020dfl,Awasthi:2021mfn}.
But, as we will demonstrate below, the values of the SM predictions for the mixing phases and amplitudes strongly depend on our choices for the input measurements, and in particular on the determination of the apex of the Unitarity Triangle (UT).
The searches for NP in $B_q$-meson mixing thus critically rely on resolving various tensions between measurements of the elements of the Cabibbo--Kobayashi--Maskawa (CKM) quark mixing matrix \cite{Cabibbo:1963yz,Kobayashi:1973fv}.
In this paper, we will demonstrate the impact that the different experimental measurements related to the CKM matrix and UT have on the parameter space for NP in $B_q^0$--$\bar B_q^0$ mixing.
In addition, we will compare the results for NP in $B_d^0$--$\bar B_d^0$ and $B_s^0$--$\bar B_s^0$ mixing with each other, highlighting the similarities as well as differences between the two systems.

The allowed parameter space for NP in $B_q$-meson mixing is impacted by two important subtleties regarding the determination of the UT apex, which are both currently at risk of being overlooked.
The first subtlety is due to existing discrepancies between the measurements of individual input parameters, while the second subtlety is related to our choice of inputs to determine the apex of the UT.
Key examples regarding the first point are the measurements of the CKM matrix elements $|V_{cb}|$ and $|V_{ub}|$.
Despite the best efforts from experts in the field, we have not yet been able to resolve the discrepancies between the inclusive and exclusive determinations of these quantities (see Ref.\ \cite{Ricciardi:2021shl} for a review).
The numerical differences between both determinations directly affect many SM predictions in flavour physics, including those related to $B_q^0$--$\bar B_q^0$ mixing.
These differences should therefore not be ignored, as would be the case when naively averaging the results from both measurement strategies.
Instead, at each step of our analysis, we will demonstrate what impact the difference between the inclusive and exclusive determinations of $|V_{cb}|$ and $|V_{ub}|$ has on the results.

Secondly, to constrain the parameter space for NP in $B_q^0$--$\bar B_q^0$ mixing, it is of utmost importance to understand how the different input measurements can be affected by NP contributions themselves.
Global analyses that assume SM expressions, like those performed by the CKMfitter \cite{Charles:2015gya} or UTfit \cite{UTfit:2006vpt} groups, combine information from all relevant input measurements to determine the UT apex.
However, each of the individual experimental inputs can, in principle, be affected by different NP contributions, clouding the resulting picture of the UT apex.
While these global fits are a valuable test of the SM, the obtained fit solutions cannot reliably be used to predict SM parameters for further NP searches.
Global fits allowing for NP effects in the inputs have been made \cite{UTfit:2007eik,Charles:2020dfl}.
However, in this work we will only focus on NP in $B_q^0$--$\bar B_q^0$ mixing.

Not all experimental inputs are equally sensitive or equally affected by NP contributions.
Observables which can be measured through decays dominated by tree-level topologies, such as the UT side $R_b$ and the UT angle $\gamma$, are typically considered less prone to NP effects, while observables affected by loop-level processes, such as the UT angle $\beta$ and the mass difference $\Delta m_s$ between the mass eigenstates of the $B_s$-meson system, are considered to be much more sensitive.
For calculations of SM parameters in the presence of unknown NP, the use of the latter category should therefore be avoided.
Since our main goal is to study NP in $B_q$-meson mixing, we especially have to make sure to use input measurements free of possible contributions from NP in $B_q^0$--$\bar B_q^0$ mixing.
For that reason, we will perform and discuss our own fits of the UT apex, which aim to be completely transparent with respect to which input measurements are used and how they could be affected by NP.

Our SM determination of the UT apex uses only two observables: the UT side $R_b$ and the UT angle $\gamma$, to keep possible contamination of the input parameters due to NP to a minimum.
Both these observables are affected by the subtleties introduced above and thus require a careful analysis.
The numerical value for $R_b$ strongly depends on our choice for the CKM matrix elements $|V_{cb}|$ and $|V_{ub}|$, propagating the unresolved discrepancies between the inclusive and exclusive determinations to the solution for the UT apex, and to the SM predictions of the $B_q$-meson mixing parameters.
For $\gamma$, on the other hand, there are multiple strategies from a variety of $B$-meson decay channels to measure its value.
The most precise direct determination of $\gamma$ is obtained from a combination of multiple CP violation measurements in the $B \to DK$ family of decays \cite{LHCb:2021dcr}.
These channels proceed via tree topologies only.
In most beyond the SM theories, the NP contributions are strongly suppressed in these topologies, and can thus still be neglected.
Nonetheless, it remains possible for NP effects to be present in these decays \cite{Brod:2014bfa,Lenz:2019lvd,Iguro:2020ndk,Cai:2021mlt,Bordone:2021cca}.
Therefore it is important to carefully compare the results from individual measurements, now and in the future, and in particular when including them in experimental averages.
In this respect, the decay-time-independent measurements using $B^+$ and $B_d^0$ decays show an interesting difference compared to the decay-time-dependent measurement in the $B_s^0\to D_s^{\mp}K^{\pm}$ channel \cite{Fleischer:2021cct,Fleischer:2021cwb}.

In addition, a second high-precision measurement of $\gamma$ can be obtained from the isospin analysis of the $B\to\pi\pi$, $\rho\pi$, $\rho\rho$ decays.
This analysis is commonly presented as a measurement of the UT angle $\alpha$, but we argue that, in view of NP searches, it is natural and more advantageous to interpret it as a measurement of $\gamma$.
In contrast to the $B\to DK$ decays, the $B\to\pi\pi$, $\rho\pi$, $\rho\rho$ modes also get contributions from loop-level penguin topologies.
The isospin relations between the three decay modes allow us to control the size of these penguin topologies, and take into account their impact on the determination of the UT angle $\gamma$.
In the literature, this relation is usually used to obtain a measurement of $\alpha$, assuming $\alpha+\beta+\gamma = 180^{\circ}$.
However, in this paper, we will not make this assumption and instead consider the isospin relations as an independent determination of $\gamma$, using experimental input from the $B_d^0$--$\bar B_d^0$ mixing phase $\phi_d$, which is given by $2 \beta$ in the SM.
It should be stressed that the underlying physics processes are vastly different from the $B \to DK$ decays, and a comparison of the results is therefore highly non-trivial.
If the two obtained values for $\gamma$ agree with each other, this would place the determination of the UT apex via $R_b$ and $\gamma$ on a more robust footing, while in case they disagree, one would have a clear indicator for possible NP.
This also illustrates that we need to be very careful when making averages of different measurements, and this should only be done after proper justification.

After a careful analysis of the UT apex focused on providing clean SM predictions for the $B_q$-meson mixing parameters, we discuss the parameter space that is still left for NP.
Here, we will consider two different scenarios.
The first uses $R_b$ and $\gamma$ as input for the SM predictions and determines the parameter space for NP separately for the $B_d$ and $B_s$ systems.
This is the most general scenario and serves as a baseline for a comparison with the other two options. 
The only assumptions are that $R_b$ and $\gamma$ are free of possible NP.
The purpose of the other scenario is to investigate what impact additional assumptions can have on the allowed parameter space for NP in $B_q$ mixing.
We will name the second scenario that we consider Flavour Universal New Physics (FUNP), where the NP in the $B_d$ and the $B_s$ system is assumed to be equal.
In this case, also the ratio between the mixing parameters $\Delta m_d$ and $\Delta m_s$ is free from NP, and can be used to determine the side $R_t$ of the UT.
Combining it with the side $R_b$, it is then possible to obtain the SM UT apex without relying on $\gamma$.
By comparing this scenario to the baseline, we can see how compatible the data are with the FUNP assumption.

The unresolved discrepancies between the measurements of the CKM matrix elements not only impact the determination of the UT apex and the NP searches in $B_q^0$--$\bar B_q^0$ mixing, but also affect other high-profile NP searches.
One particularly interesting example is the branching fraction measurement of the rare decay $B_s^0\to\mu^+\mu^-$, which originates from $b\to s\ell\ell$ flavour changing neutral current processes.
We will illustrate how our choices for $|V_{ub}|$ and $|V_{cb}|$ propagate to the allowed parameter space for the pseudo-scalar and scalar NP contributions in $B_s^0\to\mu^+\mu^-$.
The dependence on these CKM matrix elements can be minimised, as was pointed out in Refs.\ \cite{Buras:2003td,Bobeth:2021cxm,Buras:2021nns}, by taking the ratio between the $B_s^0\to\mu^+\mu^-$ branching fraction and the mixing parameter $\Delta m_s$.
Using this ratio to search for NP in the decay of $B_s^0\to\mu^+\mu^-$ does require to take into account the NP contributions in $B_q^0$--$\bar B_q^0$ mixing.
We therefore apply our results from the general, model-independent NP fit to illustrate how this ratio offers an interesting alternative constraint on the pseudo-scalar and scalar NP contributions.
All these considerations can also be applied to the $B_d^0\to\mu^+\mu^-$ decay once accurate branching fraction measurements become available in the future.

The outline of this paper is as follows:
We start our analysis in Section \ref{sec:UTapex} with a discussion on the input measurements of $\gamma$ and $R_b$, fit the UT apex, and compare our solution with the constraint following from $|\varepsilon_K|$, describing indirect CP violation in the neutral kaon system, which is highly sensitive to $|V_{cb}|$.
Next, in Section \ref{sec:BBmix} we introduce the SM $B_q^0$--$\bar B_q^0$ mixing observables, which we need as reference to constrain the parameter space for NP.
In addition, we use the ratio between the mixing parameters $\Delta m_s$ and $\Delta m_d$ to obtain an alternative solution for the UT apex that can be used for SM predictions under the assumption of FUNP. 
This alternative solution will be used in our NP search.
The two scenarios for NP in $B_q^0$--$\bar B_q^0$ mixing are discussed in Section \ref{sec:NPscenarios}, and the applications for the rare decay $B_q^0\to\mu^+\mu^-$ in Section \ref{sec:raredecays}.
In Section \ref{sec:future}, we explore and illustrate the impact of increased precision on the key input measurements in the future.
Finally, we conclude in Section~\ref{sec:conclusion}.

\newpage
\section{Determination of the UT Apex}\label{sec:UTapex}
The NP searches in $B_q^0$--$\bar B_q^0$ mixing require accurate SM predictions of the mixing parameters (see Section \ref{sec:BBmix}).
Their SM values depend on the apex $(\bar{\rho}, \bar{\eta})$ of the UT, where
\begin{equation}
    \bar\rho \equiv \left(1-\frac{\lambda^2}{2}\right) \rho\:,\qquad
    \bar\eta \equiv \left(1-\frac{\lambda^2}{2}\right) \eta\:,
\end{equation}
and $\lambda$, $\rho$ and $\eta$ are three of the Wolfenstein parameters \cite{Wolfenstein:1983yz,Buras:1994ec}.
The UT apex thus plays a crucial role in the analysis, but we cannot rely on global fits of the UT \cite{Charles:2015gya,UTfit:2006vpt} to calculate the SM values, as they include input measurements that are potentially affected by NP in $B_q^0$--$\bar{B}_q^0$ mixing, and would thus bias our results.
Instead, we will only use information on the side $R_b$ and the angle $\gamma$ to determine $\bar{\rho}$ and $\bar{\eta}$.
Using the Particle Data Group \cite{Workman:2022ynf} parametrisation of the CKM matrix, the coordinates of the UT apex are given by
\begin{equation}
    R_b \ e^{i \gamma}  = \bar{\rho} + i \bar{\eta}\:,
\end{equation}
where $R_b$ and $\gamma$ will be defined in more detail below.
The advantage of this approach is that $R_b$ and $\gamma$ can be determined from decays that proceed via tree topologies only.
In general, NP contributions are considered to be strongly suppressed in tree topologies, which are thus less sensitive to their effects.
Measurements of the ratios of semileptonic branching fractions $R(D)$ and $R(D^{*})$ \cite{HFLAV:2022pwe}
might contest this assumption \cite{Hiller:2018ijj,Fajfer:2019rjq,Bernlochner:2021vlv,Albrecht:2021tul}.
Interestingly, NP effects can be included in these decays and the corresponding determination of the CKM matrix elements \cite{Jung:2018lfu,Banelli:2018fnx,Iguro:2020cpg,Fleischer:2021yjo}.
If NP enters only through couplings to heavy $\tau$ leptons, the determinations of $|V_{cb}|$ and $|V_{ub}|$ from semileptonic $B$ decays into the light leptons $\ell=e,\mu$, which are the only determinations of these CKM matrix elements used in this paper, would be as in the SM. 
Since the determination of $\gamma$ uses non-leptonic decays, and the determination of $R_b$ does not rely on decays that involve $\tau$ leptons, they can also be considered free of NP for the analysis presented here.
Nonetheless, the measurements of $R_b$ and $\gamma$ require a careful discussion in view of existing tensions.
Let us therefore start with a detailed discussion of the individual inputs, before proceeding towards a fit of the UT apex.
\subsection[The UT Angle gamma]{The UT Angle $\gamma$}\label{sec:gamma}
The UT angle $\gamma$ is defined as 
\begin{equation}
    \gamma \equiv \arg\left(-\frac{V_{ud}^{\phantom{*}}V_{ub}^*}{V_{cd}^{\phantom{*}}V_{cb}^*}\right)\:,
\end{equation}
and can be measured in $B$ decays which are sensitive to interference between CKM-favoured $b\to c$ and CKM-suppressed $b\to u$ quark-level transitions.
The angle $\gamma$ is determined with high precision from CP violation measurements in $B\to DK$ decays. 
The label $B\to DK$ groups together a long list of decay channels and final states, including $B^+\to DK^+$, $B^+\to D^*K^+$, $B^+\to DK^{*+}$, $B_d^0\to DK^{*0}$, $B_s^0\to D_s^\mp K^\pm$ and others.
Here, $D$ represents an admixture of $D^0$ and $\bar D^0$ mesons, and similarly for $D^*$.
An overview of the various experimental measurements can be found in Refs.\ \cite{HFLAV:2022pwe,LHCb:2021dcr}.
In view of the goal of this paper, to search for NP in $B_q$-meson mixing, an important distinction needs to be made between the $B^+$ and $B_d^0$ decays on the one hand, and the $B_s^0$ decays on the other.
The sensitivity to $\gamma$ in the $B^+$ and $B_d^0$ decays comes from direct CP violation, which can be measured from decay-time-independent analyses, while the sensitivity to $\gamma$ in $B_s^0\to D_s^\mp K^\pm$ comes from mixing-induced CP violation and requires a decay-time-dependent analysis, taking into account the effects of $B_s^0$--$\bar B_s^0$ mixing.
This illustrates why we need to be careful when making averages of the different $\gamma$ measurements.

\paragraph{Decay-time-independent $B \to DK$}
For the decay-time-independent analyses, the latest average from the LHCb collaboration is given by \cite{LHCb:2021dcr}
\begin{equation}\label{eq:gamma_B2DK}
    \gamma_{B \to DK} = (64.9_{-4.5}^{+3.9})^{\circ}\:.
\end{equation}
Since the individual $\gamma$ measurements could, in principle, be affected by NP effects, it remains important to compare the results from the individual measurements and different decay channels before averaging.
NP effects may be washed out in the average, which then also becomes very challenging to interpret theoretically.
Here, we will assume that the result \eqref{eq:gamma_B2DK} is free from NP.

\paragraph{Decay-time-dependent $B^{0}_s \to D_{s}^{\mp} K^{\pm}$}
The interpretation of the results from the $B_s^0\to D_s^\mp K^\pm$ system requires more attention.
Through interference effects with the $B_s^0$--$\bar B_s^0$ mixing process, the CP asymmetry parameters in $B_s^0\to D_s^\mp K^\pm$ allow a determination of the sum $\phi_s + \gamma$ that is theoretically clean.
The latest result for the $B_s$-meson mixing phase $\phi_s$ measured in the $B_s^0\to J/\psi \phi$ channel and corrected for contributions from penguin topologies \cite{Barel:2020jvf,Barel:2022wfr}, is
\begin{equation}\label{eq:phis_JpsiPhi}
    \phi_s = -0.074_{-0.024}^{+0.025} = \left(-4.2 \pm 1.4\right)^{\circ}\:.
\end{equation}
With this experimental input, the result for $\phi_s + \gamma$ can be converted into an independent measurement of $\gamma$, regardless of possible NP in the $B_s^0$--$\bar B_s^0$ mixing phase $\phi_s$.
Any NP contributions to $\phi_s$ impact the CP asymmetry measurements in both $B_s^0\to J/\psi \phi$ and $B_s^0\to D_s^\mp K^\pm$ equally, and thus cancel in the determination of $\gamma$.
Combining the measured CP asymmetries in $B_s^0\to D_s^\mp K^\pm$ \cite{LHCb:2017hkl} with the above value for $\phi_s$, gives \cite{Fleischer:2021cct,Fleischer:2021cwb}
\begin{equation}\label{eq:gamma_Bs2DsK}
    \gamma_{B_s \rightarrow D_s K}=\left(131^{+17}_{-22}\right)^\circ \:,
\end{equation} 
which shows a difference of 3 standard deviations in comparison with the decay-time-independent result of $\gamma$ in Eq.\ \eqref{eq:gamma_B2DK}.
A dedicated discussion of this result and the puzzling tension with Eq.\ \eqref{eq:gamma_B2DK} can be found in Refs.\ \cite{Fleischer:2021cct,Fleischer:2021cwb}.
If this value is to be explained by NP, it requires additional contributions to the $B_s^0\to D_s^\mp K^\pm$ decay amplitudes.
Remarkably, Refs.\ \cite{Fleischer:2021cct,Fleischer:2021cwb} have also found puzzling patterns at the branching ratio level, as one would expect with NP effects entering the decay amplitude.
What makes the intriguing situation even more exciting, is that consistent patterns are found when complementing the analysis with decays with similar dynamics. 
In Refs.\ \cite{Fleischer:2021cct,Fleischer:2021cwb}, a model-independent strategy has been presented, showing that the data can be accommodated with NP contributions at the level of $30\%$ of the SM amplitudes. 
We note that model-dependent studies have also been discussed in Refs.\ \cite{Iguro:2020ndk,Cai:2021mlt,Bordone:2021cca}.

\paragraph{Isospin $B\to\pi\pi$, $\rho\pi$, $\rho\rho$}
The UT angle $\gamma$ can also be determined from an isospin analysis of the decays $B\to\pi\pi$, $\rho\pi$, $\rho\rho$, where $B$ is either $B^+$ or $B_d^0$, and we combine all possible charge combinations for the final states.
This method was originally proposed in Ref.\ \cite{Gronau:1990ka} for the $B\to\pi\pi$ system and has been extensively discussed in Ref.\ \cite{Charles:2017evz}.
For the remainder of this paper, we will use a shorthand notation and refer to this method as $\gamma$ from isospin relations, or $\gamma_{\text{iso}}$.
The isospin relations between the $B\to\pi\pi$, $\rho\pi$, $\rho\rho$ decays are exploited to simultaneously control the contributions from penguin topologies affecting these decays, and to determine the weak phase difference
\begin{equation}\label{eq:alpha_iso}
    \phi_d + 2\gamma = \phi_d^{\text{NP}} - 2\alpha\:,
\end{equation}
where 
\begin{equation}
    \phi_d = \phi_d^{\text{SM}} + \phi_d^{\text{NP}} = 2\beta + \phi_d^{\text{NP}}\:.
\end{equation}
In the absence of NP contributions, i.e.\ $\phi_d^{\text{NP}} = 0$, relation \eqref{eq:alpha_iso} has been used in the literature to interpret the dependence on $\phi_d + 2\gamma$ as an independent measurement of the UT angle $\alpha$, with the latest result given by \cite{HFLAV:2022pwe}
\begin{equation}\label{eq:alpha}
    \alpha = (85.2_{-4.3}^{+4.8})^{\circ}\:.
\end{equation}
However, $\phi_d^{\text{NP}}$ need not be zero, as we explore in this paper. 
The isospin analysis therefore does not provide a direct measurement of the CKM angle $\alpha$, but instead, should be interpreted as an independent measurement of the CKM angle $\gamma$, utilising external input on the $B_d^0$--$\bar B_d^0$ mixing phase $\phi_d$.
The latest result for $\phi_d$, measured in the $B_d^0\to J/\psi K^0$ channel and corrected for contributions from penguin topologies, is given as follows \mbox{\cite{Barel:2020jvf,Barel:2022wfr}:}
\begin{equation}\label{eq:phid_JpsiK}
    \phi_d = \left(44.4_{-1.5}^{+1.6}\right)^{\circ}\:.
\end{equation}
It is important to stress that NP effects in $B_d^0$--$\bar B_d^0$ mixing lead to the same $\phi_d^{\text{NP}}$ contribution in the measurement of $\phi_d$ from $B_d^0\to J/\psi K^0$ and in the determination of the weak phase $\phi_d + 2\gamma$ obtained through the isospin analysis.
This NP phase thus cancels when using the experimental input \eqref{eq:phid_JpsiK}, allowing a measurement of $\gamma$ from $B\to\pi\pi$, $\rho\pi$, $\rho\rho$.
Combining the result for $\phi_d + 2\gamma$ in Eq.\ \eqref{eq:alpha} with the measurement of $\phi_d$ in Eq.\ \eqref{eq:phid_JpsiK}, we find
\begin{equation}\label{eq:gamma_Iso}
    \gamma_{\text{iso}} = (72.6_{-4.9}^{+4.3})^{\circ}\:.
\end{equation}
In contrast to the measurement of $\gamma$ from $B\to DK$ decays, this value can be affected by possible NP entering the decay amplitudes through penguin topologies.

\paragraph{Average}
Comparing the values of $\gamma$ in Eqs.\ \eqref{eq:gamma_B2DK} and \eqref{eq:gamma_Iso}, we find a difference of $(7.5 \pm 6.7)^{\circ}$.
The two $\gamma$ determinations are consistent within 1.1 standard deviations.
Given their different origins, this is a non-trivial result.
Since both approaches also have similar precision, there is no preference for one or the other.
Because of these two reasons, we have chosen to average the results for our analysis in this paper, and will use
\begin{equation}\label{eq:gamma_Avg}
    \gamma_{\text{avg}} = (68.4 \pm 3.3)^{\circ}
\end{equation}
for the fit of the UT apex below.
However, it is important to keep in mind that the situation may well change in the future as the measurements of $\gamma$ become more precise.
With improved precision, differences between the approaches could become more pronounced, thereby hinting at NP effects either at the tree-level in $B\to DK$ decays or at the loop-level in the isospin analysis.
If such a situation arises, averaging both results would no longer be justified.
\subsection[The UT Side Rb]{The UT Side $R_b$}
The second input for the SM fit of the UT apex is the side $R_b$.
It is defined as
\begin{equation}\label{eq:Rb}
    R_b \equiv \left(1-\frac{\lambda^2}{2}\right)\frac{1}{\lambda}\left|\frac{V_{ub}}{V_{cb}}\right|
    = \sqrt{\bar\rho\,^2 + \bar\eta\,^2}\:.
\end{equation}
All three CKM matrix elements $\lambda \equiv |V_{us}|, |V_{ub}|$ and $|V_{cb}|$ can be measured in semileptonic decays involving electrons or muons, which are generally considered to be robust against NP contributions.
Nonetheless, for each of these elements, there are still unresolved discrepancies between the various theoretical and experimental approaches.
These discrepancies have a sizable impact on the coordinates of the UT apex, and thus also on the SM predictions and NP searches in $B_q$-meson mixing.
Therefore, even though the tensions have been extensively discussed in the literature (see Ref.\ \cite{Workman:2022ynf} for an overview), it is important to summarise them here again.

\paragraph{$|V_{us}|$}
The element $|V_{us}|$ is most precisely measured in semileptonic kaon decays.
The experimental average from measurements in the $K\to\pi\ell\nu_{\ell}$ decay channels, collectively referred to as $K\ell3$, is \cite{Seng:2021nar,Seng:2022wcw}  
\begin{equation}\label{eq:Vus}
    |V_{us}|= 0.22309 \pm 0.00056\qquad(\text{from }K\ell3)\:,
\end{equation}
while the experimental average from $K\to\mu\nu_{\mu}(\gamma)$ decays, referred to as $K\ell2$, is \cite{Workman:2022ynf}
\begin{equation}
    |V_{us}| = 0.2252 \pm 0.0005\qquad(\text{from }K\ell2)\:.
\end{equation}
These two results differ by three standard deviations.
The $K\ell2$ result is derived from the ratio of decay rates
\begin{equation}\label{eq:K2pi_ratio}
    \frac{\Gamma(K\to\mu\nu_{\mu}(\gamma))}{\Gamma(\pi\to\mu\nu_{\mu}(\gamma))}
    \propto \left(\frac{|V_{us}|f_{K^+}}{|V_{ud}|f_{\pi^+}}\right)^2
\end{equation}
between the $K\to\mu\nu_{\mu}(\gamma)$ and $\pi\to\mu\nu_{\mu}(\gamma)$ decay channels.
Here, $f_{K^+}$ and $f_{\pi^+}$ are the kaon and pion decay constants, respectively, whose ratio is well known from lattice calculations \cite{Aoki:2021kgd}.
Furthermore, the measurement of $|V_{us}|$ from Eq.\ \eqref{eq:K2pi_ratio} also depends on the CKM matrix element $|V_{ud}|$, whose most precise measurement \cite{Workman:2022ynf}
\begin{equation}\label{eq:Vud}
    |V_{ud}| = 0.97373 \pm 0.00031
\end{equation}
comes from superallowed nuclear beta decay.

The precise determination of $|V_{ud}|$ offers a third possibility to determine $|V_{us}|$.
Using the Wolfenstein parametrisation \cite{Wolfenstein:1983yz,Buras:1994ec} of the CKM matrix elements
\begin{equation}
    |V_{ud}| = 1 - \frac{1}{2}\lambda^2 - \frac{1}{8}\lambda^4 + \mathcal{O}(\lambda^6)\:,
    \qquad
    |V_{us}| = \lambda +  \mathcal{O}(\lambda^7)\:,
\end{equation}
it is possible to express $|V_{us}|$ in terms of $|V_{ud}|$, resulting in
\begin{equation}\label{eq:Vus_fromVud}
    |V_{us}| = 0.2277 \pm 0.0013\qquad(\text{from }|V_{ud}|)\:.
\end{equation}
Even though the uncertainty on this derived result is much larger than the direct $K\ell3$ measurement, we again find a difference of 3.3 standard deviations.
A comparison with the $K\ell2$ measurement is less straightforward as they are not independent results.
Alternatively, the measurement of $|V_{us}|$ from ${K \ell3}$ predicts a value
\begin{equation}
    |V_{ud}| = 0.97481 \pm 0.00013 \qquad(\text{from }K\ell3)\:,
\end{equation}
which differs from the result \eqref{eq:Vud} by 3.2 standard deviations.
A similar tension is found when testing the CKM matrix unitarity relation 
\begin{equation}\label{eq:normalRel}
    |V_{ud}|^2 + |V_{us}|^2 + |V_{ub}|^2 = 1\:.
\end{equation}
Combining the measurements in Eqs.\ \eqref{eq:Vud} and \eqref{eq:Vus} with the experimental results for $|V_{ub}|$ introduced below, a difference of 3.2 standard deviations with respect to 1 is found.
For this test, the impact of $|V_{ub}|$, which is only proportional to $\lambda^3$, is negligible.
Both results suggest that the tension between the $K\ell3$ and $K\ell2$ measurements of $|V_{us}|$ may find its origin in a tension between the $|V_{us}|_{K \ell3}$ and $|V_{ud}|$ measurements.

Lastly, $|V_{us}|$ has also been determined from $\tau$ lepton decays.
The latest average is \cite{Workman:2022ynf}
\begin{equation}
    |V_{us}| = 0.2221\pm 0.0013\qquad(\tau \text{ lepton decays)}\:.
\end{equation}
This result is compatible with the measurement from $K\ell3$ decays, albeit with a much larger uncertainty, and has a similar precision as the result \eqref{eq:Vus_fromVud} based on the $|V_{ud}|$ measurement.

Although the tensions between the different determinations of $|V_{us}|$ are intriguing, they turn out to have a negligible impact in the search for NP in $B_q$-meson mixing.
This is illustrated by Fig.\ \ref{fig:UT_apex_fit_comp}, which will be discussed in more detail below.
Therefore, we will only present the results for the $K\ell3$-type decays in the paper, avoiding the dependence on the measurement of $|V_{ud}|$.

\paragraph{$|V_{ub}|$ and $|V_{cb}|$}
The CKM matrix elements $|V_{ub}|$ and $|V_{cb}|$ are measured in semileptonic $B$ decays.
The latest experimental averages for the inclusive determinations are \cite{HFLAV:2022pwe,Bordone:2021oof}
\begin{equation}
    |V_{ub}|_{\text{incl}} = (4.19 \pm 0.17) \times 10^{-3}\:,\qquad
    |V_{cb}|_{\text{incl}} = (42.16 \pm 0.50) \times 10^{-3}\:,
\end{equation}
where $|V_{ub}|$ is calculated using the Gambino--Giordano--Ossola--Uraltsev (GGOU) \cite{Gambino:2007rp} approach, while the determination of $|V_{cb}|$ uses the kinematic scheme and includes the latest three-loop calculations of the total semileptonic width \cite{Bordone:2021oof}.
These results, however, differ by respectively 3.9 and 4.3 standard deviations from the exclusive determinations from the Heavy Flavour Averaging Group (HFLAV) \cite{HFLAV:2022pwe} 
\begin{equation}\label{eq:Vb_excl}
    |V_{ub}|_{\text{excl}} = (3.51 \pm 0.12) \times 10^{-3}\:,\qquad
    |V_{cb}|_{\text{excl}} = (39.10 \pm 0.50) \times 10^{-3}\:,
\end{equation}
which include constraints on the ratio $|V_{ub}|/|V_{cb}|$ from LHCb measurements of the decays $\Lambda_b^0\to p\mu^-\bar\nu_{\mu}$ \cite{Aaij:2015bfa} and $B_s^0\to K^-\mu^+\nu_{\mu}$ \cite{LHCb:2020ist}.
Contrary to the situation for $|V_{us}|$, these differences do have a significant impact on the analysis presented in this paper, and lead to different pictures for the allowed parameter spaces for NP in $B_q$-meson mixing.
Therefore, we will explicitly show all results for both the inclusive and the exclusive determinations.
This illustrates the importance of further investigating the origin of these tensions and of eventually resolving them.

\paragraph{$R_b$}
Combining the measurements of $|V_{us}|$, $|V_{ub}|$ and $|V_{cb}|$, we can calculate the UT side $R_b$.
Due to the unresolved discrepancies between the different determinations of these CKM matrix elements, the numerical value for $R_b$ will strongly depend on our choices for the individual inputs.
The two most common combinations found in the literature pair the inclusive (or exclusive) determinations of $|V_{ub}|$ and $|V_{cb}|$ with each other, resulting in a fully inclusive (or exclusive ) value for $R_b$:
\begin{equation}\label{eq:Rb_val}
    R_{b,\text{incl},K\ell3} = 0.434 \pm 0.018\:, \qquad
    R_{b,\text{excl},K\ell3} = 0.392 \pm 0.014\:.
\end{equation}
These two values differ from each other by 2.4 standard deviations.

Recent progress in the inclusive determination of $|V_{cb}|$ and associated discussions in the literature \cite{Bordone:2019guc,Ricciardi:2021shl,Bordone:2021oof,Buras:2022wpw}, aiming to understand and resolve the tension between the inclusive and exclusive values of $|V_{ub}|$ and $|V_{cb}|$, have sparked interest in a third possibility: The hybrid combination of the exclusive value for $|V_{ub}|$ with the inclusive value for $|V_{cb}|$.
This results in a value for $R_b$ of 
\begin{equation}\label{eq:Rb_hyb}
    R_{b,\text{hybrid},K\ell3} = 0.364 \pm 0.013\:,
\end{equation}
which differs from the inclusive (exclusive) $R_b$ solution by 3.7 (1.5) standard deviations.
In this paper we will consider all three options to illustrate the impact that the numerical values of $|V_{ub}|$ and $|V_{cb}|$ have on our searches for beyond the SM physics.
\subsection{The UT Apex}\label{sec:UTfit}
\begin{figure}
    \centering
    \includegraphics[width=0.49\textwidth]{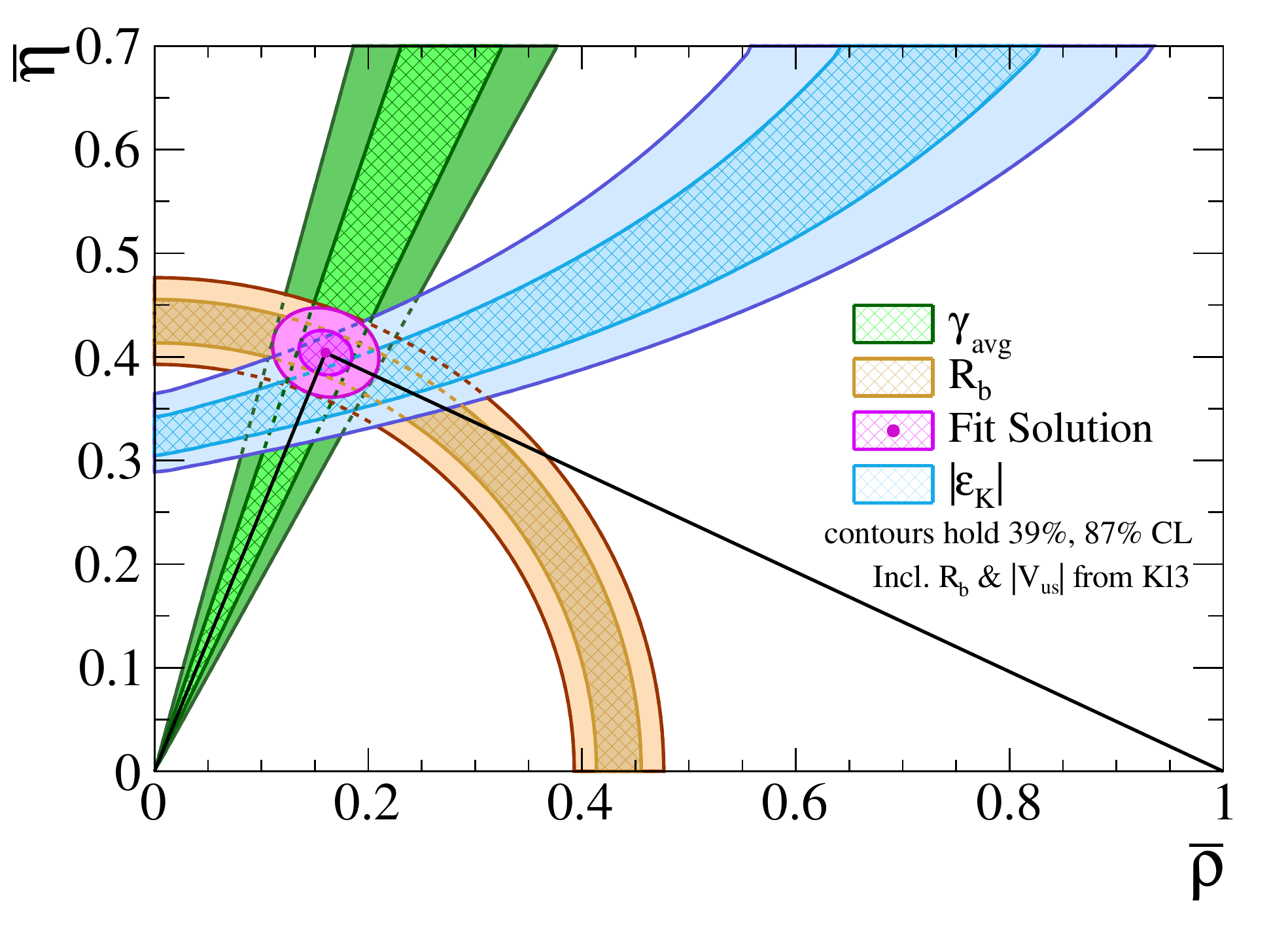}
    \includegraphics[width=0.49\textwidth]{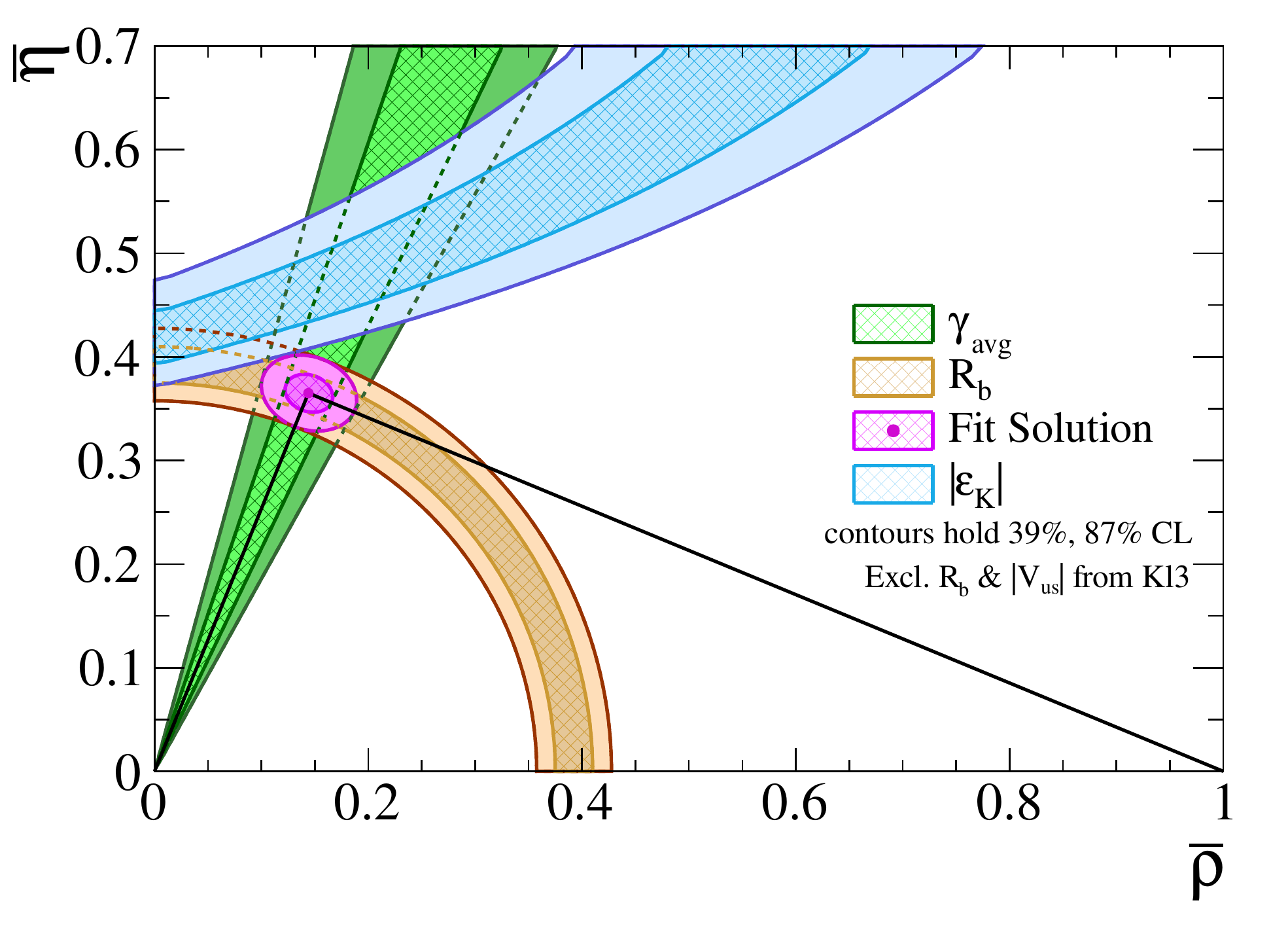}
    \includegraphics[width=0.49\textwidth]{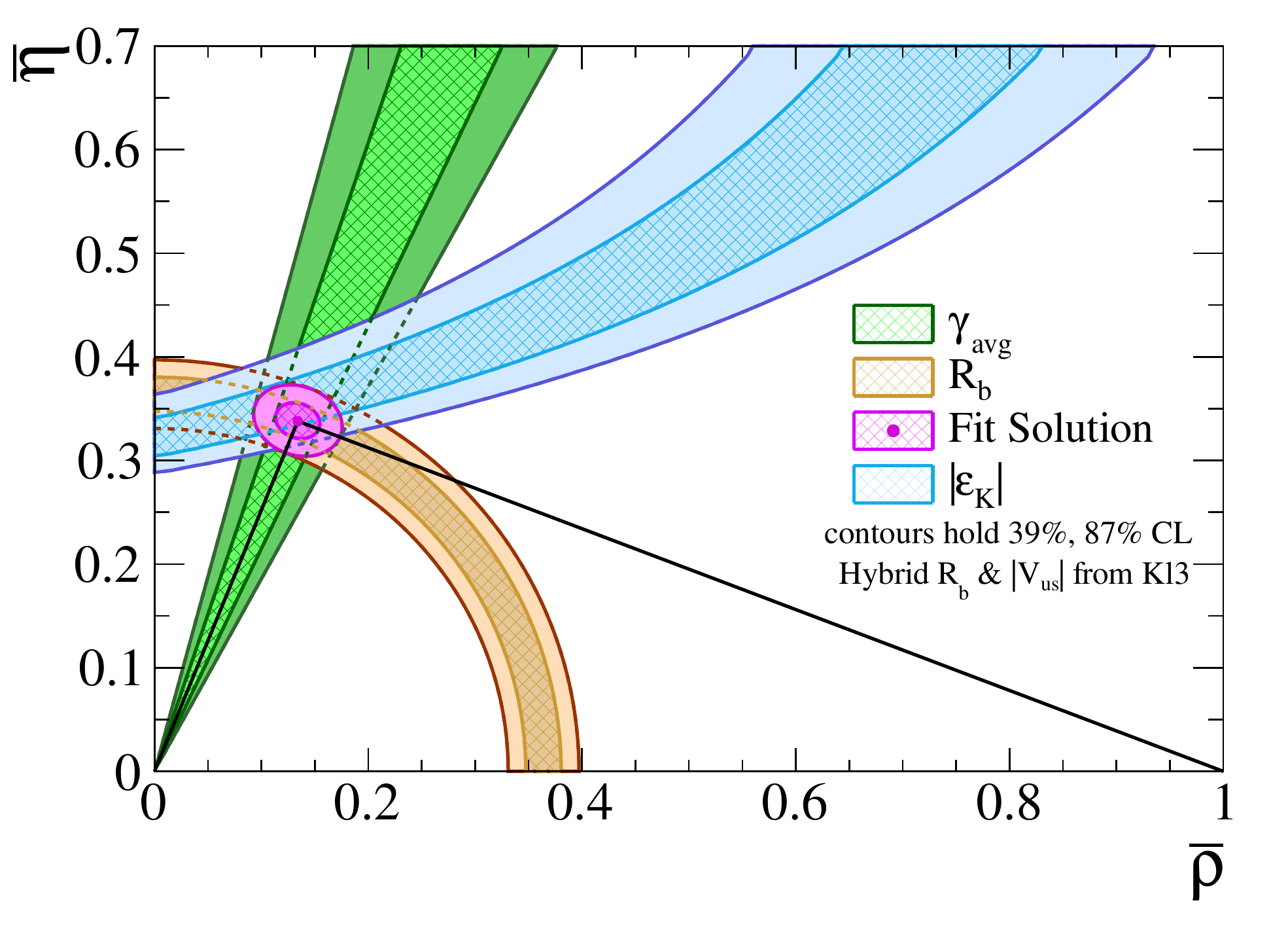}
    \caption{Determination of the coordinates $(\bar\rho,\bar\eta)$ of the UT apex from measurements of the angle $\gamma$ and the side $R_b$, with the two-dimensional confidence level contours from the GammaCombo fit overlaid.
    For comparison, the constraint from $|\varepsilon_K|$, discussed in Section \ref{sec:epsilonK}, is shown (but not included in the fit).
    The solutions based on the inclusive (Left), exclusive (Right) and hybrid (Bottom) determination of $R_b$ are shown separately.
    }
    \label{fig:UT_apex}
\end{figure}

\begin{figure}
    \centering
    \includegraphics[width=0.49\textwidth]{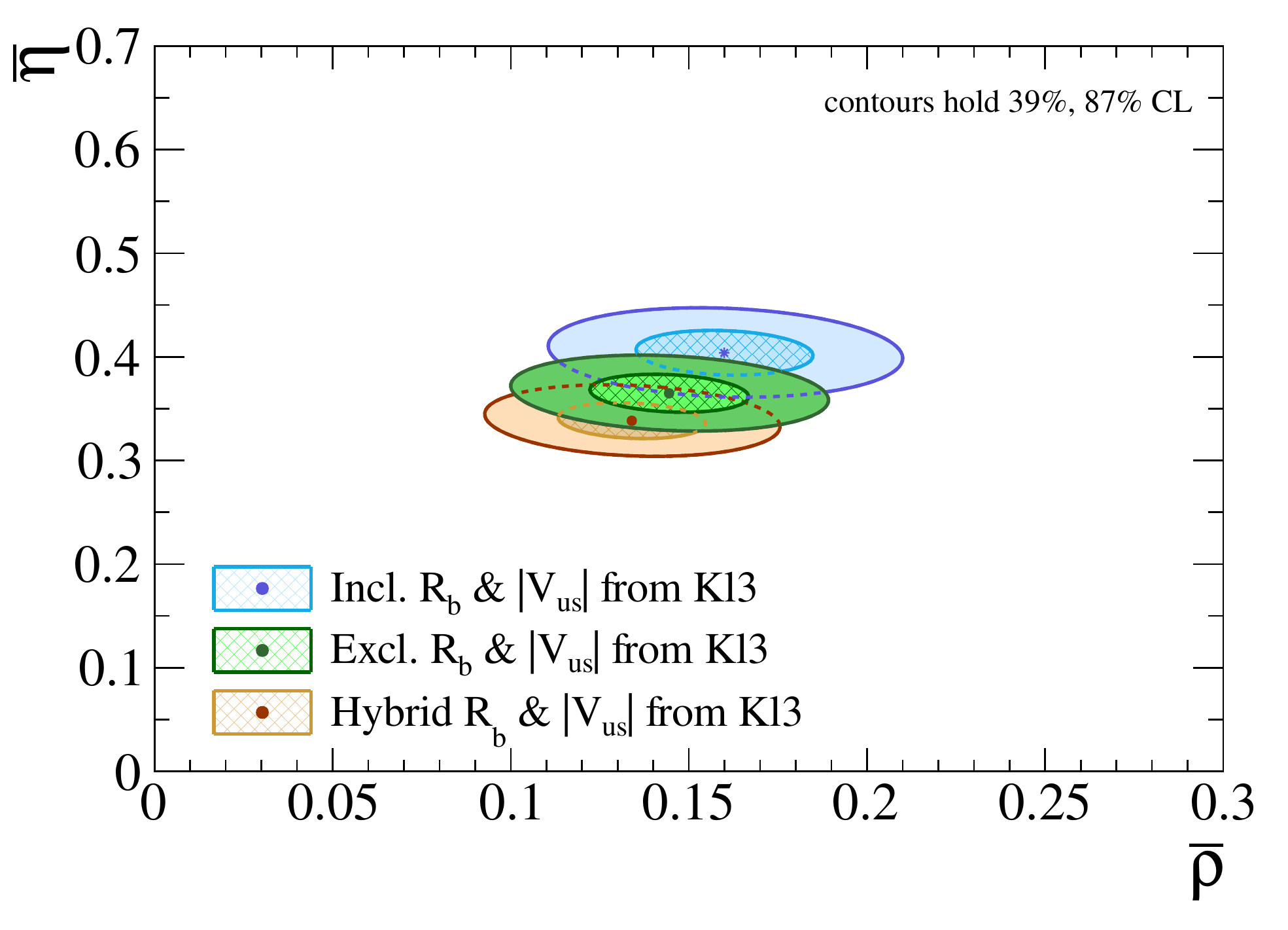}
    \includegraphics[width=0.49\textwidth]{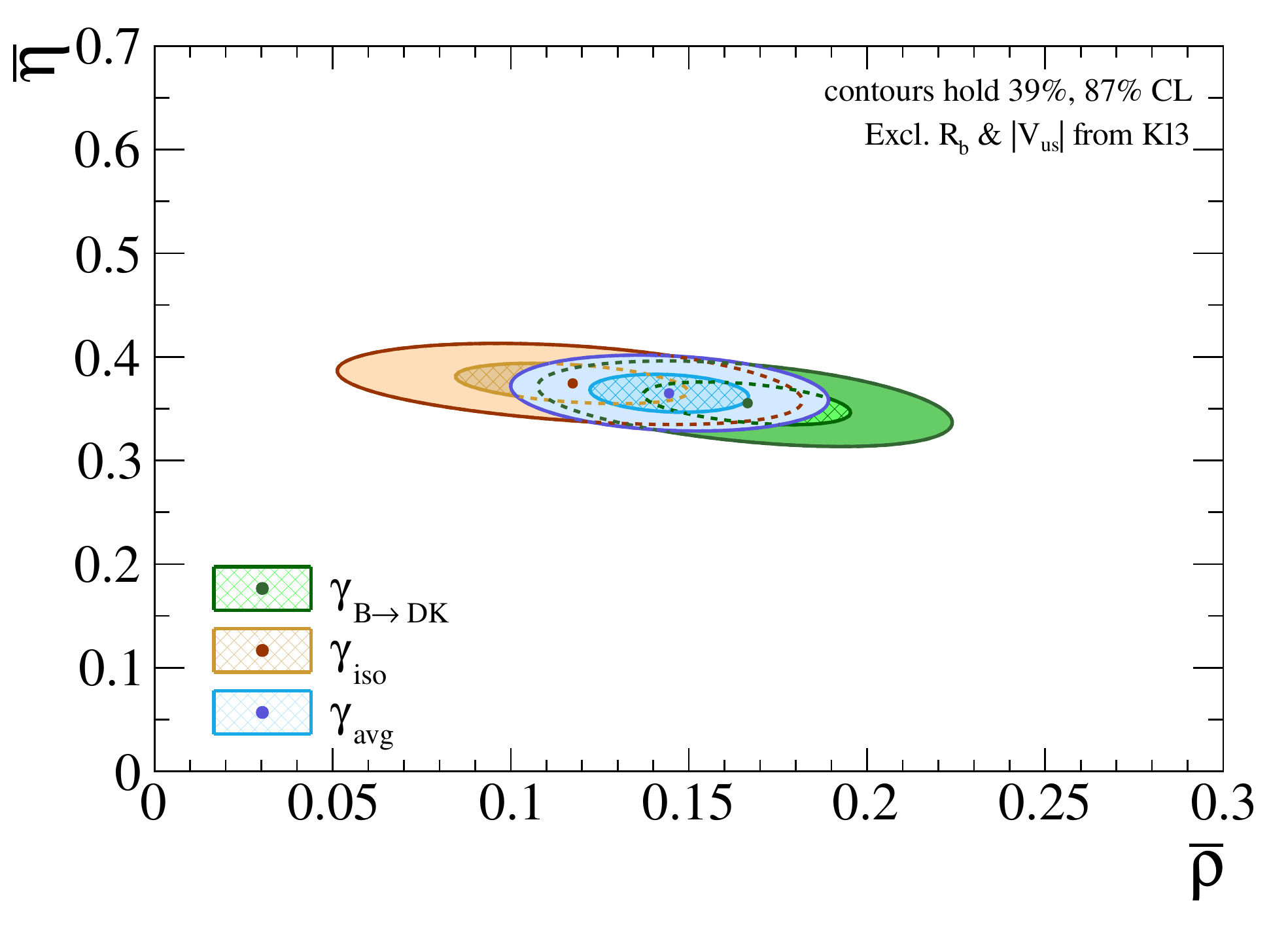}
    
    \includegraphics[width=0.49\textwidth]{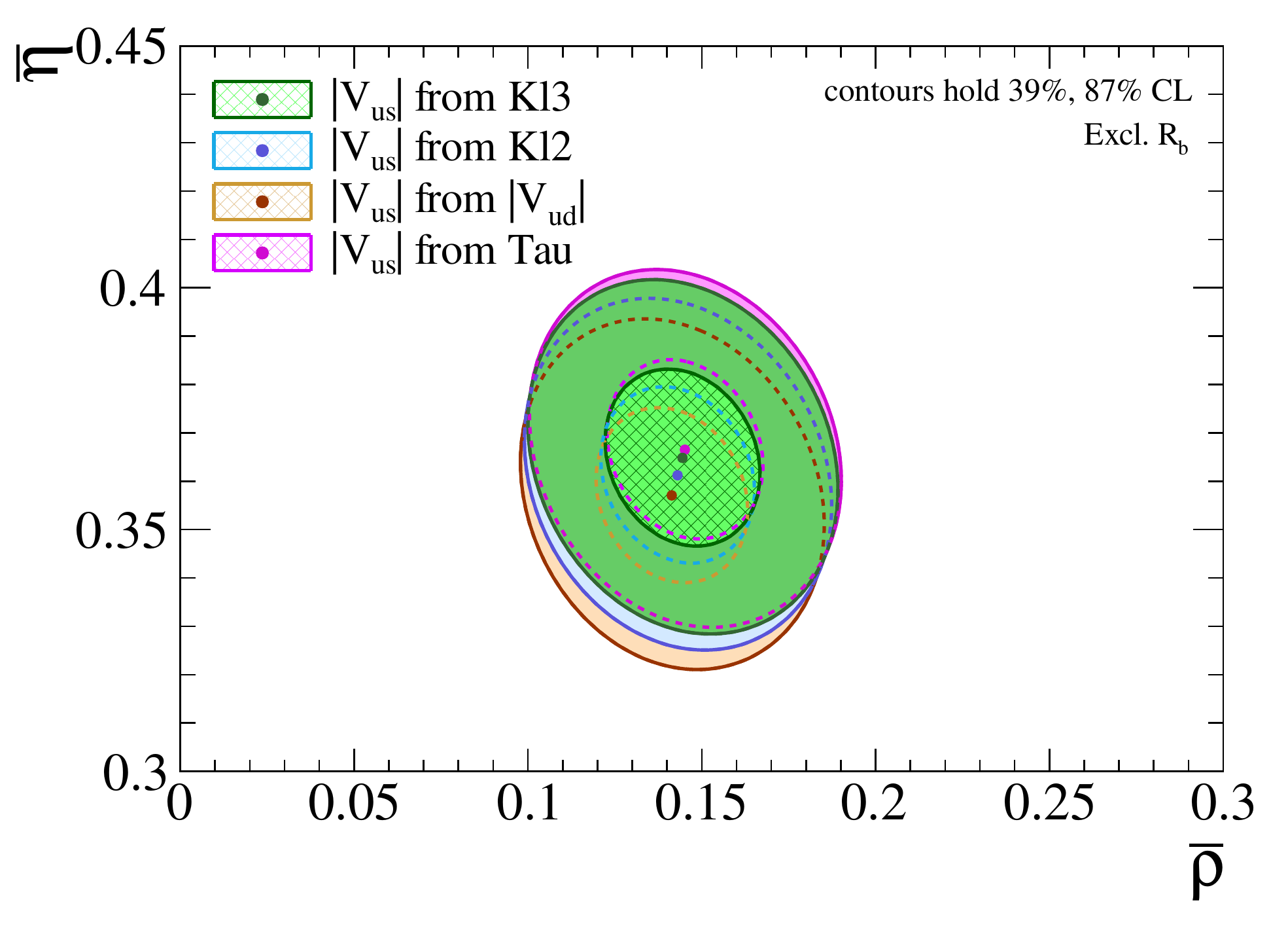}
    \caption{Two-dimensional confidence level contours for the coordinates $(\bar\rho,\bar\eta)$ of the UT apex.
    Left: Comparison between the inclusive, exclusive and hybrid scenarios.
    Right: Comparison between the different values for the angle $\gamma$ in the exclusive scenario.
    Bottom: Comparison between different values for $|V_{us}|$ in the exclusive scenario.
    }
    \label{fig:UT_apex_fit_comp}
\end{figure}

The coordinates $(\bar\rho,\bar\eta)$ of the UT apex are determined from a fit to $R_b$ and $\gamma$, based on the numerical inputs summarised in Table \ref{tab:input_UT_apex}.
The fitting procedure is implemented using the GammaCombo framework \cite{Aaij:2016kjh} and employs the frequentist method to determine the confidence interval of the fitted parameters.
It gives the following results
\begin{align}
    \text{Incl, } K\ell3 & &
    \bar\rho & = 0.160 \pm 0.025 \:, & 
    \bar\eta & = 0.404 \pm 0.022\:, \label{eq:UT_apex_I3}\\
    \text{Excl, } K\ell3 & &
    \bar\rho & = 0.144 \pm 0.022 \:, & 
    \bar\eta & = 0.365 \pm 0.018\:, \label{eq:UT_apex_E3}\\
    \text{Hybrid, } K\ell3 & &
    \bar\rho & = 0.134 \pm 0.021 \:, & 
    \bar\eta & = 0.338 \pm 0.017\:,\label{eq:UT_apex_H3}
\end{align}
with the associated two-dimensional confidence level contours shown in Fig.\ \ref{fig:UT_apex}.
The three solutions for the UT apex are compared with one other in Fig.\ \ref{fig:UT_apex_fit_comp}.

The impact of the different measurements of $|V_{us}|$ and $\gamma$ on the UT apex is illustrated in Fig.\ \ref{fig:UT_apex_fit_comp}, taking the exclusive scenario as an example.
At present, all results agree within uncertainties, but with improved precision, this may no longer be the case in the future.
We will explore this further in Section \ref{sec:future}.

\begin{table}
    \centering
    \begin{tabular}{||c|c|c|c|c||}
    \hline
    \hline
         & Inclusive & Exclusive & Hybrid & Reference  \\
    \hline
    $\alpha$ & \multicolumn{3}{c|}{$(85.2_{-4.3}^{+4.8})^{\circ}$} & \cite{HFLAV:2022pwe}\\
    $\phi_d$ & \multicolumn{3}{c|}{$\left(44.4_{-1.5}^{+1.6}\right)^{\circ}$} & \cite{Barel:2020jvf,Barel:2022wfr} \\
    $\gamma_{B \to DK}$ & \multicolumn{3}{c|}{$(64.9_{-4.5}^{+ 3.9})^\circ$} & \cite{LHCb:2021dcr} \\
    $\gamma_{\text{iso}}$ & \multicolumn{3}{c|}{$(72.6_{-4.9}^{+4.3})^\circ$} &  - \\
    $\gamma_{\text{avg}}$ & \multicolumn{3}{c|}{$(68.4 \pm 3.3)^\circ$} & - \\
    \hline
    $|V_{us}|$ & \multicolumn{3}{c|}{$0.22309 \pm 0.00056$} & \cite{Workman:2022ynf} \\
    $|V_{ub}|\times 10^{3}$ & $4.19 \pm 0.17$ & $3.51 \pm 0.12$ & $3.51 \pm 0.12$ & \cite{HFLAV:2022pwe} \\
    $|V_{cb}|\times 10^{3}$ & $42.16 \pm 0.50$ & $39.10 \pm 0.50$ & $42.16 \pm 0.50$ & \cite{Bordone:2021oof,HFLAV:2022pwe} \\
    $R_b$ & $ 0.434 \pm 0.018$ & $0.392 \pm 0.014$ & $0.364 \pm 0.013$ & - \\
    \hline
    $\bar{\rho}$ & $0.160 \pm 0.025$ & $0.144 \pm 0.022$ & $0.134 \pm 0.021$ & -  \\
    $\bar{\eta}$ & $0.404 \pm 0.022$ & $0.365 \pm 0.018$ & $0.338 \pm 0.017$ & - \\
    \hline
    \hline
    \end{tabular}
    \caption{Input parameters and results for the determination of the UT apex, split between the inclusive, exclusive and hybrid determinations of $R_b$.
    The dashes in the reference column indicate derived quantities that were computed in this paper.
    }
    \label{tab:input_UT_apex}
\end{table}
%
%
%
\subsection[Comparison with epsilonK]{Comparison with $|\varepsilon_K|$}\label{sec:epsilonK}
The apex of the UT is not only constrained by information from $B$-meson physics, but also by other flavour processes.
CP violation in the neutral kaon system constrains the UT apex through the observable $|\varepsilon_K|$, which depends on both the square and fourth power of $|V_{cb}|$ and is thus highly sensitive to its numerical value.
With sufficient precision, $|\varepsilon_K|$ could therefore be an excellent observable to explore the tension between the inclusive and exclusive determinations of $|V_{cb}|$, as it might favour one of the two.
If the tension between both determinations is resolved in the future, a discrepancy between the constraint coming from $|\varepsilon_K|$ and the fit results for the UT apex could point to NP in neutral kaon mixing.

In the SM, the dependence of $|\varepsilon_K|$ on the UT apex coordinates $\bar\rho$ and $\bar\eta$ is given by \cite{Brod:2019rzc}: 
\begin{equation}\label{eq:epsilonK}
    |\varepsilon_K| = \frac{G_{\text{F}}^2 m_W^2 m_K f_K^2}{6\sqrt{2}\pi^2 \Delta m_K}
    \kappa_{\varepsilon} \hat B_K |V_{cb}|^2\lambda^2\bar\eta
    \left[|V_{cb}|^2 (1 - \bar\rho)\, \eta_{tt}^{\text{EW}}\eta_{tt}\,\mathcal{S}(x_t) - \eta_{ut}\,\mathcal{S}(x_c, x_t)\right]\:,
\end{equation}
where $G_{\text{F}}$ is the Fermi constant, $m_W$ and $m_K$ are the $W$ and kaon mass, respectively, $f_K$ is the kaon decay constant, $\Delta m_K$ is the mass difference between the $K^0_{\text{S}}$ and $K^0_{\text{L}}$ mass eigenstates, and $\hat B_K$ is the kaon bag parameter.
The parameter $\kappa_{\varepsilon}$ is a multiplicative correction factor \cite{Buras:2010pza} due to long-distance contributions that are not included in $\hat B_K$.
The functions
\begin{equation}
    \mathcal{S}(x_t) = S_0(x_t) + S_0(x_c) - 2 S_0(x_c, x_t)\:,
    \qquad
    \mathcal{S}(x_c, x_t) = S_0(x_c) - S_0(x_c, x_t)
\end{equation}
are combinations of the following two Inami--Lim functions \cite{Inami:1980fz}:
\begin{align}\label{eq:inamilim}
    S_0(x_i) = & x_i \left[ \frac{1}{4} + \frac{9}{4(1-x_i)} - \frac{3}{2(1-x_i)^2} - \frac{3 x_i^2 \ln{x_i}}{2 (1-x_i)^3} \right] \:,\\
    S_0(x_i,x_j) = & \frac{x_i x_j}{x_i-x_j} \left[ \frac{1}{4} + \frac{3}{2(1-x_i)} - \frac{3}{4(1-x_i)^2} \right] \ln{x_i} \nonumber \\
    & + \frac{x_j x_i}{x_j-x_i} \left[ \frac{1}{4} + \frac{3}{2(1-x_j)} - \frac{3}{4(1-x_j)^2} \right] \ln{x_j}  - \frac{3 x_i x_j}{4(1-x_i)(1-x_j)} \:,
\end{align}
where
\begin{equation}
    x_i \equiv \left[\frac{\bar m_i(\bar m_i)}{m_W}\right]^2 \qquad \text{with }i = c,t\:, 
\end{equation}
and $\bar m_i(\bar m_i)$ is the mass of quark $i$ in the $\overline{\text{MS}}$ scheme.
For the top quark, we use the RunDec tool \cite{Chetyrkin:2000yt,Schmidt:2012az} to convert the latest experimental average of the pole mass \cite{Workman:2022ynf} into
\begin{equation}
    \bar m_t(\bar m_t) = (162.19 \pm 0.30) \: \text{GeV}\:,
\end{equation}
ignoring light-quark mass effects and using additional experimental input for the $Z$ boson mass $m_Z$ and the QCD coupling constant $\alpha_s(m_Z)$.
Finally, $\eta_{tt}^{\text{EW}}$, $\eta_{tt}$ and $\eta_{ut}$ are correction factors to the modified Inami--Lim functions $\mathcal{S}(x_t)$ and $\mathcal{S}(x_c,x_t)$ due to next-to-leading order (NLO) electroweak and QCD contributions \cite{Brod:2019rzc,Brod:2021qvc}.
The factors $\eta_{tt}$ and $\eta_{ut}$ parametrise the QCD corrections and are known to next-to-leading-logarithmic (NLL) and next-next-to-leading-logarithmic (NNLL) precision, respectively \cite{Brod:2019rzc}.
The factor $\eta_{tt}^{\text{EW}}$ estimates the impact of NLO electroweak corrections based on the first calculations of two-loop electroweak contributions to $\varepsilon_K$ \cite{Brod:2021qvc}.
All numerical values for the observables in Eq.\ \eqref{eq:epsilonK} are listed in Table \ref{tab:input_epsK}.

\begin{table}
    \centering
    \begin{tabular}{||c|c|c|c|c||}
    \hline
    \hline
    & \multicolumn{2}{c|}{Value} & Unit & Reference \\
    \hline
    $G_{\text{F}}$ & \multicolumn{2}{c|}{$1.1663787 \times 10^{-11}$} & MeV$^{-2}$ & \cite{Workman:2022ynf}\\
    $m_W$ & \multicolumn{2}{c|}{$80\,377 \pm 12$} & MeV & \cite{Workman:2022ynf}$^{\dagger}$\\
    $m_K$ & \multicolumn{2}{c|}{$497.611 \pm 0.013$} & MeV & \cite{Workman:2022ynf}\\
    $\bar m_c(\bar m_c)$ & \multicolumn{2}{c|}{$1.278 \pm 0.013$} & GeV & \cite{Aoki:2021kgd}\\
    \hline
    $m_t$ & \multicolumn{2}{c|}{$172.69 \pm 0.30$} & GeV & \cite{Workman:2022ynf}\\
    $m_Z$ & \multicolumn{2}{c|}{$91.1876 \pm 0.0021$} & GeV & \cite{Workman:2022ynf}\\
    $\alpha_s(m_Z)$ & \multicolumn{2}{c|}{$0.1179 \pm 0.0009$} & & \cite{Workman:2022ynf} \\
    $\bar m_t(\bar m_t)$ & \multicolumn{2}{c|}{$162.19 \pm 0.30$} & GeV & - \\
    \hline
    $\Delta m_K$ & \multicolumn{2}{c|}{$0.005289 \pm 0.000010$} & ps$^{-1}$ & \cite{Workman:2022ynf} \\
    $f_K$ & \multicolumn{2}{c|}{$155.7 \pm 0.3$} & MeV & \cite{Aoki:2021kgd} \\
    $\hat{B}_{K}$ & \multicolumn{2}{c|}{$0.7625 \pm 0.0097$} & & \cite{Aoki:2021kgd} \\
    $\kappa_{\varepsilon}$ & \multicolumn{2}{c|}{$0.94 \pm 0.02$} & & \cite{Buras:2010pza} \\
    \hline
    $\eta_{tt}^{\text{EW}}$ & \multicolumn{2}{c|}{$0.990 \pm 0.004$} & & \cite{Brod:2021qvc} \\
    $\eta_{tt}$ & \multicolumn{2}{c|}{$0.550 \pm 0.023$} & & \cite{Brod:2019rzc} \\
    $\eta_{ut}$ & \multicolumn{2}{c|}{$0.402 \pm 0.005$} & & \cite{Brod:2019rzc}$^{\ddagger}$ \\
    \hline
    \hline
    \multicolumn{5}{c}{~} \\
    \hline
    \hline
    & Inclusive \& Hybrid & Exclusive & Unit & Reference \\
    \hline
    $A$ & $(6.94 \pm 0.40)\times 10^{-3}$ & $(5.36 \pm 0.32)\times 10^{-3}$ & & - \\
    $B$ & $(5.11 \pm 0.35)\times 10^{-3}$ & $(3.78 \pm 0.27)\times 10^{-3}$ & & - \\
    \hline
    $|\varepsilon_{K}|_{\text{SM}}$ & $(2.54 \pm 0.22)\times 10^{-3}$ & $(1.74 \pm 0.15)\times 10^{-3}$ & & - \\
    $|\varepsilon_{K}|_{\text{exp}}$ & \multicolumn{2}{c|}{$(2.228 \pm 0.011)\times 10^{-3}$} & &  \cite{Workman:2022ynf} \\
    \hline
    \hline
    \end{tabular}
    \caption{Input parameters and results for the SM calculation of $|\varepsilon_K|$, split between the inclusive and exclusive determinations of $|V_{cb}|$.
    The dashes in the reference column indicate derived quantities computed in this paper.\protect\linebreak
    $^{\dagger}$ This average does not yet include the new measurement from the CDF II experiment \cite{CDF:2022hxs}, but the difference with the current world average has no significant impact on the numerical prediction for $|\varepsilon_K|_{\text{SM}}$.\protect\linebreak
    $^{\ddagger}$ Ref.\ \cite{Brod:2022har} has computed a correction factor for $\eta_{ut}$ of 1.005, which is not yet included here but has a negligible impact on $|\varepsilon_K|$.
    }
    \label{tab:input_epsK}
\end{table}

In the SM, the measured value of $|\varepsilon_K|$ describes a hyperbola
\begin{equation}
    \bar{\eta} = \frac{|\varepsilon_K|}{A - B \bar{\rho}}
\end{equation}
in the $\bar{\rho}\--\bar{\eta}$ plane.
Using our solutions for the UT apex in Eqs.\ \eqref{eq:UT_apex_I3}--\eqref{eq:UT_apex_H3}, we find the SM predictions
\begin{align}
    \text{Incl, } K\ell3 & &
    |\varepsilon_K|_{\text{SM}} & = (2.54 \pm 0.22)\times 10^{-3} \:, \label{eq:epsK_incl} \\
    \text{Excl, } K\ell3 & &
    |\varepsilon_K|_{\text{SM}} & = (1.74 \pm 0.15)\times 10^{-3} \:, \label{eq:epsK_excl}
\end{align}
where the inclusive value also covers the hybrid scenario, as $|\varepsilon_K|_{\text{SM}}$ only depends on $|V_{cb}|$.
The values of $A$ and $B$ can be found in Table \ref{tab:input_epsK}.
The differences between the SM predictions and the experimental result \cite{Workman:2022ynf}
\begin{equation}\label{eq:epsK_exp}
    |\varepsilon_K| = (2.228 \pm 0.011)\times 10^{-3}
\end{equation}
are 1.4 and 3.2 standard deviations for the inclusive/hybrid and exclusive scenarios, respectively.
This shows a preference for the inclusive value, but the uncertainty on the SM predictions is not yet sufficient to draw any further conclusions.

The same conclusion also arises from the comparison between the contours following from the measurement of $|\varepsilon_K|$ and the fit solution for the UT apex fixed through $\gamma$ and $R_b$, shown in Fig.\ \ref{fig:UT_apex}.
Besides finding the same small tensions as between Eqs.\ \eqref{eq:epsK_incl}, \eqref{eq:epsK_excl} and \eqref{eq:epsK_exp}, it is interesting that the contour for $|\varepsilon_K|$ from the exclusive determination lies completely above the fit solution from $\gamma$ and $R_b$, while the contour from the inclusive determination lies mainly below the corresponding fit solution.
The contour from the hybrid determination falls in between the inclusive and exclusive determinations, and overlaps with the fit solution.
Out of the three scenarios, it gives the most consistent picture of the UT apex.
This illustrates the strong dependence on the value of $|V_{cb}|$, which in the future could help to understand the puzzle between the inclusive and exclusive scenarios, assuming NP in kaon mixing can be controlled or ignored.

\section[B-Bbar Mixing in the Standard Model]{$B_q^0$--$\bar{B}_q^0$ Mixing in the Standard Model}\label{sec:BBmix}
In the SM, mixing between the neutral $B_q^0$ and $\bar B_q^0$ mesons originates from box diagrams.
Since these are loop processes and thus strongly suppressed, the mixing phenomenon is an excellent place for NP to enter. 
The determination of the parameter space of possible NP contributions to $B_q^0$--$\bar{B}_q^0$ mixing will rely on the results for the UT apex from Section \ref{sec:UTfit}, which are needed to compute the SM predictions of the mixing parameters characterising neutral $B_q$-meson mixing.
Comparing their SM predictions to the corresponding experimental values will allow us to constrain possible NP effects in $B_q^0$--$\bar{B}_q^0$ mixing, as we will show in Section \ref{sec:NPscenarios}.
\subsection{The Mixing Parameters in the Standard Model}\label{sec:BBmix_SM}
The SM prediction of the mixing angle $\phi_d$ is given by:
\begin{equation}
    \phi_{d}^{\text{SM}} = 2 \beta = 2 \text{arg} \left( - \frac{V_{cd} V_{cb}^{*}}{V_{td} V^{*}_{tb}} \right)= 2 \tan^{-1} \left(\frac{\bar{\eta}}{1 - \bar{\rho}}\right)\:.
\end{equation}
Using Eqs.\ \eqref{eq:UT_apex_I3}--\eqref{eq:UT_apex_H3}, we obtain the following results:
\begin{align}
    \text{Incl, } K\ell3 & &
    \phi_d^{\text{SM}} & = (51.4 \pm 2.8)^{\circ}\:, \label{eq:phidSM_I3} \\
    \text{Excl, } K\ell3 & &
    \phi_d^{\text{SM}} & = (46.2 \pm 2.3)^{\circ}\:, \label{eq:phidSM_E3} \\ 
    \text{Hybrid, } K\ell3 & &
    \phi_d^{\text{SM}} & = (42.6 \pm 2.2)^{\circ}\:. \label{eq:phidSM_H3}
\end{align}
These can be compared to the experimental measurement of $\phi_d$ that was already given in Eq.\ \eqref{eq:phid_JpsiK}.

The SM expression for $\phi_s$ is given by:
\begin{equation}\label{eq:phisSM}
    \phi_{s}^{\text{SM}} = -2 \delta \gamma = - 2 \lambda^2 \bar\eta + \mathcal{O}(\lambda^4)\:,
\end{equation}
which at this order in the Wolfenstein parametrisation \cite{Wolfenstein:1983yz, Buras:1994ec} can be expressed in terms of either $\eta$ or $\bar\eta$ interchangeably.
Note that in contrast to the expression for $\phi_d^{\text{SM}}$, the dependence on the UT apex is doubly Cabibbo-suppressed.
The numerical predictions based on Eqs.\ \eqref{eq:UT_apex_I3}--\eqref{eq:UT_apex_H3} are:
\begin{align}
    \text{Incl, } K\ell3 & &
    \phi_s^{\text{SM}} & = -0.0402 \pm 0.0022 = (-2.30 \pm 0.13)^{\circ}\:, \label{eq:phisSM_I3} \\
    \text{Excl, } K\ell3 & &
    \phi_s^{\text{SM}} & = -0.0363 \pm 0.0018 = (-2.08 \pm 0.10)^{\circ}\:, \label{eq:phisSM_E3} \\
    \text{Hybrid, } K\ell3 & &
    \phi_s^{\text{SM}} & = -0.0336 \pm 0.0017 = (-1.93 \pm 0.10)^{\circ}\:. \label{eq:phisSM_H3}
\end{align}
These SM predictions are a factor 2.5 to 3 less precise than the value
\cite{Charles:2015gya}
\begin{equation}
    \phi_s^{\text{SM}} = -0.03682_{-0.00060}^{+0.00086} = \left(-2.110_{-0.034}^{+0.049}\right)^{\circ}\:, 
\end{equation}
which was obtained from a global fit of the UT, and is commonly referenced in the literature. 
However, the global UT fits rely for their input on information from $B_q^0$--$\bar B_q^0$ mixing, without accounting for possible contributions from NP that may introduce biases.
We will therefore not use these results here and trade the loss in precision for a better control of the possible NP effects that could enter this determination.
The experimental measurement of $\phi_s$ is given in Eq.\ \eqref{eq:phis_JpsiPhi}.

The mass difference between the heavy and light mass eigenstates of the neutral $B_q$-meson system is given by
\begin{equation}
    \Delta m_q = 2 |M_{12}^q| + \mathcal{O}\left( \left| \frac{\Gamma_{12}^q}{M_{12}^q}\right|^2\right)\:,
\end{equation}
where $M_{12}^q - \frac{i}{2}\Gamma_{12}^q$ is the off-diagonal element of the effective Hamiltonian describing $B_q^0$--$\bar B_q^0$ mixing.
In the SM, the element $M_{12}^q$ is given as
\begin{equation}\label{eq:M12_SM}
    \left|M_{12}^q\right|_{\text{SM}} = \frac{G_{\text{F}}^2 m_W^2}{12 \pi^2} m_{B_q} \left|V_{tq}V_{tb}\right|^2 \: S_0(x_t)\: \eta_{2B}\hat B_{B_q} f_{B_q}^2\:,
\end{equation}
where $m_{B_q}$ is the $B_q$ mass, $f_{B_q}$ is the $B_q$ decay constant, $\hat B_{B_q}$ is the renormalisation group invariant bag parameter, $S_0 (x_t)$ is the Inami--Lim function given in Eq.\ \eqref{eq:inamilim}, and $\eta_{2B} = 0.551$ is a short-distance QCD correction factor \cite{Buras:1990fn,Buchalla:1995vs}.
The decay constants and bag parameters have been calculated using lattice QCD methods.
The latest computation using $N_f = 2 + 1 + 1$ quark flavours gives the following results \cite{Dowdall:2019bea,Aoki:2021kgd}:
\begin{equation}\label{eq:fBq}
    f_{B_d} \sqrt{\hat B_{B_d}} = (210.6 \pm 5.5)\:\text{MeV}\:, \qquad
    f_{B_s} \sqrt{\hat B_{B_s}} = (256.1 \pm 5.7)\:\text{MeV}\:.
\end{equation}
The uncertainty on $\eta_{2B}$ is $\mathcal{O}(1\%)$ \cite{Buchalla:1995vs}, and can therefore still be neglected with the current precision.
Eq.\ \eqref{eq:M12_SM} does not take into account next-to-leading-order electroweak effects, which have been calculated in Ref.\ \cite{Gambino:1998rt} to be at the level of 1\%.
Hence, they can also be omitted given the current precision.
The values of all parameters needed to compute $\Delta m_{q}^{\text{SM}}$ are summarised in Table \ref{tab:DMqinput}.

\begin{table}
    \centering
    \begin{tabular}{||c|c|c|c|c|c||}
    \hline
    \hline    
     & Inclusive & Exclusive & Hybrid & Unit & Reference\\
    \hline
    $m_{B_s}$ & \multicolumn{3}{c|}{$5366.92 \pm 0.10$} & MeV & \cite{Workman:2022ynf}\\
    $m_{B_d}$ & \multicolumn{3}{c|}{$5279.66 \pm 0.12$} & MeV &  \cite{Workman:2022ynf}\\
    $f_{B_d} \sqrt{\hat B_{B_d}}$ & \multicolumn{3}{c|}{$210.6 \pm 5.5$} & MeV & \cite{Dowdall:2019bea}\\
    $f_{B_s} \sqrt{\hat B_{B_s}}$ & \multicolumn{3}{c|}{$256.1 \pm 5.7$} & MeV & \cite{Dowdall:2019bea}\\
    $\xi$ & \multicolumn{3}{c|}{$1.212 \pm 0.016$} & & \cite{Dowdall:2019bea}\\
    \hline
    $\Delta m_d^{\text{SM}}$ & $0.513 \pm 0.040$ & $0.439 \pm 0.033$ & $0.510 \pm 0.037$ & $\text{ps}^{-1}$ & - \\
    $\Delta m_s^{\text{SM}}$ & $17.23 \pm 0.87$ & $14.80 \pm 0.76$ & $17.19 \pm 0.87$ & $\text{ps}^{-1}$ & - \\
    \hline
    $\Delta m_d$ & \multicolumn{3}{c|}{$0.5065 \pm 0.0019$} & $\text{ps}^{-1}$ & \cite{HFLAV:2022pwe}\\
    $\Delta m_s$ & \multicolumn{3}{c|}{$17.7656 \pm 0.0057$} & $\text{ps}^{-1}$ & \cite{HFLAV:2022pwe,Aaij:2021jky}\\
    \hline
    $\phi_d^{\text{SM}}$ & $51.4 \pm 2.8$ & $46.2 \pm 2.3$ & $42.6 \pm 2.2$ & Degrees & - \\
    $\phi_s^{\text{SM}}$ & $-2.30 \pm 0.13$ & $-2.08 \pm 0.10$ & $-1.93 \pm 0.10$ & Degrees & - \\
    \hline
    $\phi_d$ & \multicolumn{3}{c|}{$44.4_{-1.5}^{+1.6}$} & Degrees & \cite{Barel:2020jvf,Barel:2022wfr}\\
    $\phi_d$ & \multicolumn{3}{c|}{$-4.2 \pm 1.4$} & Degrees & \cite{Barel:2020jvf,Barel:2022wfr}\\
    \hline
    \hline
    \end{tabular}
    \caption{Input values and results for the determination of the $B_q^0$--$\bar B_q^0$ mixing parameters $\Delta m_q$ and $\phi_q$.
    The dashes in the reference column indicate derived quantities that were computed in this paper.}
    \label{tab:DMqinput}
\end{table}

Using the Wolfenstein parametrisation of the CKM matrix \cite{Wolfenstein:1983yz, Buras:1994ec}, we can rewrite the CKM matrix elements appearing in the SM expression for the mass differences $\Delta m_q$ in terms of the experimental inputs and the coordinates of the UT apex that were discussed in Section \ref{sec:UTapex}:
\begin{align}
    |V_{td}V_{tb}| & = \lambda |V_{cb}| \sqrt{(1-\bar\rho)^2 + \bar\eta\,^2} + \mathcal{O}\left(\lambda^7\right)\:, \label{eq:VtdVtb_SM} \\
    & = \lambda |V_{cb}| \sqrt{1 - 2R_b\cos\gamma + R_b^2} + \mathcal{O}\left(\lambda^7\right)\:, \\
    |V_{ts}V_{tb}|  
    & = \phantom{\lambda}|V_{cb}|\left[1-\frac{\lambda^2}{2} \left(1-2\bar\rho\right)\right] + \mathcal{O}\left(\lambda^6\right)\:, \label{eq:VtsVtb_SM} \\
    & = \phantom{\lambda}|V_{cb}|\left[1-\frac{\lambda^2}{2} \left(1-2 R_b \cos \gamma\right)\right] + \mathcal{O}\left(\lambda^6\right)\:.
\end{align}
We see that $|V_{td}V_{tb}|$, and thus $\Delta m_d^{\text{SM}}$, depends at leading order on $\bar{\rho}$ and $\bar{\eta}$, while for $|V_{ts}V_{tb}|$ and $\Delta m_s^{\text{SM}}$ the dependence on the UT apex only enters at next-to-leading order in $\lambda$.
In these expressions, the difference between $\rho$ and $\bar\rho$ is of the same order as the neglected terms.

\begin{figure}
    \centering
    \includegraphics[width=0.49\textwidth]{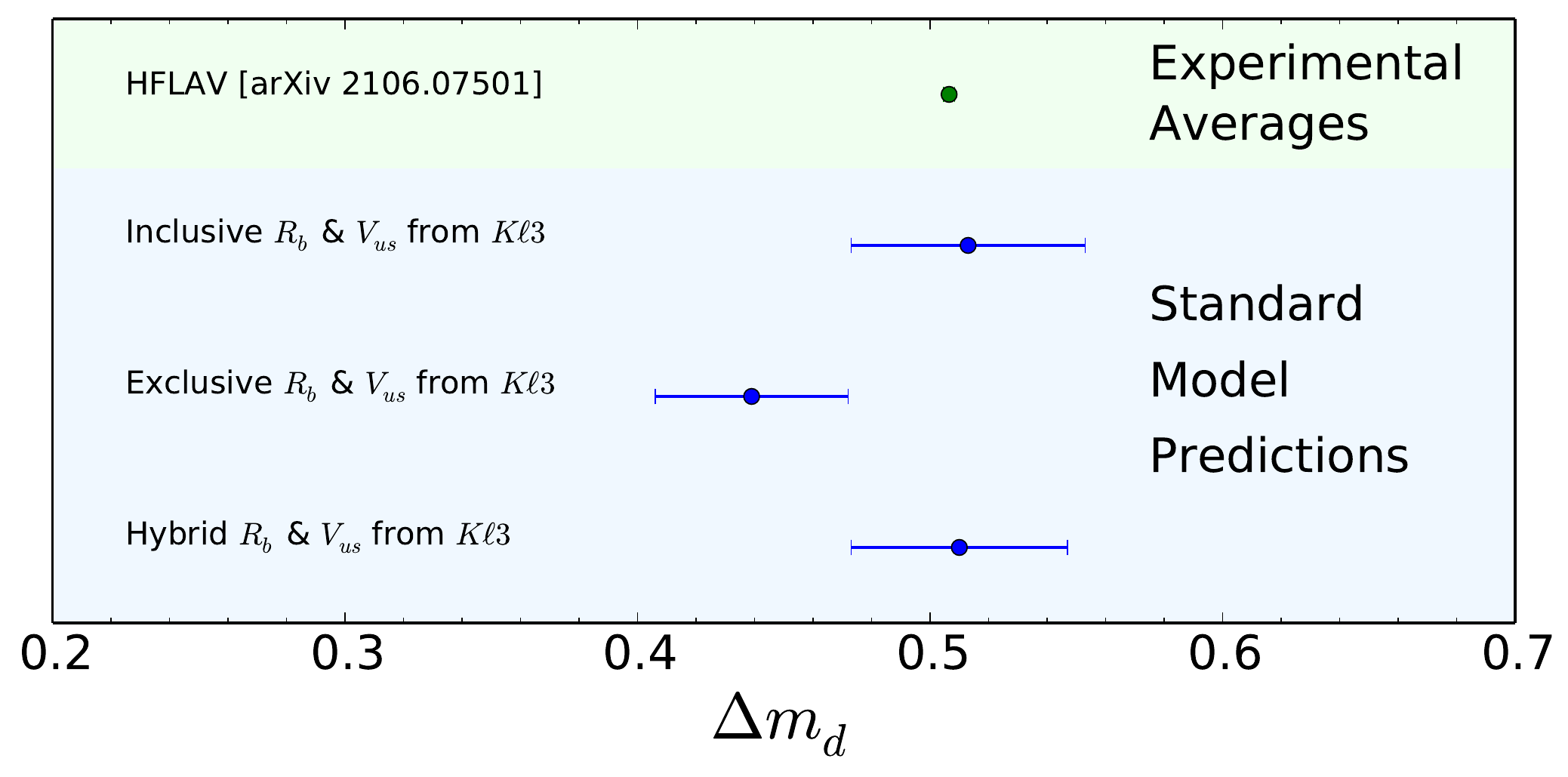}
    \hfill
    \includegraphics[width=0.49\textwidth]{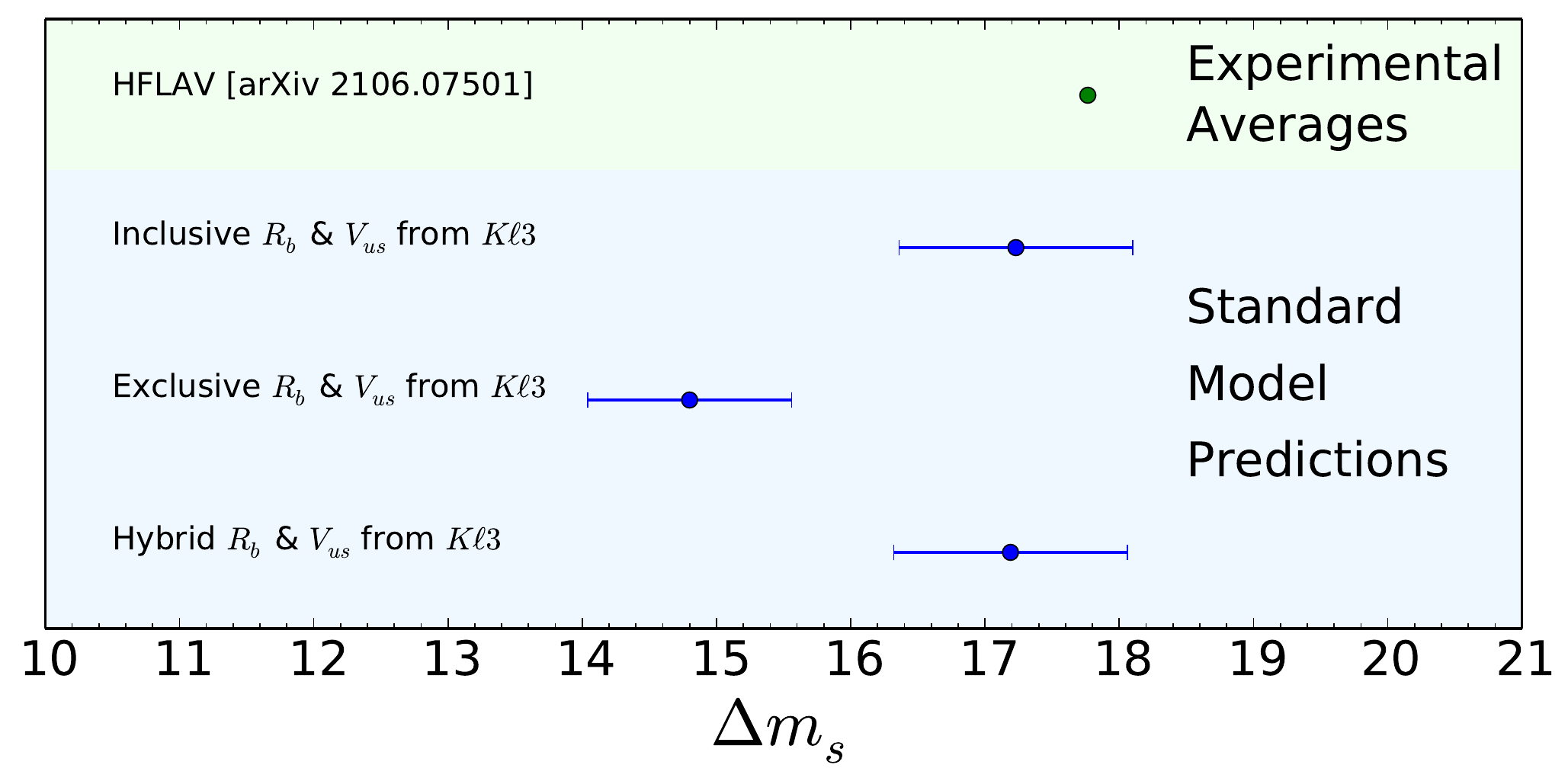}
    \caption{Comparison of the experimental averages and SM predictions for the $B_q^0$--$\bar B_q^0$ mixing parameters $\Delta m_d$ (Left) and $\Delta m_s$ (Right).}
    \label{fig:Dmq}
\end{figure}

We find the following predictions for the mass differences $\Delta m_q$:
\begin{align}
    \text{Incl, } K\ell3 & &
    \Delta m_d^{\text{SM}} & = (0.513 \pm 0.040)\:\text{ps}^{-1}\:, & 
    \Delta m_s^{\text{SM}} & = (17.23 \pm 0.87)\:\text{ps}^{-1}\:, \label{eq:Dmq_SM_I3} \\
    \text{Excl, } K\ell3 & &
    \Delta m_d^{\text{SM}} & = (0.439 \pm 0.033)\:\text{ps}^{-1}\:, &  
    \Delta m_s^{\text{SM}} & = (14.80 \pm 0.76)\:\text{ps}^{-1}\:, \\
    \text{Hybrid, } K\ell3 & &
    \Delta m_d^{\text{SM}} & = (0.510 \pm 0.037)\:\text{ps}^{-1}\:, & 
    \Delta m_s^{\text{SM}} & = (17.19 \pm 0.87)\:\text{ps}^{-1}\:.
\end{align}
These SM predictions can be compared with the experimental measurements \cite{HFLAV:2022pwe,Aaij:2021jky}:
\begin{align}
    \Delta m_d & = (0.5065 \pm 0.0019)\:\text{ps}^{-1}\:, \label{eq:Dmd_exp}\\
    \Delta m_s & = (17.7656 \pm 0.0057)\:\text{ps}^{-1} \label{eq:Dms_exp}\:,
\end{align}
which are already one to two orders of magnitude more precise.
Similar results have been obtained from a combined analysis of lattice and light-cone QCD sum rules (LQSR) results in Ref.\ \cite{DiLuzio:2019jyq}.
The experimental values and the SM predictions can be visually compared with each other as in Fig.\ \ref{fig:Dmq}, or numerically through the ratio
\begin{equation}
    \rho_q \equiv \frac{\Delta m_q }{\Delta m_{q}^{\text{SM}}}\:,
\end{equation}
which equals 1 in the SM. 
With the results in Eqs.\ \eqref{eq:Dmq_SM_I3}--\eqref{eq:Dms_exp}, we obtain:
\begin{align}
    \text{Incl, } K\ell3 & &
    \rho_d & = 0.989 \pm 0.078\:, & 
    \rho_s & = 1.031 \pm 0.052\:, \\
    \text{Excl, } K\ell3 & &
    \rho_d & = 1.153 \pm 0.088\:, & 
    \rho_s & = 1.200 \pm 0.062\:, \\
    \text{Hybrid, } K\ell3 & &
    \rho_d & = 0.993 \pm 0.073\:, & 
    \rho_s & = 1.034 \pm 0.052\:.
\end{align}
For the inclusive and hybrid scenarios, the ratios are compatible with 1, and thus the SM, while those from the exclusive scenario differ by (15--20)\%.
In the exclusive scenario, the SM predictions for $\Delta m_s$ and $\Delta m_d$ differ by respectively 3 and 2 standard deviations from their experimentally measured values.
The smaller central values for $\Delta m_s^{\text{SM}}$ and $\Delta m_d^{\text{SM}}$ in the exclusive scenario compared to the inclusive and hybrid scenarios, and hence the smaller discrepancies with the experimental measurements, are due to the value of $|V_{cb}|$, which shows the same pattern.

%
%
%
\subsection[Determining the UT Apex from Rb and Mixing]{Determining the UT Apex from $R_b$ and Mixing}\label{sec:Rt}
If we assume the SM expressions for the $B_q^0$--$\bar B_q^0$ mixing parameters $\Delta m_{d}$ and $\Delta m_{s}$, it is possible to determine the UT apex without having to rely on information from $\gamma$.
Although this scenario is not considered as the baseline, due to the strong assumption that needs to be made, it can be a useful alternative when discrepancies arise between the various measurements of $\gamma$.
It is therefore interesting to explore this option as well.

In the scenario without $\gamma$, the coordinates of the UT apex are fixed by the values of the sides $R_b$ and $R_t$.
The UT side $R_t$ is defined as
\begin{equation}\label{eq:Rt}
    R_t \equiv \left| \frac{V_{td}V_{tb}}{V_{cd}V_{cb}} \right| 
    = \sqrt{(1 - \bar\rho)^2 + \bar\eta\,^2} 
    = \frac{1}{\lambda}\left| \frac{V_{td}}{V_{ts}} \right| \left[1-\frac{\lambda^2}{2}\left(1-2\bar\rho\right)\right] + \mathcal{O}\left(\lambda^4\right)\:,
\end{equation}
where the ratio of CKM matrix elements
\begin{equation}\label{eq:VtdVts}
    \left| \frac{V_{td}}{V_{ts}} \right| = \xi \sqrt{\frac{m_{B_s} \Delta m_{d}^{\text{SM}}}{m_{B_d} \Delta m_{s}^{\text{SM}}}}
\end{equation}
can be determined from the $B_q^0$--$\bar B_q^0$ mixing parameters.
Here,
\begin{equation}
    \xi \equiv \frac{f_{B_s}\sqrt{\hat{B}_{B_s}}}{f_{B_d} \sqrt{\hat{B}_{B_d}}}
\end{equation}
is the ratio of bag parameters and decay constants of the $B_d$ and the $B_s$ systems.
This ratio has been calculated on the lattice, with the latest value given by \cite{Dowdall:2019bea,Aoki:2021kgd}:
\begin{equation}
    \xi = 1.212 \pm 0.016\:.
\end{equation}
Compared to the individual results for the bag parameters and decay constants in Table \ref{tab:DMqinput}, it is known with much higher precision as many uncertainties cancel in the ratio.
Combining this result with the experimental values of $\Delta m_d$ and $\Delta m_s$ in Eqs.\ \eqref{eq:Dmd_exp} and \eqref{eq:Dms_exp} we get
\begin{equation}
    \left| \frac{V_{td}}{V_{ts}} \right| = 0.2063 \pm 0.0004 \pm 0.0027\:,
\end{equation}
where the first uncertainty is due to the experimental measurements and the second due to the lattice input.

\begin{figure}
    \centering
    \includegraphics[width=0.49\textwidth]{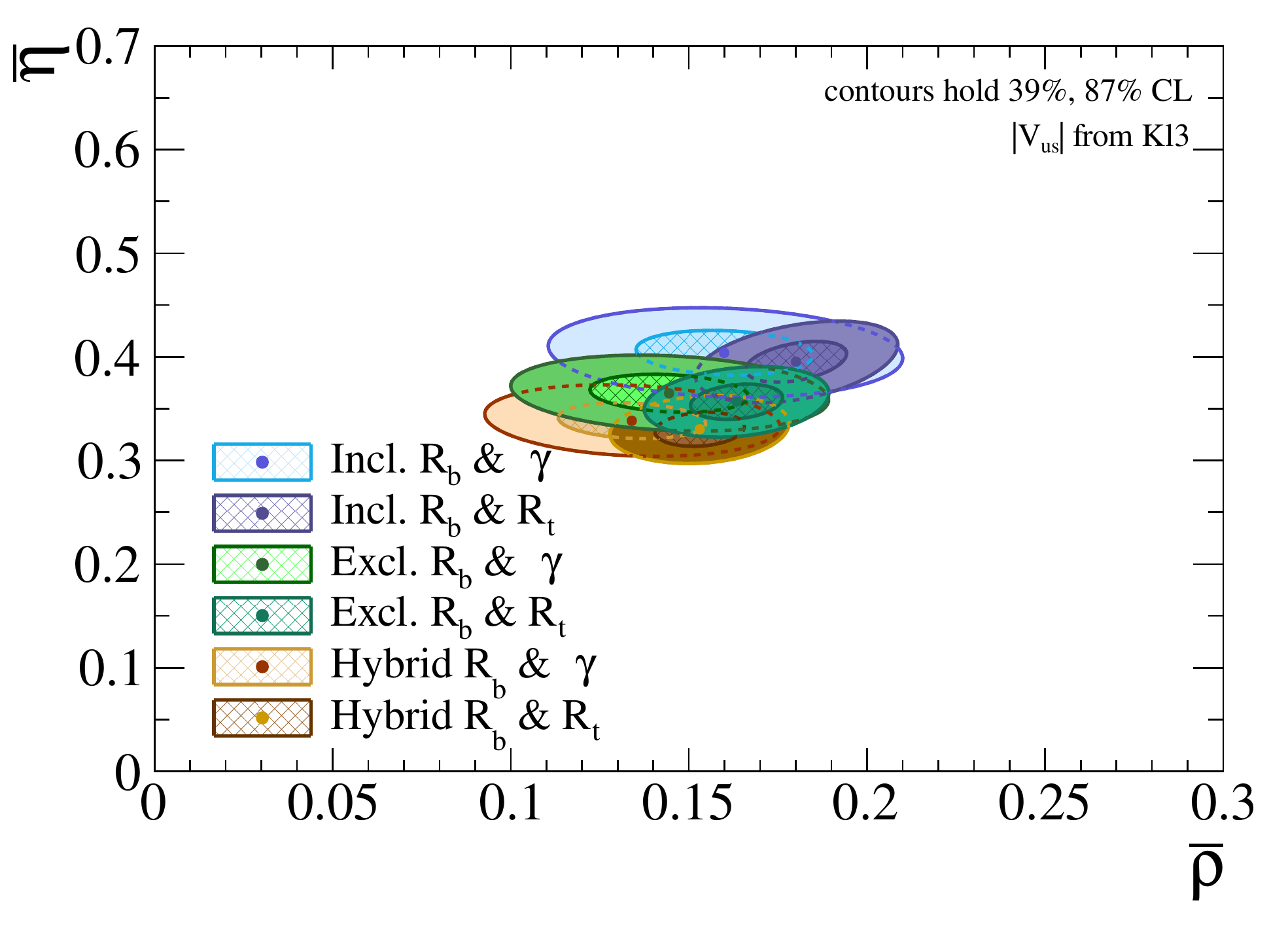}
    \caption{Comparison of the two-dimensional confidence regions for the UT apex $(\bar\rho,\bar\eta)$ determined through $\gamma$ and $R_b$, and through $R_b$ and $R_t$.}
    \label{fig:Rb_Rt_Apex}
\end{figure}

The coordinates of the UT apex, determined from a fit to the sides $R_b$ and $R_t$ are
\begin{align}
    \text{Incl, } K\ell3 & &
    \bar\rho & = 0.180 \pm 0.014 \:, & 
    \bar\eta & = 0.395 \pm 0.020\:, \label{eq:Rt_apex_I3} \\
    \text{Excl, } K\ell3 & &
    \bar\rho & = 0.163 \pm 0.013 \:, & 
    \bar\eta & = 0.357 \pm 0.017\:, \label{eq:Rt_apex_E3} \\
    \text{Hybrid, } K\ell3 & &
    \bar\rho & = 0.153 \pm 0.013 \:, & 
    \bar\eta & = 0.330 \pm 0.016\:, \label{eq:Rt_apex_H3}
\end{align}
with the two-dimensional confidence level contours shown in Fig.\ \ref{fig:Rb_Rt_Apex}.
Compared to the fit results based on $R_b$ and $\gamma$ in Eqs.\ \eqref{eq:UT_apex_I3}--\eqref{eq:UT_apex_H3}, we find a similar precision for $\bar\eta$, while the precision on $\bar\rho$ is a factor 2 better.
The solutions \eqref{eq:Rt_apex_I3}--\eqref{eq:Rt_apex_H3} correspond to the values
\begin{align}
    \text{Incl, } K\ell3 & & R_{t} & = 0.910 \pm 0.012\:, \label{eq:Rt_Dms_I3} \\
    \text{Excl, } K\ell3 & & R_{t} & = 0.909 \pm 0.012\:, \label{eq:Rt_Dms_E3} \\
    \text{Hybrid, } K\ell3 & & R_{t} & = 0.909 \pm 0.012\:, \label{eq:Rt_Dms_H3}
\end{align}
which can be compared to the values for $R_t$ calculated from Eqs.\ \eqref{eq:UT_apex_I3}--\eqref{eq:UT_apex_H3}:
\begin{align}
    \text{Incl, } K\ell3 & & R_t & = 0.932 \pm 0.024\:, \label{eq:Rt_classic_I3} \\
    \text{Excl, } K\ell3 & & R_t & = 0.930 \pm 0.021\:, \label{eq:Rt_classic_E3} \\
    \text{Hybrid, } K\ell3 & & R_t & = 0.930 \pm 0.021\:. \label{eq:Rt_classic_H3}
\end{align}
Also here the scenarios with $\gamma$ are a factor 2 less precise than the scenarios without $\gamma$.

\begin{figure}
    \centering
    \includegraphics[width=0.49\textwidth]{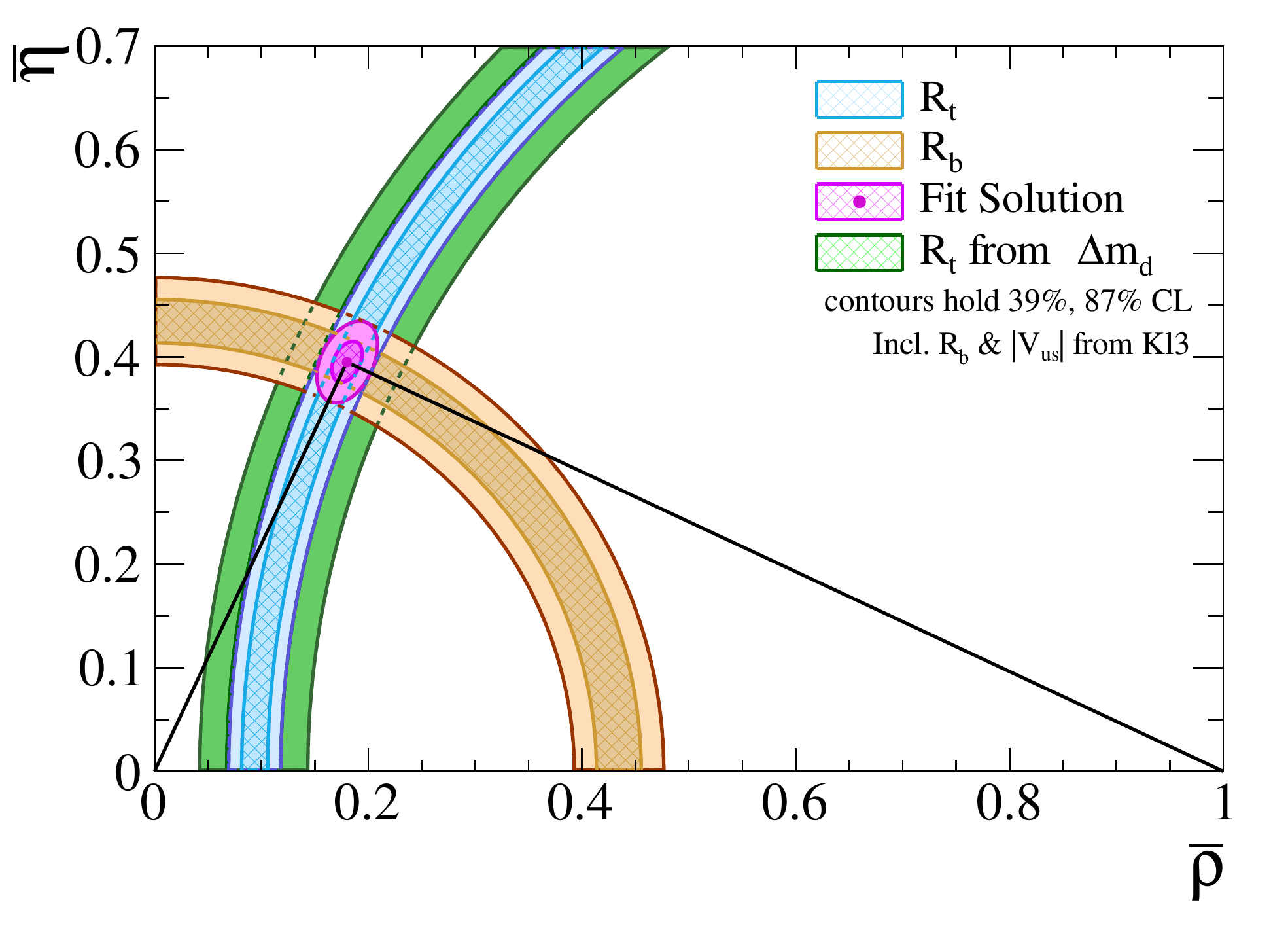}
    \includegraphics[width=0.49\textwidth]{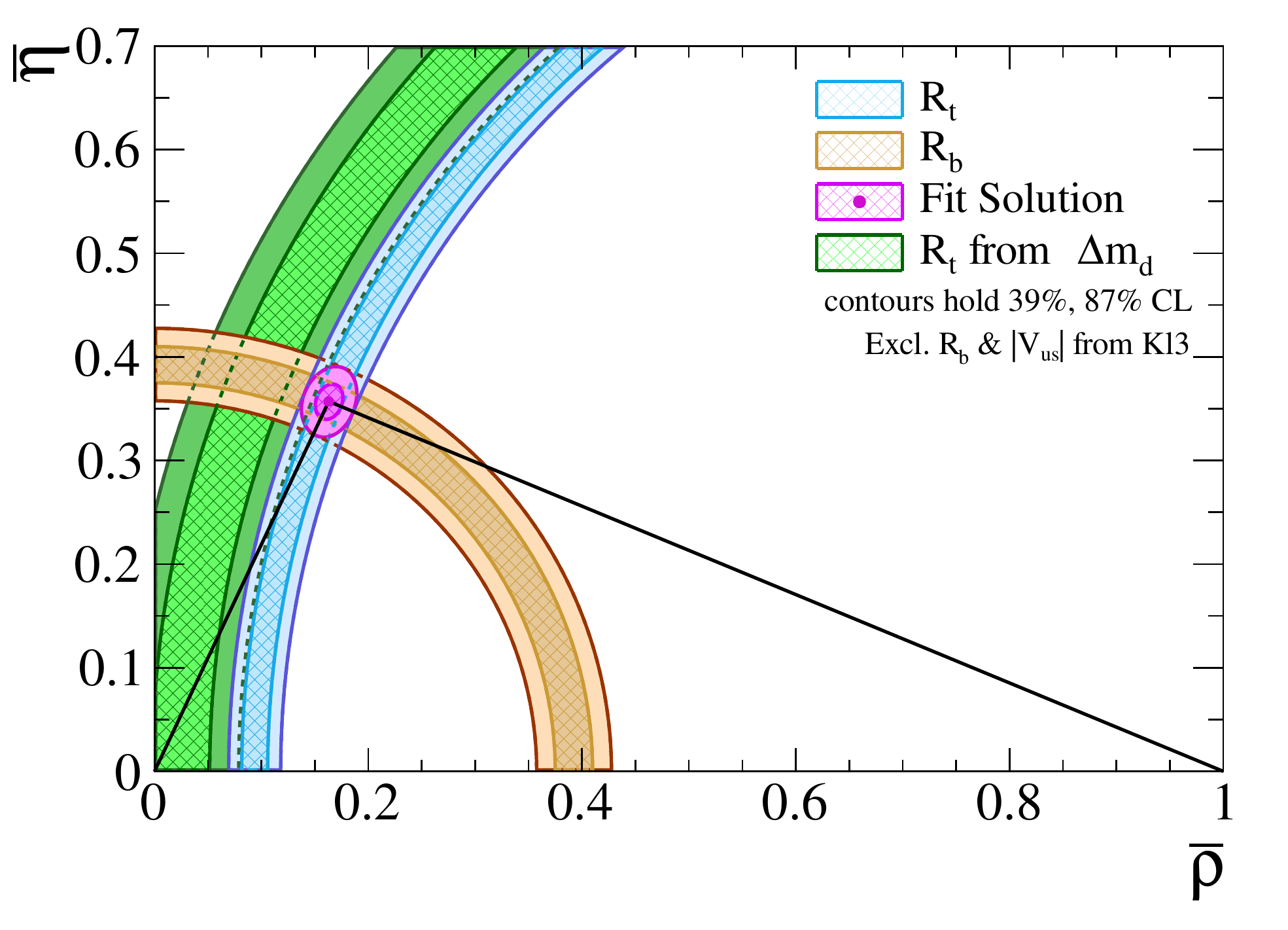}

    \includegraphics[width=0.49\textwidth]{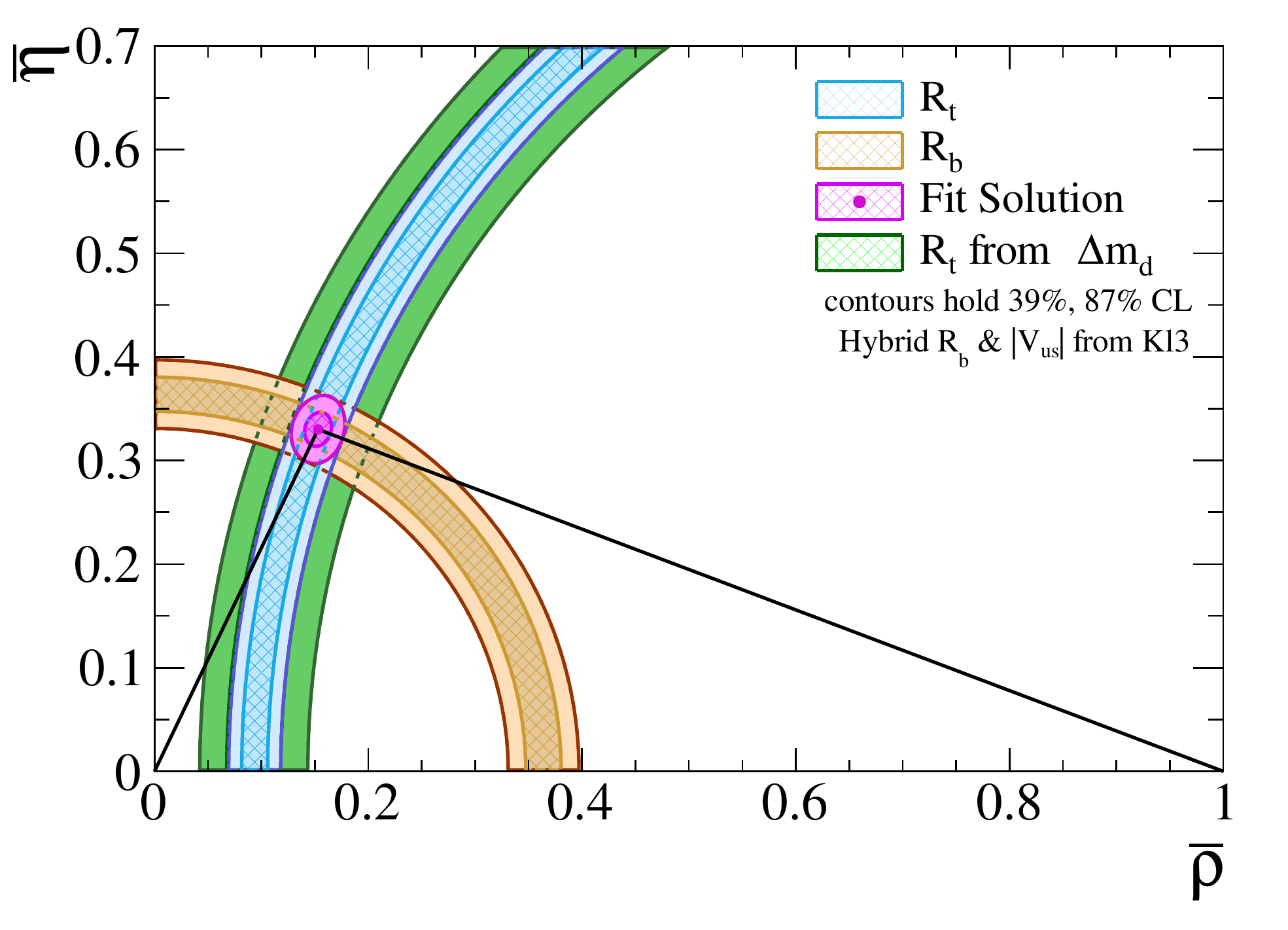}
    \caption{Determination of the UT apex $(\bar\rho,\bar\eta)$ from the measurements of the sides $R_b$ and $R_t$.
    For comparison, also the constraint from $\Delta m_d$ alone is shown (but not included in the fit).
    The solutions based on the inclusive (Left), exclusive (Right) and hybrid (Bottom) determination of $R_b$ are shown separately.}
    \label{fig:SM_Apex_noGam}
\end{figure} 

Alternatively, $R_t$ can also be determined without relying on information from the $B_s$ system, which has the advantage that its value will not be affected by NP in $B_s^0$--$\bar B_s^0$ mixing.
The comparison with the results in Eqs.\ \eqref{eq:Rt_Dms_I3}--\eqref{eq:Rt_Dms_H3} then theoretically provides another way to constrain NP effects in the $B_s$-meson system.
The disadvantage of this approach is that we cannot use the ratio $\xi$ of bag parameters and decay constants, which will result in larger associated uncertainties.
In this alternative method, $R_t$ is calculated from the definition in Eq.\ \eqref{eq:Rt}, using the inclusive or exclusive value of $|V_{cb}|$, and obtaining $|V_{td}V_{tb}|$ from the measurement of $\Delta m_d$ through Eq.\ \eqref{eq:M12_SM}.
The approach using only the measurement of $\Delta m_d$, and avoiding $\Delta m_s$, gives
\begin{align}
    \text{Incl, } K\ell3 & & R_{t} & = 0.926 \pm 0.027\:, \\
    \text{Excl, } K\ell3 & & R_{t} & = 0.999 \pm 0.029\:,
\end{align}
with the hybrid scenario matching the inclusive result as it does not depend on $|V_{ub}|$.
The corresponding contour is also shown in Fig.\ \ref{fig:SM_Apex_noGam}.
Given the current precision, the results for $R_t$ determined through only $\Delta m_d$ are not competitive with the results for $R_t$ determined through $\xi$. 
Hence, we will only use $R_t$ determined through $\xi$ in our analysis.
For the inclusive and hybrid scenarios the two determinations of $R_t$ overlap completely, while a slight discrepancy is visible for the exclusive scenario.
Combined with the relatively large errors on $R_t$ determined through $\Delta m_d$, it does not yet allow us to draw any conclusions regarding NP in $B_s$ mixing.

From the comparison in Fig.\ \ref{fig:Rb_Rt_Apex}, we can see that the determination of the apex from $R_b$ and $R_t$ is more precise than the one utilising $R_b$ and $\gamma$.
However, $R_t$ has been determined assuming SM expressions for $\Delta m_d$ and $\Delta m_s$, and hence ignores contributions from possible NP in $B_q^0$--$\bar B_q^0$ mixing. 
Possible NP will contaminate the determination of $R_t$, except in the special case of FUNP.
In such scenarios, possible NP in $B_q^0$--$\bar B_q^0$ mixing drops out in the ratio between $\Delta m_d$ and $\Delta m_s$. 
To determine NP in $B_q^0$--$\bar B_q^0$ mixing in a general scenario, we cannot use the above determination of $R_t$, but we need the UT apex determination of $R_b$ and $\gamma$ discussed in Section \ref{sec:UTfit}.
In the next Section we will investigate in more detail what impact the different determinations of the UT apex have on NP searches in $B_q^0$--$\bar{B}_q^0$ mixing.

\section[New Physics in B-Bbar Mixing]{New Physics in $B_q^0$--$\bar{B}_q^0$ Mixing}\label{sec:NPscenarios}
We can use the SM predictions for the mixing parameters $\Delta m_q$ and $\phi_q$, together with their experimental values, to constrain the parameter space of possible NP in $B_q^0$--$\bar{B}_q^0$ mixing.
We introduce the NP parameters $\kappa_q$ and $\sigma_q$ by modifying the mixing parameters in the same way as in Ref.\ \cite{Ball:2006xx}:
\begin{align}
    \Delta m_{q} & = \Delta m_{q}^{\text{SM}} \left|1 + \kappa_{q} e^{i\sigma_{q}}\right|\:, \label{eq:NP_mix_Dmq}\\
    \phi_{q} & = \phi_{q}^{\text{SM}} + \phi_{q}^{\text{NP}} = \phi_{q}^{\text{SM}}  + \arg\left(1 + \kappa_{q} e^{i\sigma_{q}}\right)\:. \label{eq:NP_mix_phiq}
\end{align}
Here, the size of the NP effects is described by $\kappa_q$, while $\sigma_q$ is a complex phase that accounts for additional CP-violating effects.
This parametrisation is model-independent; we are not making any assumptions regarding the origin of the NP. 
We will use Eqs.\ \eqref{eq:NP_mix_Dmq} and \eqref{eq:NP_mix_phiq} to explore two different NP scenarios, which are summarised in Table \ref{tab:twoNPscen}.

\begin{table}
    \centering
{\renewcommand{\arraystretch}{1.5}
    \begin{tabular}{||c|c|c|c||}
        \hline
        \hline
        & Scenario I & Scenario II\\
        \hline
        Assumptions & no NP in $\gamma$ & FUNP  \\
        \hline
        UT apex fit &  $R_b$ and $\gamma$ & $R_b$ and $R_t$ \\
        \hline
        NP parameters  & $(\kappa_d, \sigma_d)$ and $(\kappa_s, \sigma_s)$  & $(\kappa_d, \sigma_d) = (\kappa_s, \sigma_s)$  \\
        \hline
        \hline
    \end{tabular}
}
    \caption{Two different scenarios for which we will determine the NP parameters.}
    \label{tab:twoNPscen}
\end{table}

Scenario I represents the most general case, and has the least assumptions.
We utilise the UT apex determination in Section \ref{sec:UTfit} for the SM predictions of the mixing parameters $\Delta m_q$ and $\phi_q$, together with the numerical values given in Section \ref{sec:BBmix_SM}, to determine the NP parameters $(\kappa_d, \sigma_d)$ and $(\kappa_s, \sigma_s)$ independently from each other.

For the second scenario we explore other options that require additional assumptions.
In Scenario II we consider the case where NP contributions are equal in the $B_d$ and the $B_s$ systems, i.e.\ FUNP.
Such a scenario could occur if NP predominantly couples to the third generation quarks and leptons, and is explicitly realised in NP models with a $U(2)$ symmetry \cite{Barbieri:2011ci,Barbieri:2012uh,Barbieri:2012bh,Charles:2013aka}.
Note that FUNP is different from the well-known Minimal Flavour Violation scenario, which assumes no CP-violating NP phase at all, as was already noted in Ref.\ \cite{Ball:2006xx,Charles:2013aka}, where the same scenario was studied.
Assuming the FUNP framework, it is possible to employ the UT apex determination that only relies on $R_b$ and the mixing parameters described in Section \ref{sec:Rt}, without requiring additional information on $\gamma$. 
In this way, possible NP in $\gamma$ will not affect the results.
However, assuming that NP affects the $B_d$ and $B_s$ systems equally is a strong assumption. 
A comparison of the FUNP scenario with the general scenario provides a test of the FUNP assumption and gives a measure of the impact that these assumptions have on the constraints on the parameter space of NP in $B_q^0$--$\bar{B}_q^0$ mixing.

The last scenario we will discuss is an interpolation between Scenario I and II. 
We will assume FUNP to determine the UT apex, using only information from $R_b$ and the mixing parameters, but will relax this assumption when determining $(\kappa_d, \sigma_d)$ and $(\kappa_s, \sigma_s)$. 
This scenario has less stringent assumptions on the form of the NP than for the full FUNP scenario.
Comparing it with Scenario I and II allows us both to estimate the impact of the FUNP assumption on the fit to the NP parameters, and to explore the impact of the different UT apex determinations.
\subsection{Scenario I: General New Physics}
\begin{figure}
    \centering
    \includegraphics[width=0.49\textwidth]{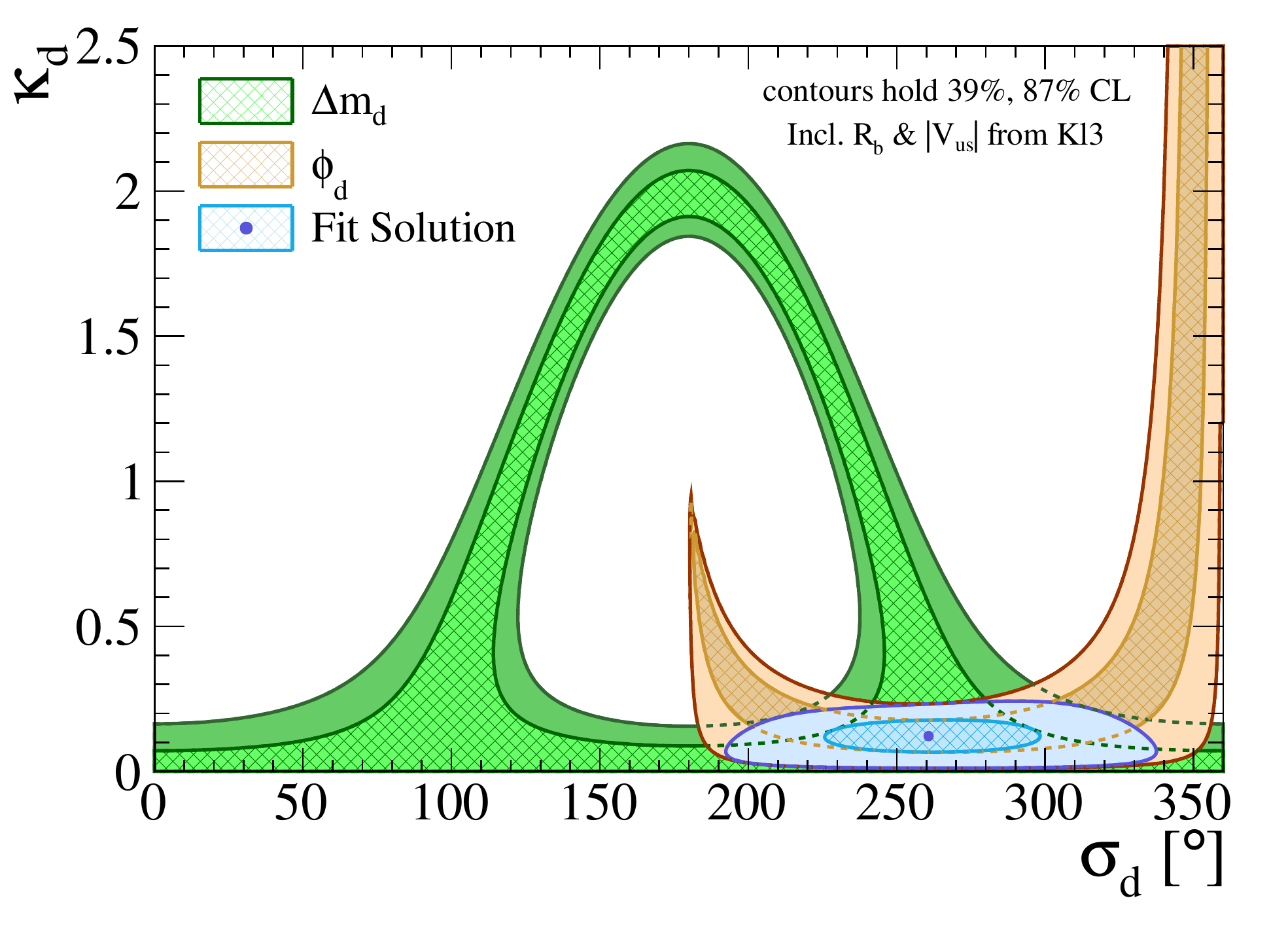}
    \hfill
    \includegraphics[width=0.49\textwidth]{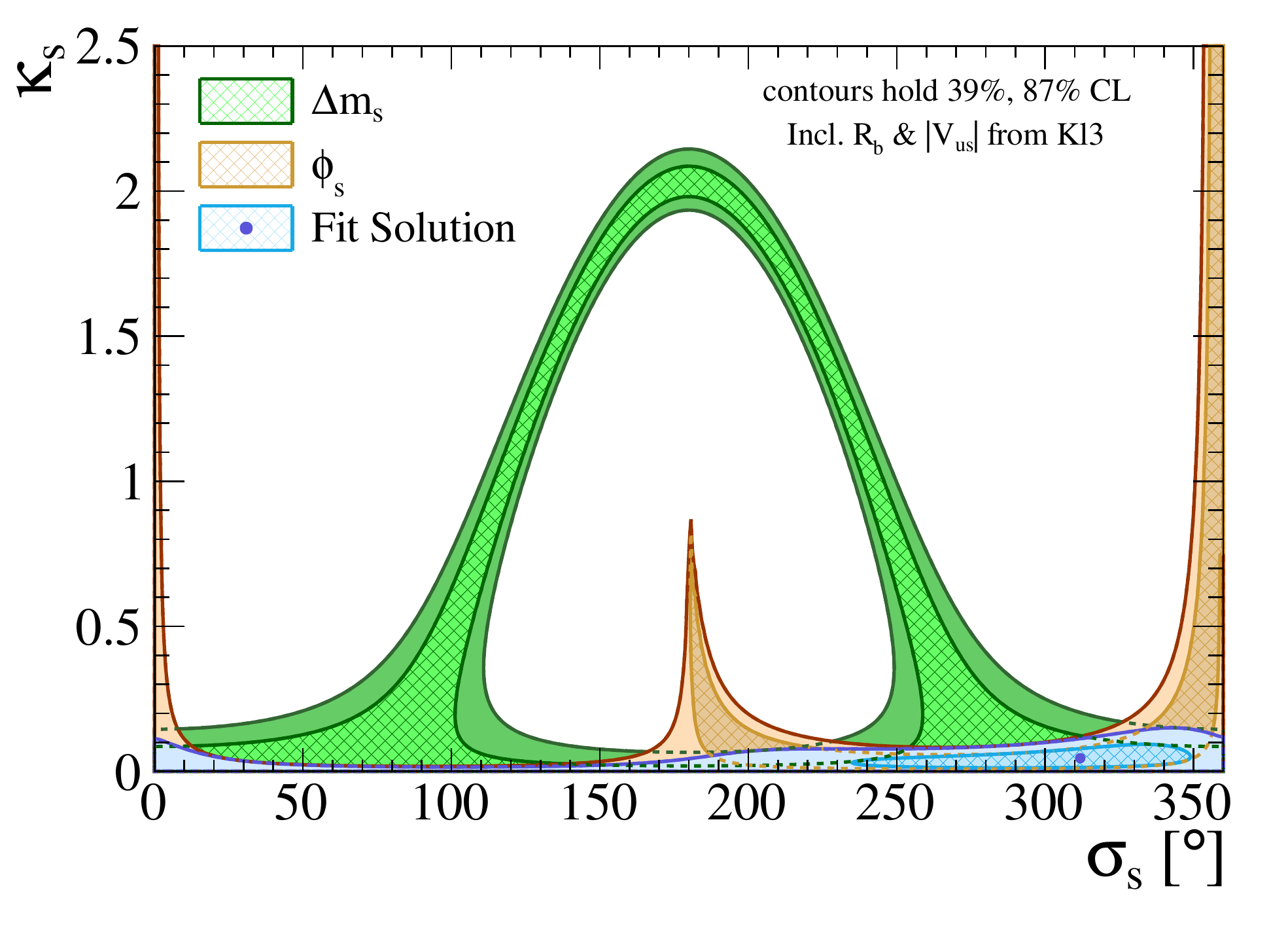}
    
    \includegraphics[width=0.49\textwidth]{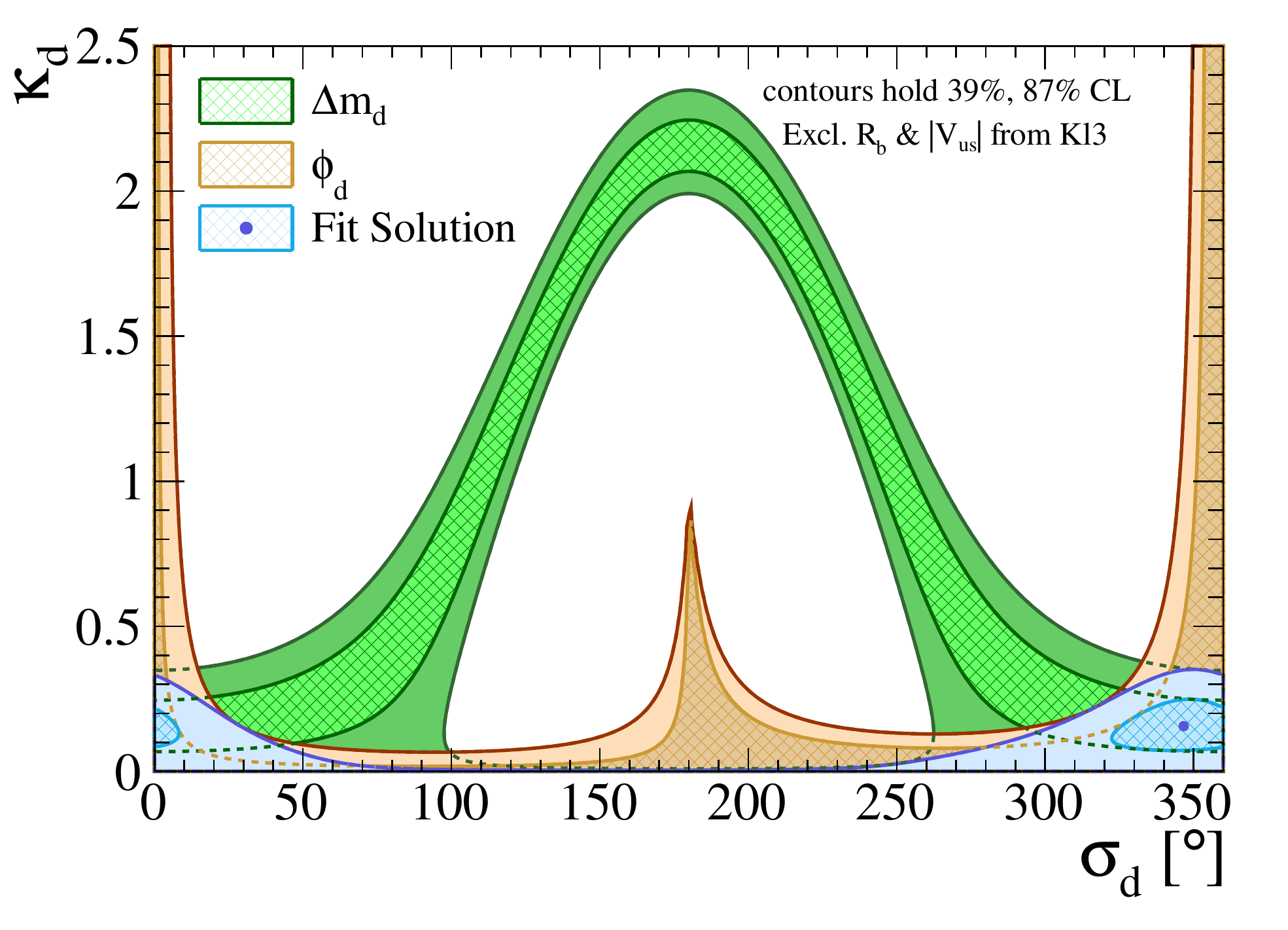}
    \hfill
    \includegraphics[width=0.49\textwidth]{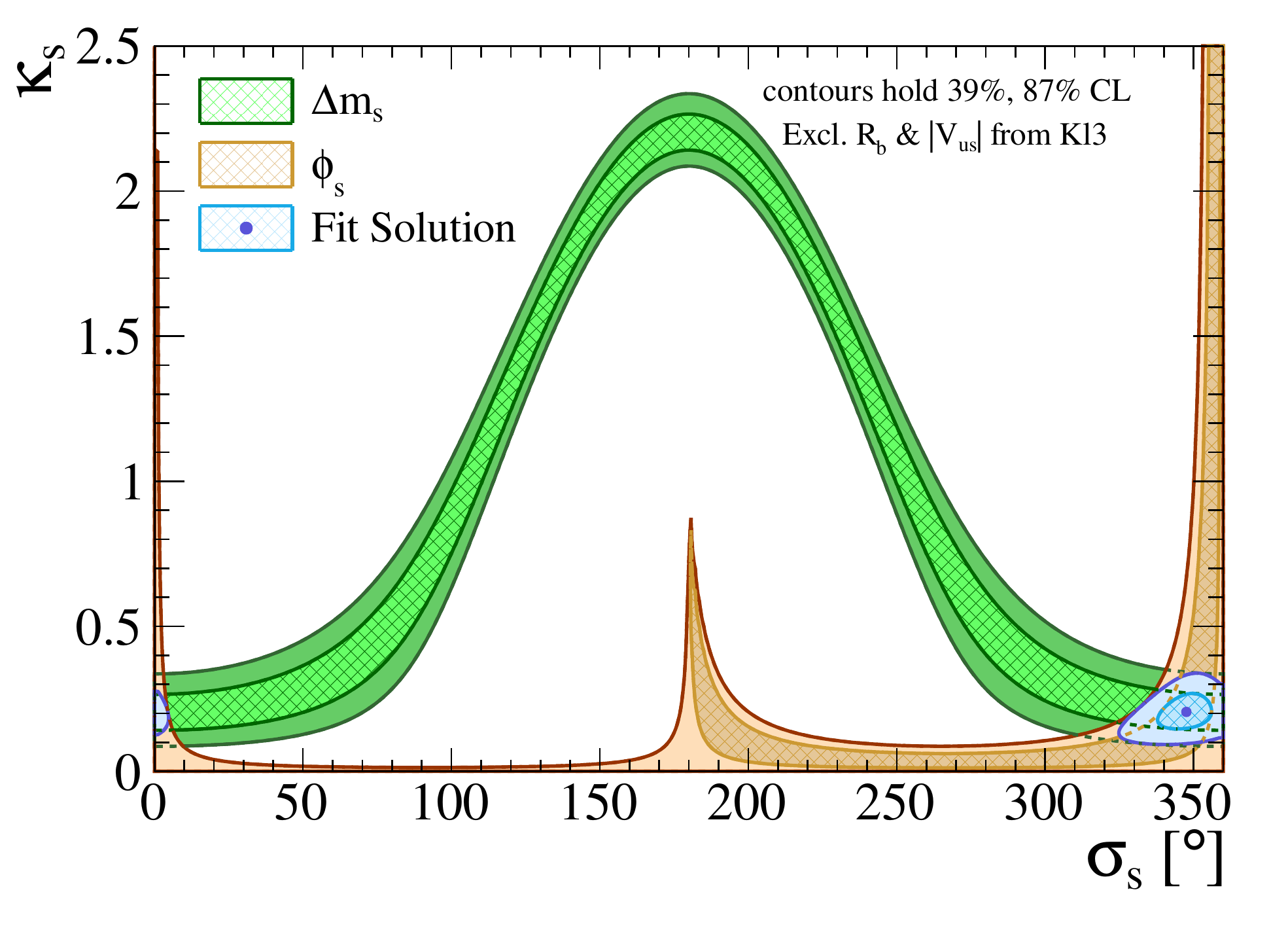}
    
    \includegraphics[width=0.49\textwidth]{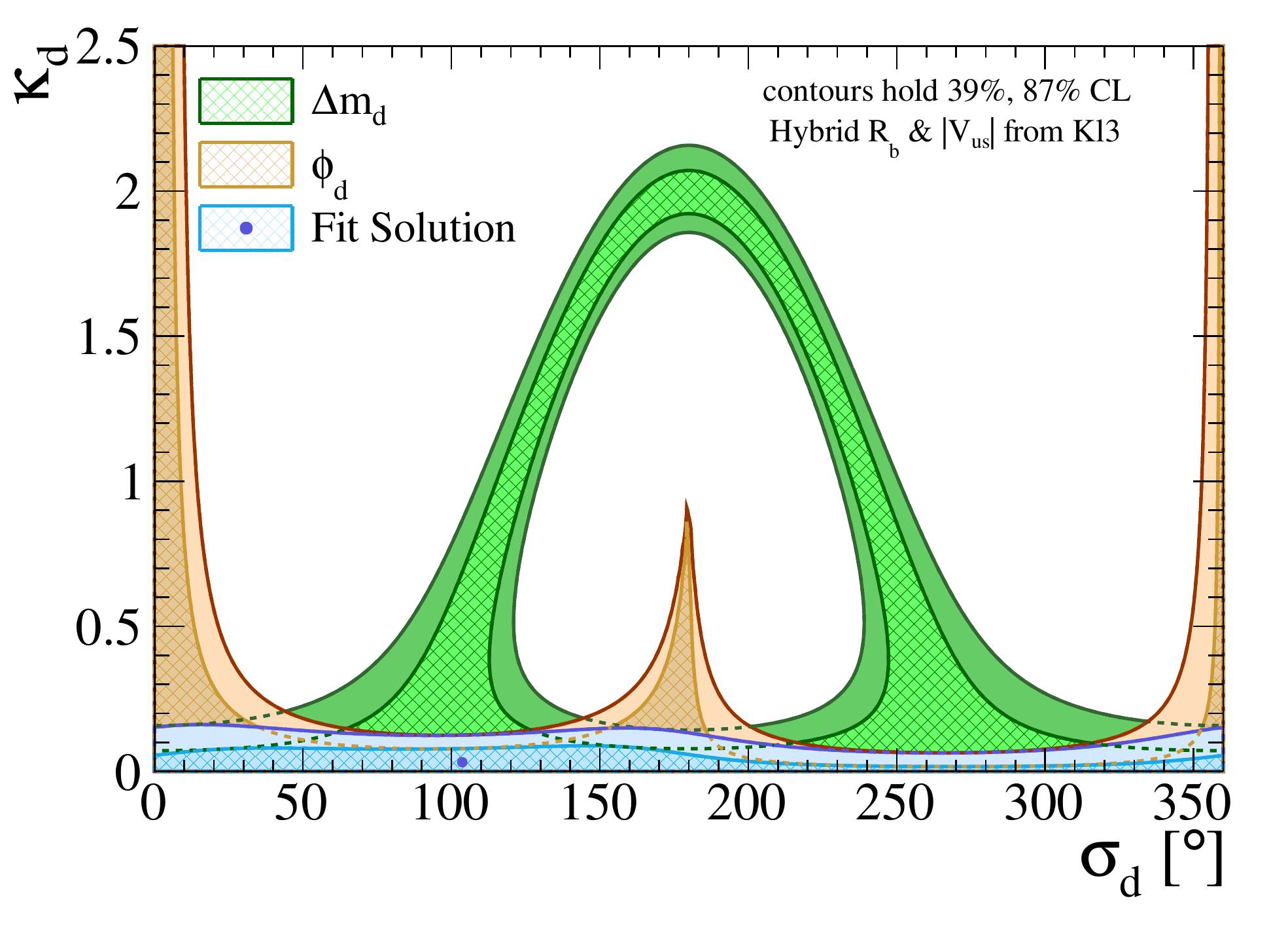}
    \hfill
    \includegraphics[width=0.49\textwidth]{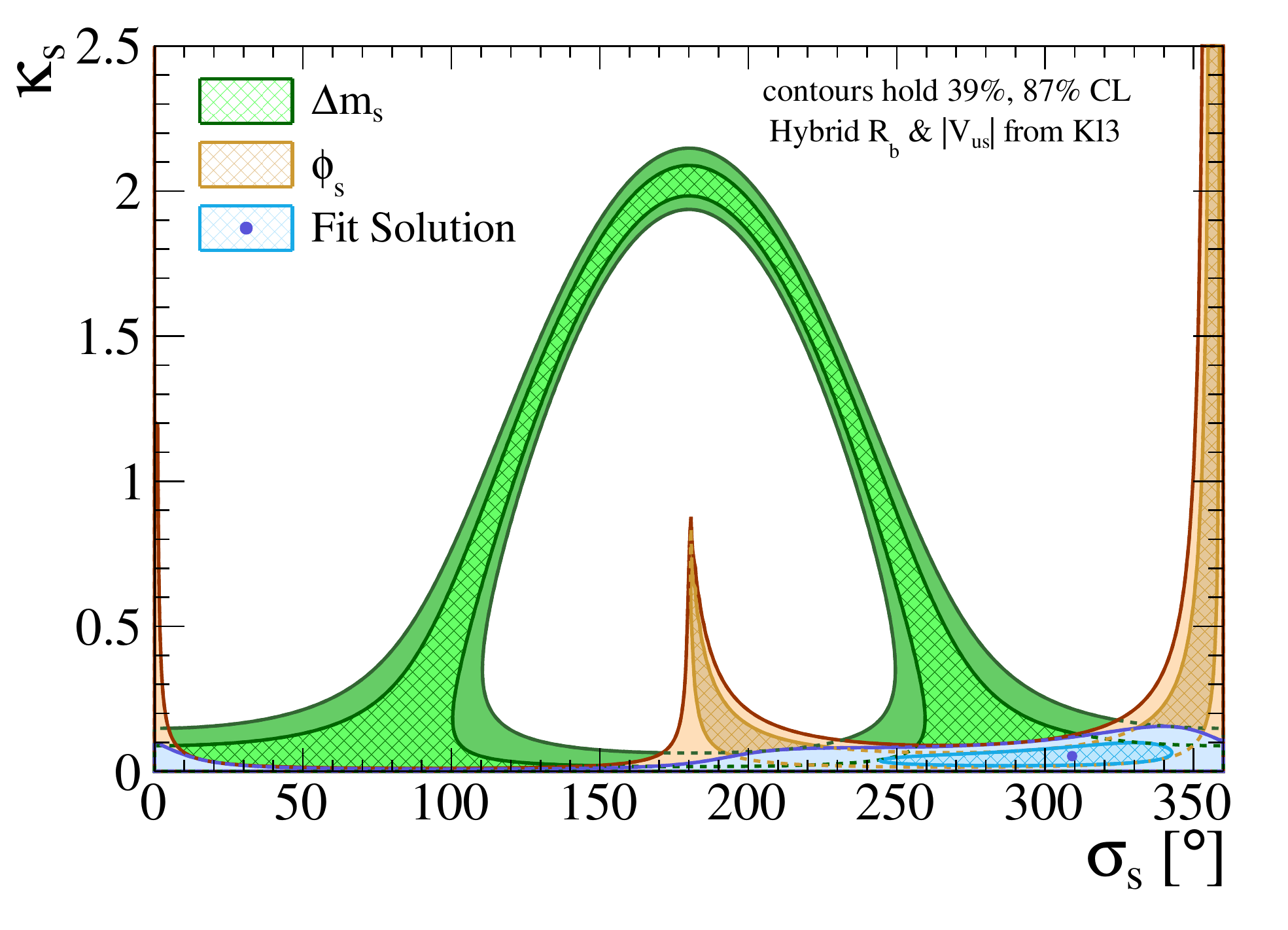}
    \caption{Two-dimensional confidence regions of the fit for $\kappa_d$ and $\sigma_d$ (left), and $\kappa_s$ and $\sigma_s$ (right), which parametrise NP contributions in $B_q^0$--$\bar B_q^0$ mixing.
    For illustration, also the individual constraints from $\phi_q$ and $\Delta m_q$ are shown.
    Top: Inclusive scenario. 
    Middle: Exclusive scenario.
    Bottom: Hybrid scenario.}
    \label{fig:NP_Scen1}
\end{figure}

Scenario I is a general determination of the NP parameters in $B_q^0$--$\bar B_q^0$ mixing, given in Eqs.\ \eqref{eq:NP_mix_Dmq} and \eqref{eq:NP_mix_phiq}.
We will use our results from Section \ref{sec:UTapex} including information from $R_b$ and $\gamma$, together with the SM expressions and values for the mixing parameters, summarised in Table \ref{tab:DMqinput}, to determine the NP parameters $(\kappa_q, \sigma_q)$.
The only assumption that enters this scenario is the absence of NP in $\gamma$ and the semileptonic decays needed to determine $R_b$.
The results are shown in Fig.\ \ref{fig:NP_Scen1} and given as follows:
\begin{align}
    \text{Incl, } K\ell3 & &
    \kappa_d & = 0.121_{-0.055}^{+0.056}\:, & 
    \sigma_d & = \left(261_{-35}^{+37}\right)^{\circ}\:, \label{eq:NP_Scen1_Bd_I3}\\
    & &
    \kappa_s & = 0.045_{-0.033}^{+0.048}\:, & 
    \sigma_s & = \left(312_{-77}^{+37}\right)^{\circ}\:, \\
    \text{Excl, } K\ell3 & &
    \kappa_d & = 0.156_{-0.084}^{+0.093}\:, & 
    \sigma_d & = \left(347_{-25}^{+21}\right)^{\circ}\:, \\
    & &
    \kappa_s & = 0.205_{-0.059}^{+0.064}\:, & 
    \sigma_s & = \left(347.6_{-9.8}^{+8.5}\right)^{\circ}\:, \\
    \text{Hybrid, } K\ell3 & &
    \kappa_d & = 0.031_{-0.031}^{+0.057}\:, & 
    \sigma_d & = \left(104_{-104}^{+256}\right)^{\circ}\:, \\
    & &
    \kappa_s & = 0.053_{-0.034}^{+0.046}\:, & 
    \sigma_s & = \left(309_{-65}^{+34}\right)^{\circ}\:. \label{eq:NP_Scen1_Bs_H3}
\end{align}
We observe in Fig.\ \ref{fig:NP_Scen1} that the individual constraints from $\phi_d$ in the inclusive scenario and $\Delta m_s$ in the exclusive scenario are in tension with the SM, while all other constraints are compatible with $\kappa_q = 0$.

\begin{figure}
    \centering
    \includegraphics[width=0.49\textwidth]{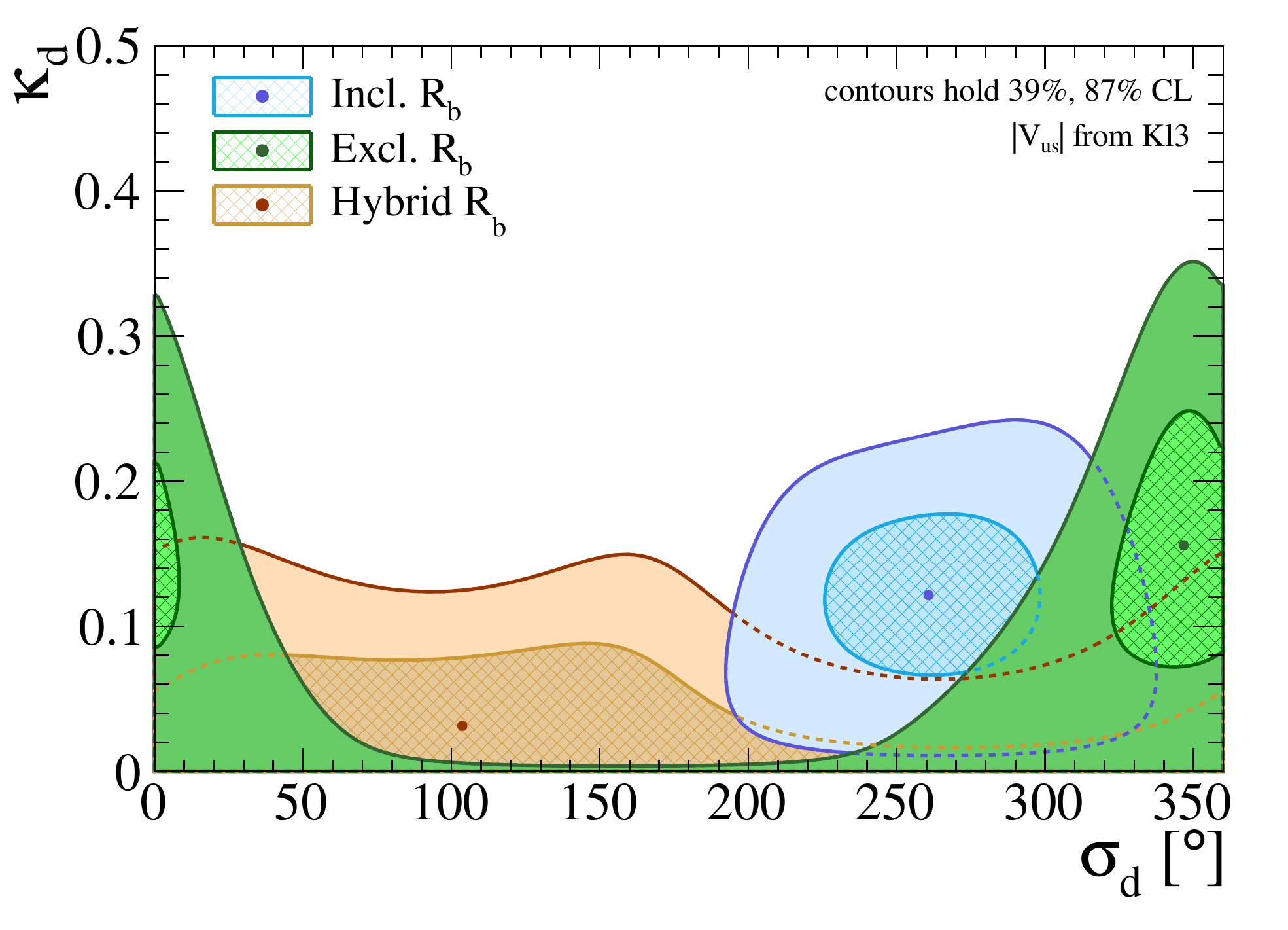}
    \hfill
    \includegraphics[width=0.49\textwidth]{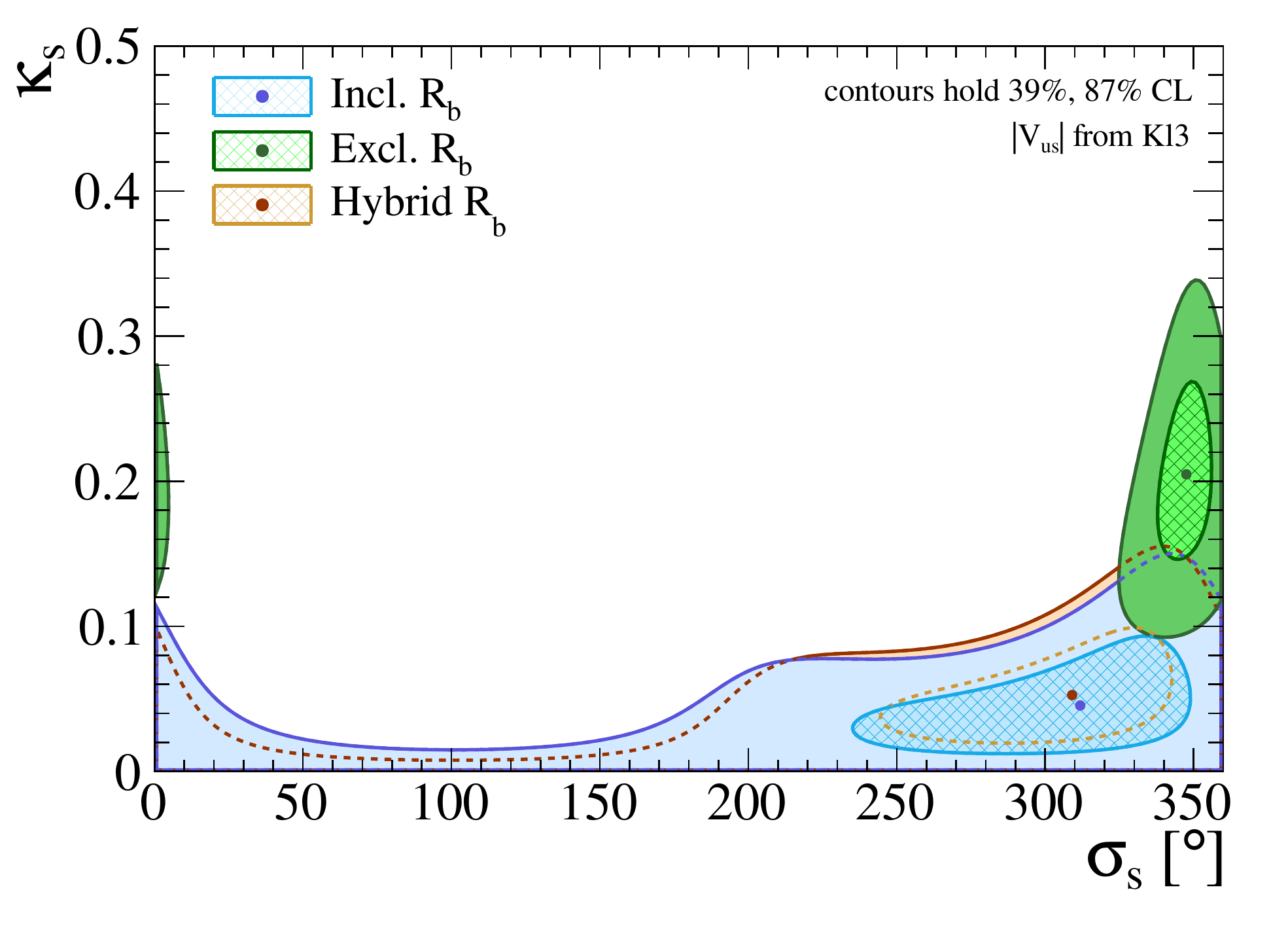}
    \caption{Comparison between the inclusive, exclusive and hybrid scenario of the two-dimensional confidence regions of the fits for $\kappa_d$ and $\sigma_d$ (left), and $\kappa_s$ and $\sigma_s$ (right), which parametrise NP contributions in $B_q^0$--$\bar B_q^0$ mixing.}
    \label{fig:NP_Scen1_comp}
\end{figure} 
\begin{figure}
    \centering
    \includegraphics[width=0.49\textwidth]{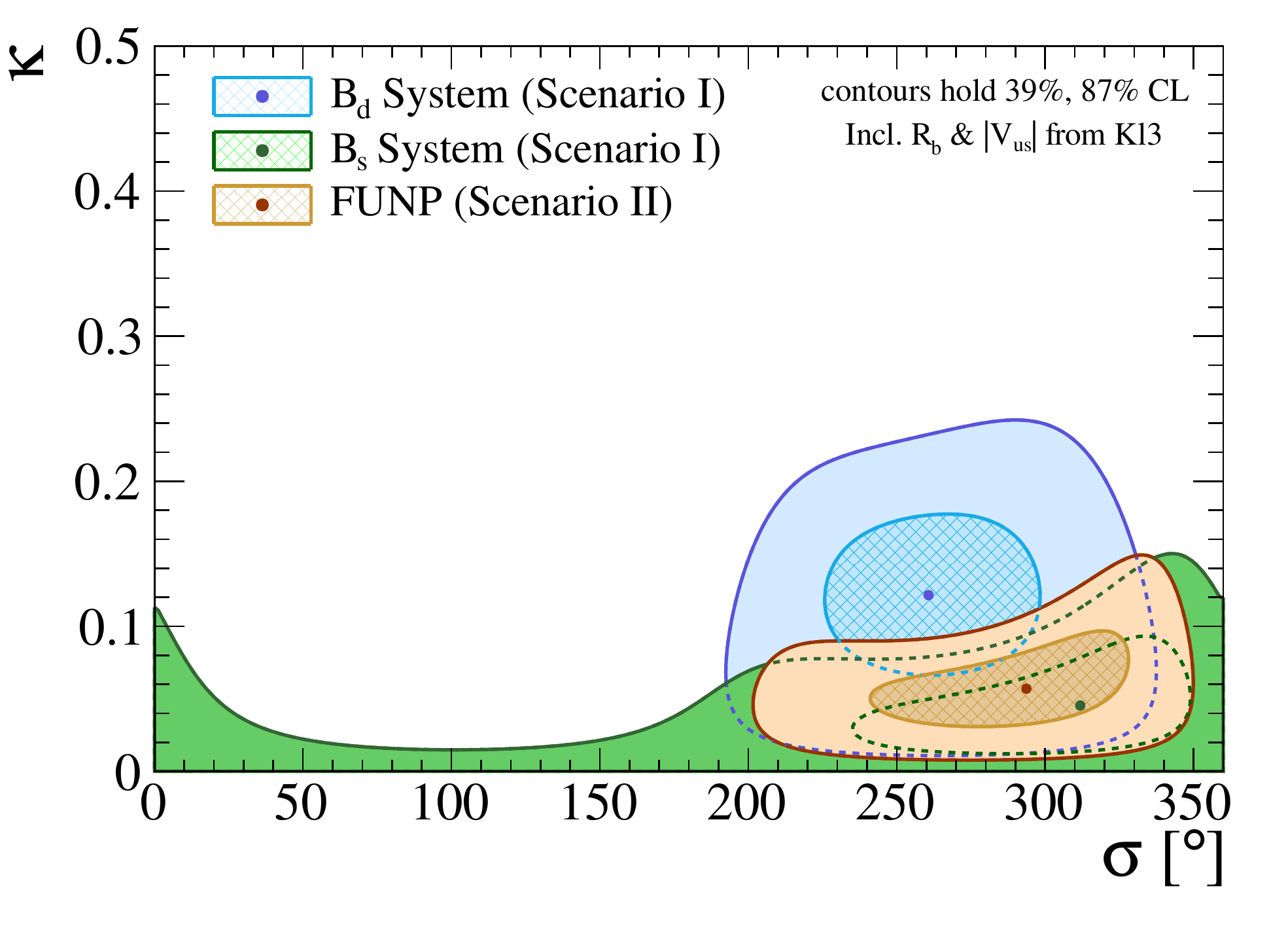}
    \includegraphics[width=0.49\textwidth]{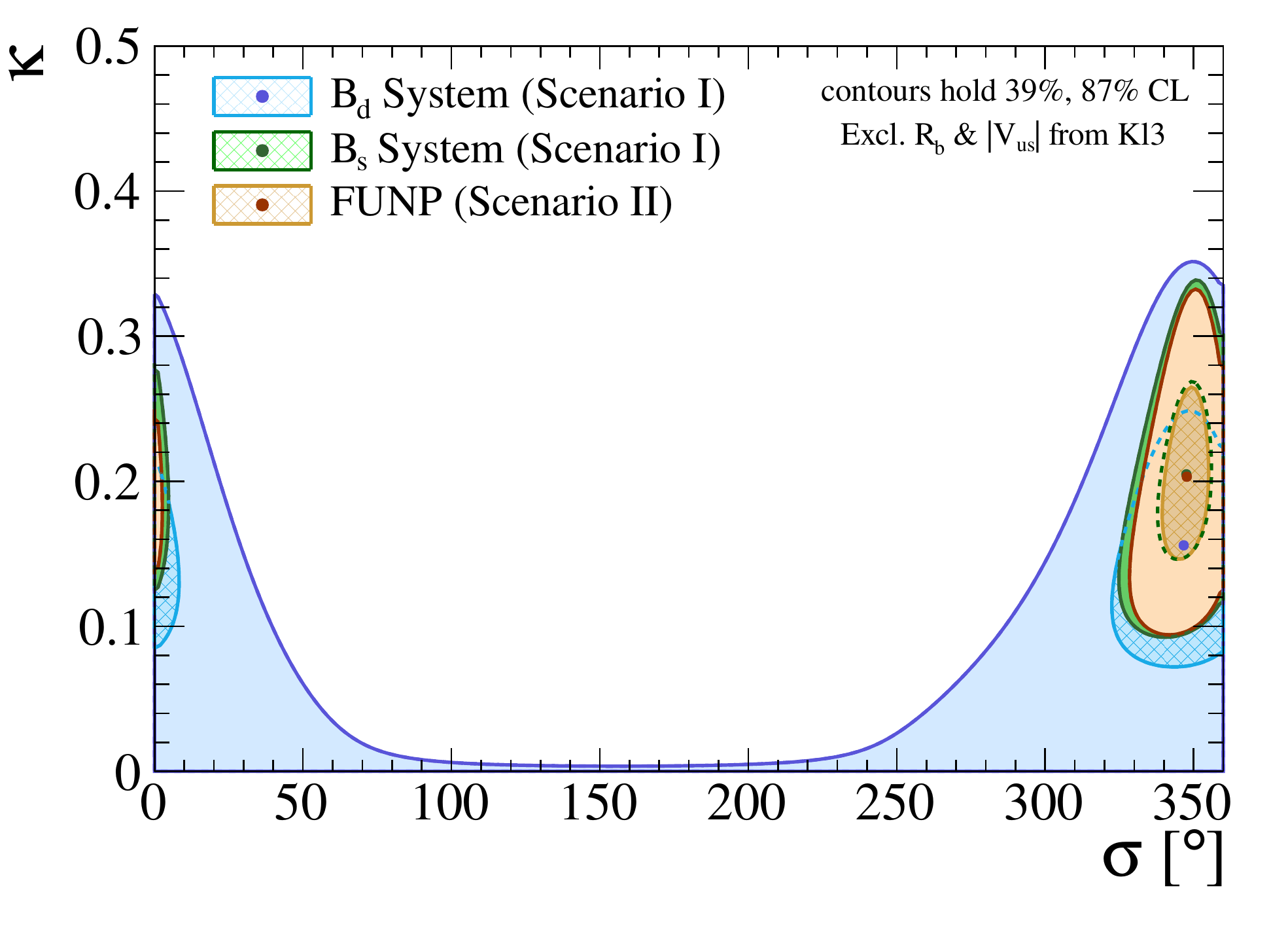}
    
    \includegraphics[width=0.49\textwidth]{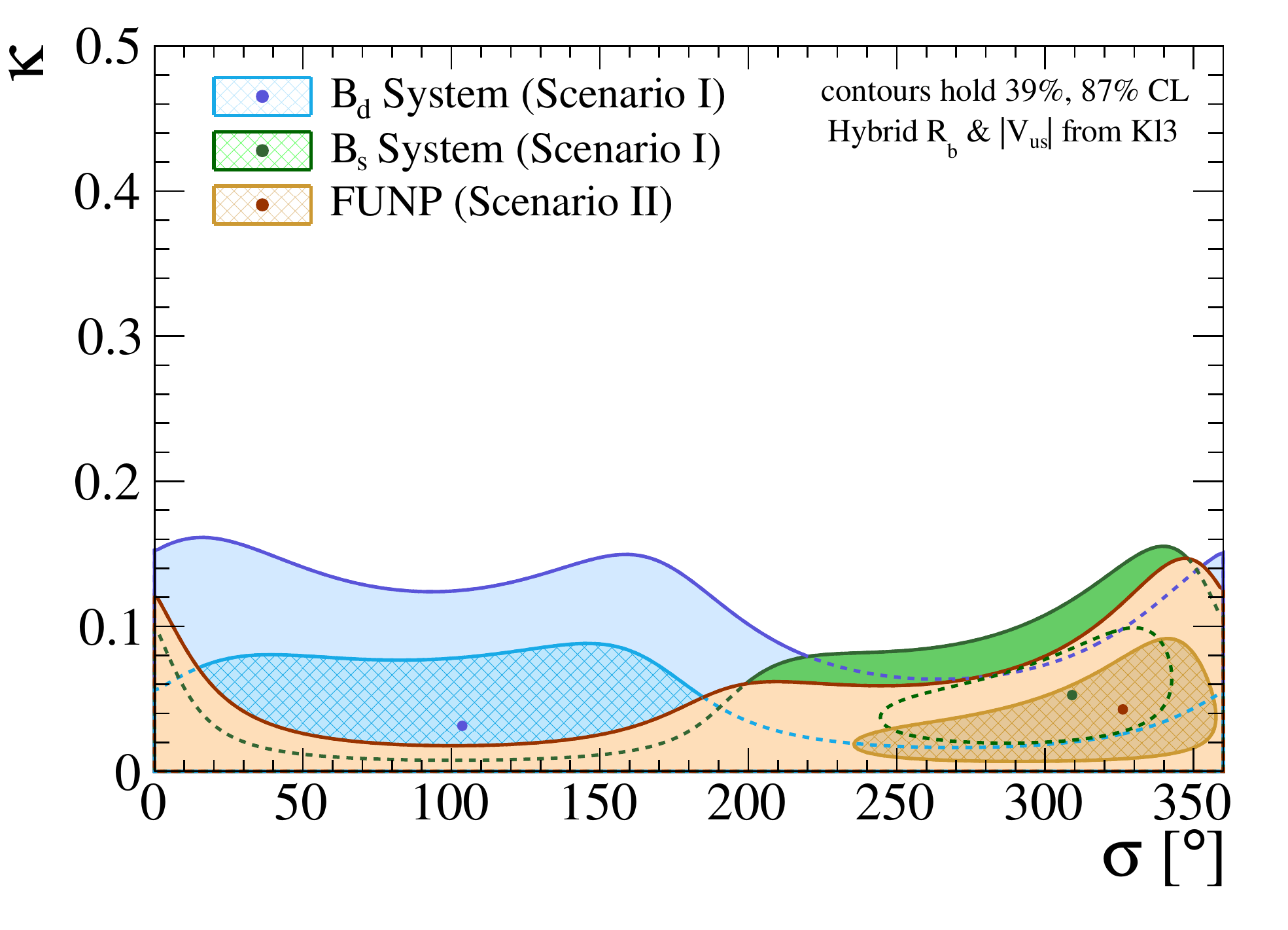}
    \caption{Comparison between the two-dimensional confidence regions of the Scenario I and Scenario II fits for $\kappa_q$ and $\sigma_q$, which parametrise NP contributions in $B_q^0$--$\bar B_q^0$ mixing.
    Left: Inclusive scenario.
    Right: Exclusive scenario.
    Bottom: Hybrid scenario.}
    \label{fig:NP_Scen1_vs_Scen2}
\end{figure} 
\begin{figure}
    \centering
    \includegraphics[width=0.49\textwidth]{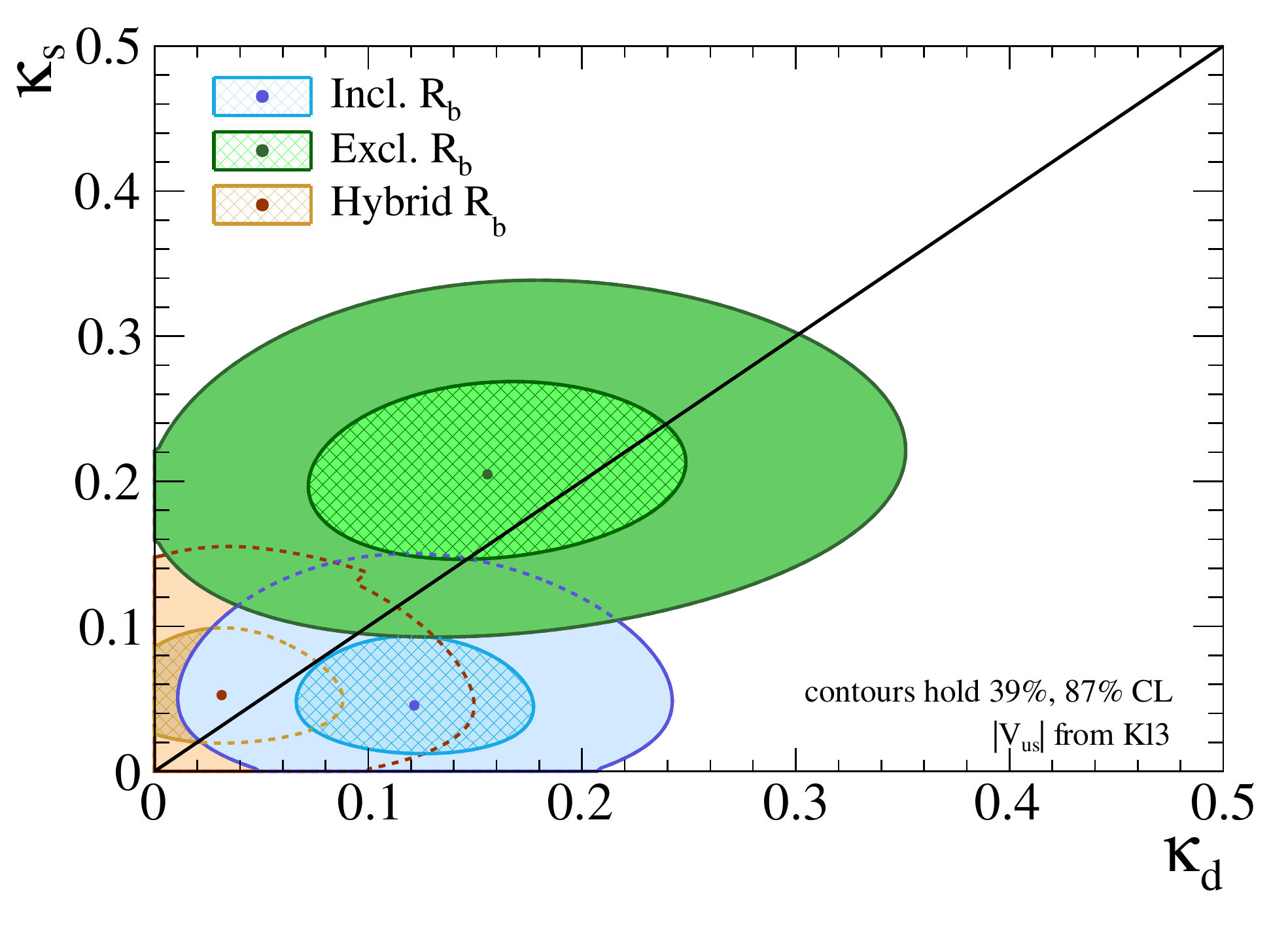}
    \caption{Comparison between the inclusive, exclusive and hybrid scenarios of the two-dimensional confidence regions of a combined fit for the NP parameters $\kappa_d$ and $\kappa_s$ describing NP contributions in $B_q^0$--$\bar B_q^0$ mixing.
    The black diagonal line represents the FUNP scenario, where $\kappa_d = \kappa_s$.}
    \label{fig:NP_comparisons_kappa}
\end{figure} 

A comparison of the two-dimensional confidence regions between the three fit solutions based on different values of $R_b$ is given in Fig.\ \ref{fig:NP_Scen1_comp}, and a comparison between the fit solutions for the $B_d$ and $B_s$ systems in Fig.\ \ref{fig:NP_Scen1_vs_Scen2}.
The correlation between $\kappa_d$ and $\kappa_s$ is also shown in Fig.\ \ref{fig:NP_comparisons_kappa}.
As illustrated in Fig.\ \ref{fig:NP_comparisons_kappa}, all three $R_b$ solutions are compatible with the assumption $\kappa_d = \kappa_s$, which is used in the FUNP scenario.
However, the conclusions regarding the presence of NP are different for the three scenarios:
The inclusive one is compatible with $\kappa_s = 0$, but suggests $\kappa_d\neq 0$;
the exclusive fit result is compatible with $\kappa_d = 0$, but finds $\kappa_s$ different from zero with a significance of 3.5 standard deviations;
and the hybrid case is compatible with both $\kappa_d = 0$ and $\kappa_s = 0$ and hence finds no tension with the SM.

\subsection{Scenario II: Flavour Universal New Physics}\label{sec:equalNP}
In the second scenario, we will assume FUNP, where the size and phase of the NP contributions, $\kappa_q$ and $\sigma_q$, are equal for the $B_d$ and $B_s$ systems:
\begin{equation}\label{eq:FUNPconstr}
    \kappa_d = \kappa_s \equiv \kappa\:, \quad \sigma_d = \sigma_s \equiv \sigma\:.
\end{equation}
An important consequence is that the NP effects drop out in the ratio $\Delta m_d/\Delta m_s$. 
This means that the side $R_t$ of the UT, defined in Eq.\ \eqref{eq:Rt}, will receive no contributions from NP at order $\lambda^2$. 
We can use the results for the UT apex coming from information from $R_b$ and $R_t$, as given in Eqs.\ \eqref{eq:Rt_apex_I3}--\eqref{eq:Rt_apex_H3}, to determine the SM values of $\Delta m_q$ and $\phi_q$ without additional information on $\gamma$.
Comparing the SM predictions to their experimental values, we get constraints on $\kappa$ and $\sigma$.
This results in the following fit values for $\kappa$ and $\sigma$:
\begin{align}
    \text{Incl, } K\ell3 & &
    \kappa & = 0.057_{-0.026}^{+0.040}\:, & 
    \sigma & = \left(294_{-53}^{+34}\right)^{\circ}\:, \\
    \text{Excl, } K\ell3 & &
    \kappa & = 0.203_{-0.057}^{+0.062}\:, & 
    \sigma & = \left(347.7_{-8.3}^{+7.4}\right)^{\circ}\:, \\
    \text{Hybrid, } K\ell3 & &
    \kappa & = 0.043_{-0.036}^{+0.049}\:, & 
    \sigma & = \left(326_{-90}^{+32}\right)^{\circ}\:.
\end{align}
These results are very similar to those arising from the fit for the $B_s$ system from $R_b$ and $\gamma$ in Eqs.\ \eqref{eq:NP_Scen1_Bd_I3}--\eqref{eq:NP_Scen1_Bs_H3}, implying that the $B_s$ system dominates these fits.
In Fig.\ \ref{fig:NP_Scen1_vs_Scen2}, we can see that in the exclusive case, the results for FUNP almost completely overlap with those of the $B_s$ system for Scenario I. 
In the inclusive and hybrid scenarios, the shape of the contours for the $B_d$ and the $B_s$ system are very different, indicating that FUNP might not be a correct assumption.
However, the solutions still overlap within the error margin, and one can see that the FUNP scenario interpolates between the $B_d$ and $B_s$ systems, with the $B_s$ system still dominating.
In all cases, we cannot rule out FUNP, as it is still compatible with the results from the general scenario for both $B_d$ and $B_s$.

\section[Applications for the Rare Decays Bq2mumu]{Applications for the Rare Decays $B_q^0\to\mu^+\mu^-$}\label{sec:raredecays}
%
%
%
\begin{table}
    \centering
    \begin{tabular}{||c|c|c|c||}
        \hline
        \hline
        Parameter & Value & Unit & Reference\\
        \hline
        $\hat{B}_s$ & $1.232 \pm 0.053$ & & \cite{Dowdall:2019bea, Aoki:2021kgd} \\
        $\hat{B}_s/\hat{B}_d$ & $1.008 \pm 0.025$ & & \cite{Dowdall:2019bea, Aoki:2021kgd}  \\
        $2 y_s \equiv \Delta \Gamma_s/ \Gamma_s$ & $0.128 \pm 0.007$ & & \cite{HFLAV:2022pwe} \\
        $2 y_d \equiv \Delta \Gamma_d/ \Gamma_d$ & $0.001 \pm 0.010$ & & \cite{HFLAV:2022pwe} \\
        \hline
        $\tau_{\mu\mu}^{s}$ & $2.07 \pm 0.29$ & ps  & \cite{LHCb:2021vsc} \\
        $\tau_{B_s}$ & $1.520 \pm 0.005$ & ps & \cite{HFLAV:2022pwe} \\
        \hline
        $f_{B_s}$ & $230.3 \pm 1.3$ & MeV & \cite{Aoki:2021kgd} \\
        $\tau_{B_s,H}$ & $1.624 \pm 0.009$ & ps & \cite{HFLAV:2022pwe} \\
        $\eta_Y$ & 1.0113 & & \cite{Buras:2012ru} \\
        \hline
        \hline
    \end{tabular}
    \caption{Input values for the SM computation of the $B_q^0\to\mu^+\mu^-$ branching fractions.
    $\tau_{B_s, H}$ is the lifetime of the heavy $B_s$ mass eigenstate and can be expressed in terms of $\tau_{B_s}$ and $y_s$ as $\tau_{B_s}/(1 - y_s)$}.
    \label{tab:raredecays}
\end{table}

The NP searches in the rare decays $B_q^0\to\mu^+\mu^-$ are also affected by our choices for the CKM matrix elements.
The SM predictions for the branching fractions are, using unitarity of the CKM matrix, proportional to the square of $|V_{cb}|$, and thus differ between the inclusive and exclusive determinations.
Refs.\ \cite{Buras:2003td,Buras:2021nns,Bobeth:2021cxm} suggested to take the ratio with $\Delta m_q$ in order to eliminate this dependence on $|V_{cb}|$.
However, as we have seen above, $\Delta m_q$ may be affected by NP effects.
Using our analyses we are able to take these effects into account, and show the resulting constraints on NP in $B_s^0\to\mu^+\mu^-$.
In addition, we will also explore the option of using the ratio of $B_d^0\to\mu^+\mu^-$ and $B_s^0\to\mu^+\mu^-$ branching fractions to determine the UT side $R_t$, providing us with a second opportunity to determine the UT apex without relying on the angle $\gamma$.
\subsection[BsMuMu in the Standard Model]{$B_s^0\to\mu^+\mu^-$ in the Standard Model}
At lowest order in the SM, the ``theoretical" branching fraction of the decay $B_s^0\to\mu^+\mu^-$ is given as \cite{Buras:2013uqa}
\begin{equation}\label{eq:Bs2mumu_theo}
    \mathcal{B}(B_s\to\mu^+\mu^-) =
    \frac{\tau_{B_s} G_{\text{F}}^4m_W^4\sin^4\theta_W}{8\pi^5}\left|C_{10}^{\text{SM}}\right|^2
    |V_{ts}^{\phantom{*}}V_{tb}^*|^2 f_{B_s}^2
    m_{B_s}m_{\mu}^2\sqrt{1-\frac{4m_{\mu}^2}{m_{B_s}^2}}\:,
\end{equation}
where $\theta_W$ is the weak mixing angle, and $\tau_{B_s}$ is the $B_s$-meson lifetime.
The theoretical branching fraction is defined at decay time $t = 0$ and thus does not include effects coming from $B_s^0$--$\bar B_s^0$ mixing.
This is not the case for the experimentally measured time-integrated branching fraction, which is related to the theoretical branching fraction as follows\cite{DeBruyn:2012wk}:
\begin{equation}\label{eq:Bs2mumu_TI}
    \bar{\mathcal{B}}(B_s\to\mu^+\mu^-) = \frac{1 + \mathcal{A}_{\Delta \Gamma}^{\mu\mu}  y_s}{1 - y_s^2} \mathcal{B}(B_s\to\mu^+\mu^-)\:,
\end{equation}
where $\mathcal{A}_{\Delta \Gamma}^{\mu\mu} = 1$ in the SM.
The parameter $y_s$ is defined as
\begin{equation}
    y_s \equiv \frac{\tau_{B_s}}{2} \Delta\Gamma_s\:,
\end{equation}
where $\Delta\Gamma_s$ is the decay width difference between the $B_s$ mass eigenstates.
It is now clear that $B_q^0$--$\bar B_q^0$ mixing effects will affect the time-integrated branching fraction.
The Wilson coefficient $C_{10}$ can be expressed as
\begin{equation}
    C_{10}^{\text{SM}} = \frac{\eta_Y Y_0(x_t)}{\sin^2\theta_W}\:,
\end{equation}
where $\eta_Y$ is a QCD correction factor \cite{Buras:2012ru}, and $Y_0$ is the Inami--Lim function \cite{Inami:1980fz}
\begin{equation}
    Y_0(x) = \frac{x}{8}\left[\frac{4-x}{1-x} + \frac{3x}{(1-x)^2}\ln x\right]\:.
\end{equation}
It is interesting to note that in this specific parametrisation the dependence on $\sin^2\theta_W$ drops out from the theoretical prediction \eqref{eq:Bs2mumu_theo}.
All numerical inputs are summarised in Table \ref{tab:raredecays}.

\begin{figure}
    \centering
    \includegraphics[width=0.6\textwidth]{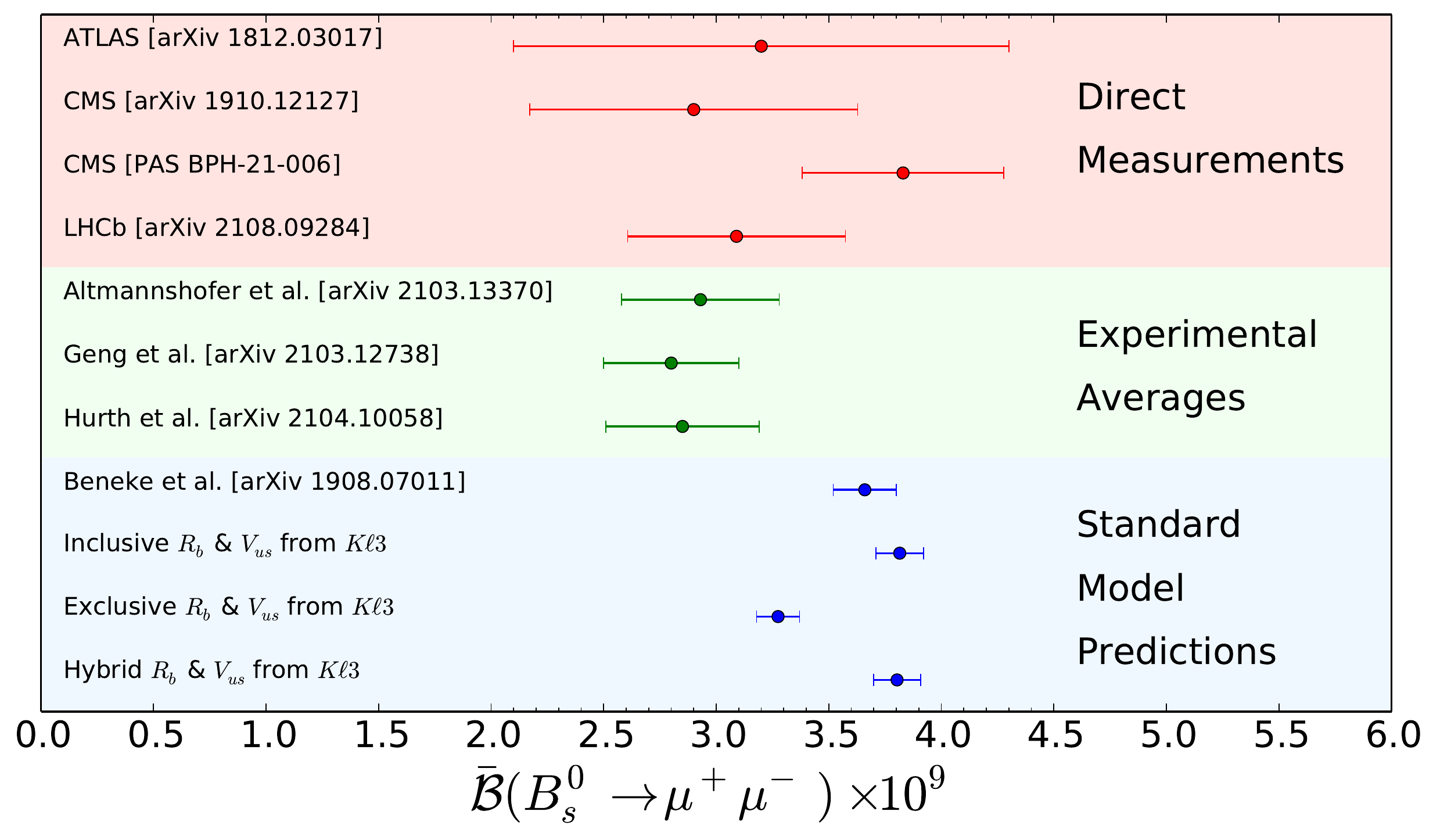}
    \caption{Comparison of the experimental measurements, averages and SM predictions for the time-integrated branching fraction $\bar{\mathcal{B}}( B_s^0\to\mu^+\mu^-$)}
    \label{fig:BR_BsMuMu}
\end{figure}

The SM branching fraction depends both on the value of $|V_{cb}|$ and on the solution of the UT apex through the relation \eqref{eq:VtsVtb_SM} for the CKM matrix elements $|V_{ts}^{\phantom{*}}V_{tb}^*|$, although the latter dependence enters only through higher order corrections.
This results in different predictions for the three scenarios considered in this paper:
\begin{align}
    \text{Incl, } K\ell3 & &
    \bar{\mathcal{B}}(B_s\to\mu^+\mu^-) & = (3.81 \pm 0.11)\times 10^{-9}\:, \label{eq:Bs2MuMu_I3}\\
    \text{Excl, } K\ell3 & &
    \bar{\mathcal{B}}(B_s\to\mu^+\mu^-) & = (3.27 \pm 0.10)\times 10^{-9}\:, \label{eq:Bs2MuMu_E3}\\
    \text{Hybrid, } K\ell3 & &
    \bar{\mathcal{B}}(B_s\to\mu^+\mu^-) & = (3.80 \pm 0.10)\times 10^{-9}\:. \label{eq:Bs2MuMu_H3}
\end{align}
In comparison with the latest prediction \cite{Beneke:2019slt}
\begin{equation}\label{eq:Bs2MuMu_NLO}
    \bar{\mathcal{B}} (B_s \to \mu^{+} \mu^{-})|_{\text{SM}} = (3.66 \pm 0.14) \times 10^{-9}
\end{equation}
found in the literature, our uncertainties are slightly smaller.
This is primarily because the experimental measurements for the CKM matrix elements and top quark mass have improved compared to what was used in Ref.\ \cite{Beneke:2019slt}.
It is important to point out that the result \eqref{eq:Bs2MuMu_NLO} includes NLO electroweak and QCD corrections, while our predictions in Eqs.\ \eqref{eq:Bs2MuMu_I3}-\eqref{eq:Bs2MuMu_H3} are based on the LO expression only.
We have not included NLO corrections since our goal is to illustrate the dependence of the numerical prediction on the choice for the CKM matrix elements and UT apex.
This introduces a spread, as illustrated in Fig.\ \ref{fig:BR_BsMuMu}, that is still much larger and thus more important than the NLO corrections.

On the experimental side, the $B_s^0\to\mu^+\mu^-$ branching fraction has been measured by the LHCb \cite{LHCb:2021vsc}, ATLAS \cite{ATLAS:2018cur} and CMS \cite{CMS:2019bbr,CMS:2022dbz} collaborations.
Their results are compared to the SM predictions in Fig.\ \ref{fig:BR_BsMuMu}.
In the context of the discussion on the $b\to s\ell\ell$ anomalies, multiple groups \cite{Geng:2021nhg,Altmannshofer:2021qrr,Hurth:2021nsi} have performed averages for these results, which are also shown in Fig.\ \ref{fig:BR_BsMuMu}.
The recent update \cite{CMS:2022dbz} from CMS is not yet included in these averages.
For the analysis below, we choose the result from Ref.\ \cite{Hurth:2021nsi}:
\begin{equation}\label{eq:Bsmumu_Avg}
    \bar{\mathcal{B}}(B_s\to\mu^+\mu^-) = (2.85_{-0.31}^{+0.34})\times 10^{-9}\:.
\end{equation}
This average differs from the SM prediction \eqref{eq:Bs2MuMu_E3} by one standard deviation, and from the SM predictions \eqref{eq:Bs2MuMu_I3} and \eqref{eq:Bs2MuMu_H3} by 2.4 standard deviations.
Individual measurements of the $B_d^0\to\mu^+\mu^-$ branching fraction are not yet significantly different from zero.
The LHCb \cite{LHCb:2021vsc}, ATLAS \cite{ATLAS:2018cur} and CMS \cite{CMS:2019bbr} results therefore only provide an upper limit, with the most stringent upper bound given by \cite{LHCb:2021vsc}
\begin{equation}
     \bar{\mathcal{B}}(B_d \to \mu^{+} \mu^{-}) < 0.26 \times 10^{-9}\:.
\end{equation}
Nonetheless, based on the combined analysis with $B_s^0\to\mu^+\mu^-$, Ref.\ \cite{Altmannshofer:2021qrr} does give the average
\begin{equation}\label{eq:Bdmumu_Avg}
    \bar{\mathcal{B}}(B_d\to\mu^+\mu^-)
    = (0.56 \pm 0.70) \times 10^{-10}\:.
\end{equation}
This result is still compatible with zero within one standard deviation. 
\subsection[Determining NP in Bs2mumu]{Determining NP in $ \bar{\mathcal{B}}(B_s\to\mu^+\mu^-)$}
\begin{figure}
    \centering
    \includegraphics[width=0.49\textwidth]{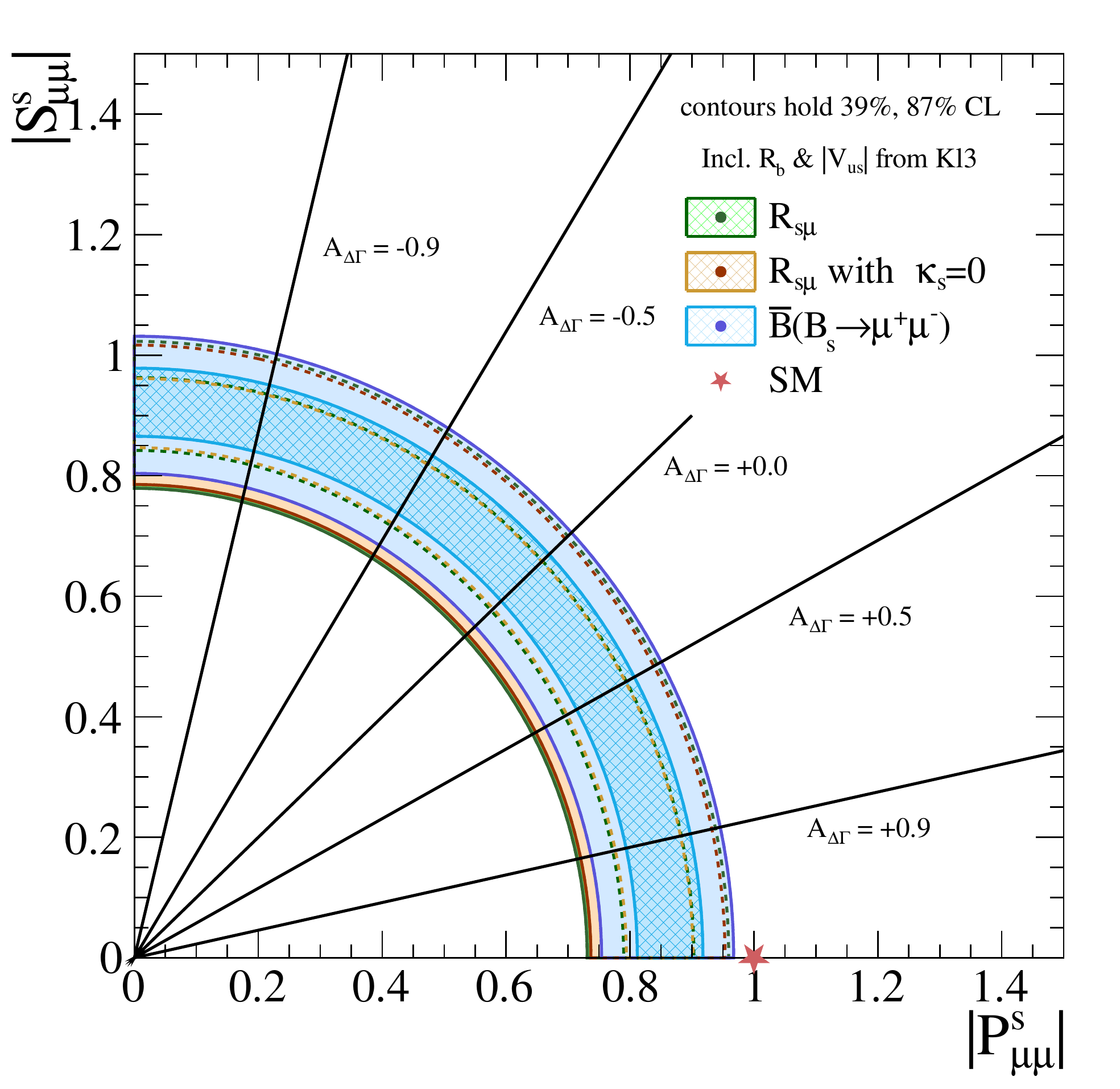}
    \hfill
    \includegraphics[width=0.49\textwidth]{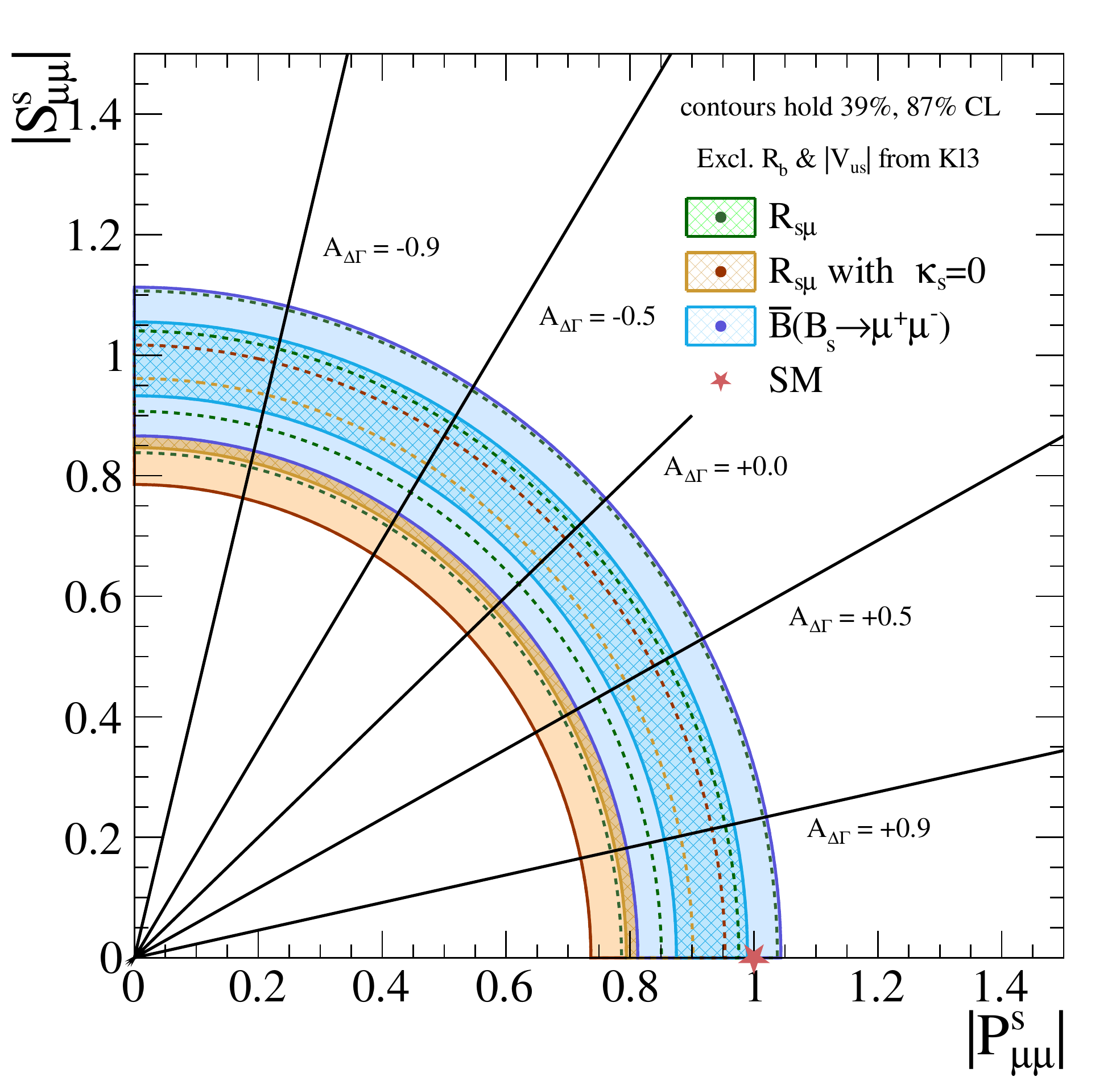}
    
    \includegraphics[width=0.49\textwidth]{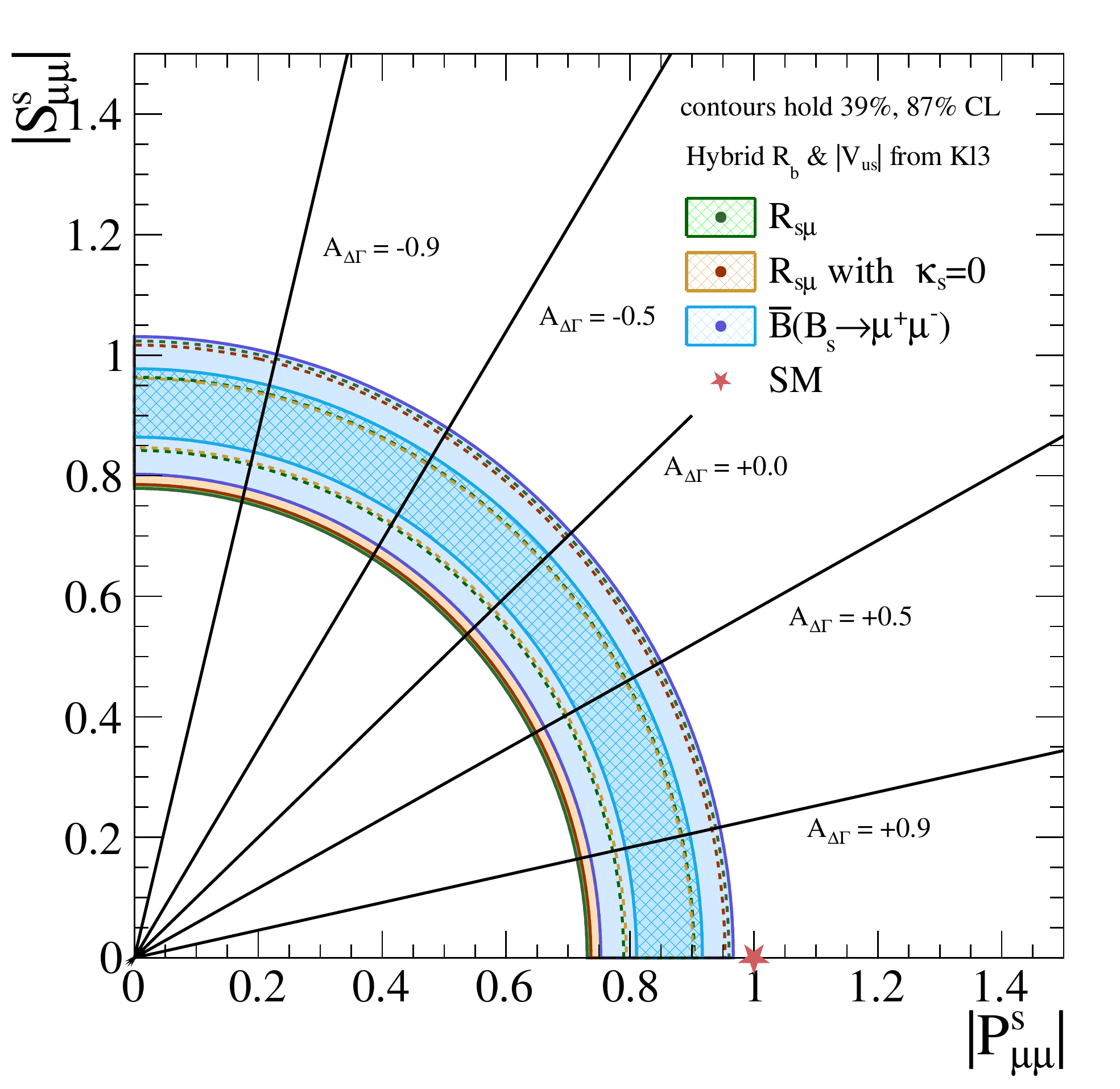}
    \caption{Two-dimensional confidence regions for the constraints from $\mathcal{B}(B_s\to\mu^+\mu^-)$ and $R_{s\mu}$ on $|P_{\mu\mu}^s|$ and $|S_{\mu\mu}^s|$, assuming $\varphi_P=\varphi_S=0$.
    Left: Inclusive scenario.
    Right: Exclusive scenario.
    Bottom: Hybrid scenario.}
    \label{fig:NP_Bsmumu_PS}
\end{figure}

NP can modify the $B_s^0\to\mu^+\mu^-$ branching fraction through pseudo-scalar (P) or scalar (S) contributions, or through $B_s^0$--$\bar B_s^0$ mixing.
The measured $B_s^0\to\mu^+\mu^-$ branching fraction is given by
\begin{equation}
    \bar{\mathcal{B}}(B_s\to\mu^+\mu^-) =
    \bar{\mathcal{B}}(B_s\to\mu^+\mu^-)^{\text{SM}} \times
    \frac{1 + \mathcal{A}^{\mu\mu}_{\Delta \Gamma_s} y_s}{1 + y_s}
    \left(|P_{\mu\mu}^{s}|^2 + |S_{\mu\mu}^{s}|^2\right)\:,
\end{equation}
where $\mathcal{A}^{\mu\mu}_{\Delta \Gamma_s}$ depends on $P_{\mu\mu}^{s} \equiv |P_{\mu\mu}^{s}|e^{i\varphi_P}$, $S_{\mu\mu}^{s} \equiv |S_{\mu\mu}^{s}|e^{i\varphi_S}$ and $\phi_s^{\text{NP}}$ as
\begin{equation}\label{eq:Amumu}
    \mathcal{A}_{\Delta \Gamma}^{\mu\mu} = 
    \frac{|P_{\mu\mu}^s|^2\cos(2\varphi_P - \phi_s^{\text{NP}}) - |S_{\mu\mu}^s|^2\cos(2\varphi_S - \phi_s^{\text{NP}})}{|P_{\mu\mu}^s|^2 + |S_{\mu\mu}^s|^2}\:.
\end{equation}
In the SM, we have
\begin{equation}
    P^{s, \text{SM}}_{\mu\mu} = 1\:,\qquad S^{s, \text{SM}}_{\mu\mu} = 0\:.
\end{equation}
Comparing the experimental average \eqref{eq:Bsmumu_Avg} with the SM predictions \eqref{eq:Bs2MuMu_I3}-\eqref{eq:Bs2MuMu_H3}, we can constrain the parameters $|P_{\mu\mu}^{s}|$ and $|S_{\mu\mu}^{s}|$, as proposed in Ref.\ \cite{DeBruyn:2012wk}.
For the scenario where the NP phases for the pseudo-scalar and scalar contributions are zero, i.e.\ $\varphi_P=\varphi_S=0$, the results are shown in Fig.\ \ref{fig:NP_Bsmumu_PS}.

The dependence of the NP searches with $\bar{\mathcal{B}}(B_s\to\mu^+\mu^-)$ on the CKM matrix element $|V_{cb}|$ and the UT apex, which is clearly visible when comparing the blue contours in Fig.\ \ref{fig:NP_Bsmumu_PS}, can be minimised by constructing the ratio with the $B_s$ mass difference $\Delta m_s$ \cite{Buras:2003td,Buras:2021nns,Bobeth:2021cxm}:
\begin{equation}
    \mathcal{R}_{s\mu} \equiv \left|\frac{\bar{\mathcal{B}}(B_s\to\mu^+\mu^-)}{\Delta m_s} \right|\:.
\end{equation}
At lowest order, the SM contribution is given by
\begin{equation}
    \mathcal{R}_{s\mu}^{\text{SM}} = 
    \frac{\tau_{B_s}}{1-y_s}
    \frac{3 G_{\text{F}}^2 m_{W}^2 \sin^4\theta_W}{4 \pi^3}
    \frac{|C_{10}^{\text{SM}}|^2}{S_0(x_t)\eta_{2B} \hat{B}_{B_s}}
        m_{\mu}^2 \sqrt{1 - 4 \frac{m_{\mu}^2}{m_{B_s}^2}}\:,\label{eq:RsmuSM}
\end{equation}
resulting in the prediction
\begin{equation}
    \mathcal{R}_{s\mu}^{\text{SM}} = (2.22 \pm 0.10)\times 10^{-10}\: \text{ps}\:.
\end{equation}
Including NLO corrections Ref.\ \cite{Bobeth:2021cxm} finds a similar result
\begin{equation}
    \mathcal{R}_{s\mu}^{\text{SM}} = (2.042_{-0.058}^{+0.083})\times 10^{-10}\: \text{ps}\:,
\end{equation}
which differs from ours due to the choice for the bag parameter $\hat{B}_{B_s}$.
Also here, the choices made for the input parameters still have a bigger impact than the NLO corrections.

Including NP effects in both $\bar{\mathcal{B}}(B_s\to\mu^+\mu^-)$ and $\Delta m_s$, the expression for the ratio $\mathcal{R}_{s\mu}$ is generalised as follows:
\begin{equation}\label{eq:Rsmu}
    \mathcal{R}_{s\mu} =
    \mathcal{R}_{s\mu}^{\text{SM}} \times
    \frac{1 + \mathcal{A}^{\mu\mu}_{\Delta \Gamma_s} y_s}{1 + y_s}
    \frac{|P_{\mu\mu}^{s}|^2 + |S_{\mu\mu}^{s}|^2}{\sqrt{1 + 2 \kappa_s \cos\sigma_s + \kappa_s^2}}\:.
\end{equation}
It is important to stress that although the leading dependence on the CKM matrix elements drops out in Eq.\ \eqref{eq:RsmuSM} because we take the ratio with respect to $\Delta m_s$, the ratio in Eq.\ \eqref{eq:Rsmu} introduces a dependence on the CKM matrix elements through the NP parameters $(\kappa_s, \sigma_s)$.
Hence, when taking into account NP effects in $B_q^0$--$\bar B_q^0$ mixing, one should stay careful with the choice of the input of the CKM matrix elements.

Combining the experimental results in Eqs.\ \eqref{eq:Bsmumu_Avg} and \eqref{eq:Dms_exp}, we get
\begin{equation}
    \mathcal{R}_{s\mu} = (1.60 \pm 0.19)\times 10^{-10}\:.
\end{equation}
Comparing this with the SM value, we can again obtain contours in the $|P^{s}_{\mu\mu}|$--$|S^{s}_{\mu\mu}|$ plane.
These are also shown in Fig.\ \ref{fig:NP_Bsmumu_PS} for the scenario where the NP phases for the pseudo-scalar and scalar contributions are zero.
\subsection[Using B2mumu to Extract Rt]{Using $ \bar{\mathcal{B}}(B_q\to\mu^+\mu^-)$ to Extract $R_t$}\label{sec:Rt_RD}
In Section \ref{sec:Rt}, we introduced an alternative determination of the SM UT apex based on the sides $R_b$ and $R_t$, avoiding the use of the angle $\gamma$.
This second UT fit scenario becomes interesting when searching for potential NP contributions in the measurements of $\gamma$.
However, its main limitation is the calculation of the side $R_t$ based on the ratio of CKM matrix elements $|V_{td}/V_{ts}|$, which is susceptible to NP contributions as well.
In Eq.\ \eqref{eq:VtdVts}, this ratio was determined using the $B_q^0$--$\bar B_q^0$ mixing parameters.
Here we would like to explore the determination of the ratio $|V_{td}/V_{ts}|$ using the rare decays $B_q^0\to\mu^+\mu^-$.
In the SM and FUNP scenario, the result would be consistent with the strategy based on $\Delta m_s$ and $\Delta m_d$, which can be tested experimentally.

Based on the SM expressions in Eqs.\ \eqref{eq:Bs2mumu_theo} and \eqref{eq:Bs2mumu_TI}, the ratio of branching fractions between $B_d^0\to\mu^+\mu^-$ and $B_s^0\to\mu^+\mu^-$ is given by
\begin{equation}
    \left.\frac{\bar{\mathcal{B}}(B_d\to\mu^+\mu^-)}{\bar{\mathcal{B}}(B_s\to\mu^+\mu^-)}\right|_{\text{SM}}
    = \lambda^2 R_t^2 \left[1 + \frac{\lambda^2}{2}\left(1-2\bar\rho\right)\right]^2
    \frac{\sqrt{m_{B_d}^2 - 4 m_{\mu}^2}}{\sqrt{m_{B_s}^2 - 4 m_{\mu}^2}}
    \frac{f_{B_d}^2}{f_{B_s}^2} \frac{1-y_s}{1-y_d}\:,
\end{equation}
where we used Eq.\ \eqref{eq:Rt} to rewrite the ratio $|V_{td}/V_{ts}|$ in terms of the UT side $R_t$.
Combining this with the SM values for $R_t$ given in Eqs.\ \eqref{eq:Rt_apex_I3}--\eqref{eq:Rt_apex_H3}, which are calculated using $R_b$ and $\gamma$, and using $f_{B_s}/f_{B_d} = 1.209 \pm 0.005$ \cite{Aoki:2021kgd}, we get the predictions:
\begin{align}
    \text{Incl, } K\ell3 & &
    \left.\frac{\bar{\mathcal{B}}(B_d\to\mu^+\mu^-)}{\bar{\mathcal{B}}(B_s\to\mu^+\mu^-)}\right|_{\text{SM}}
    & = 0.0282 \pm 0.0019\:,\\
    \text{Excl, } K\ell3 & &
    \left.\frac{\bar{\mathcal{B}}(B_d\to\mu^+\mu^-)}{\bar{\mathcal{B}}(B_s\to\mu^+\mu^-)}\right|_{\text{SM}}
    & = 0.0281 \pm 0.0018\:,\\
    \text{Hybrid, } K\ell3 & &
    \left.\frac{\bar{\mathcal{B}}(B_d\to\mu^+\mu^-)}{\bar{\mathcal{B}}(B_s\to\mu^+\mu^-)}\right|_{\text{SM}}
    & = 0.0281 \pm 0.0017\:.
\end{align}
The difference between the three $R_b$ scenarios is almost negligible given the current precision.
This suggests a high-precision measurement of the $B_q^0\to\mu^+\mu^-$ branching fractions will be necessary to constrain $R_t$ through this method.

In contrast, the current experimental result for the ratio of branching fractions, based on Ref.\ \cite{Altmannshofer:2021qrr} that gives the average for $\bar{\mathcal{B}}(B_d\to\mu^+\mu^-)$ in Eq.\ \eqref{eq:Bdmumu_Avg}, is
\begin{equation}
    \frac{\bar{\mathcal{B}}(B_d\to\mu^+\mu^-)}{\bar{\mathcal{B}}(B_s\to\mu^+\mu^-)}
    = 0.019 \pm 0.024\:,
\end{equation}
and has an uncertainty of 125\%.
This corresponds to a value for the UT side $R_t$ of
\begin{equation}
    R_t = 0.77 \pm 0.48\:.
\end{equation}
In comparison with the determinations in Eqs.\ \eqref{eq:Rt_Dms_I3}--\eqref{eq:Rt_classic_H3} the uncertainty is a factor 20 to 40 larger.
However, with the third data taking period of the LHC just started, more precise measurements of the $B_q^0\to\mu^+\mu^-$ branching fractions can be expected in the near future, thereby making this an interesting option to explore, and useful addition to the study of NP effects in $B_q^0$--$\bar B_q^0$ mixing.

\section{What Will the Future Bring?}\label{sec:future}
\subsection[Improved Precision on the NP Parameters kappa and sigma]{Improved Precision on the NP Parameters $\kappa_q$ and $\sigma_q$}
\begin{figure}
    \centering
    \includegraphics[width=0.49\textwidth]{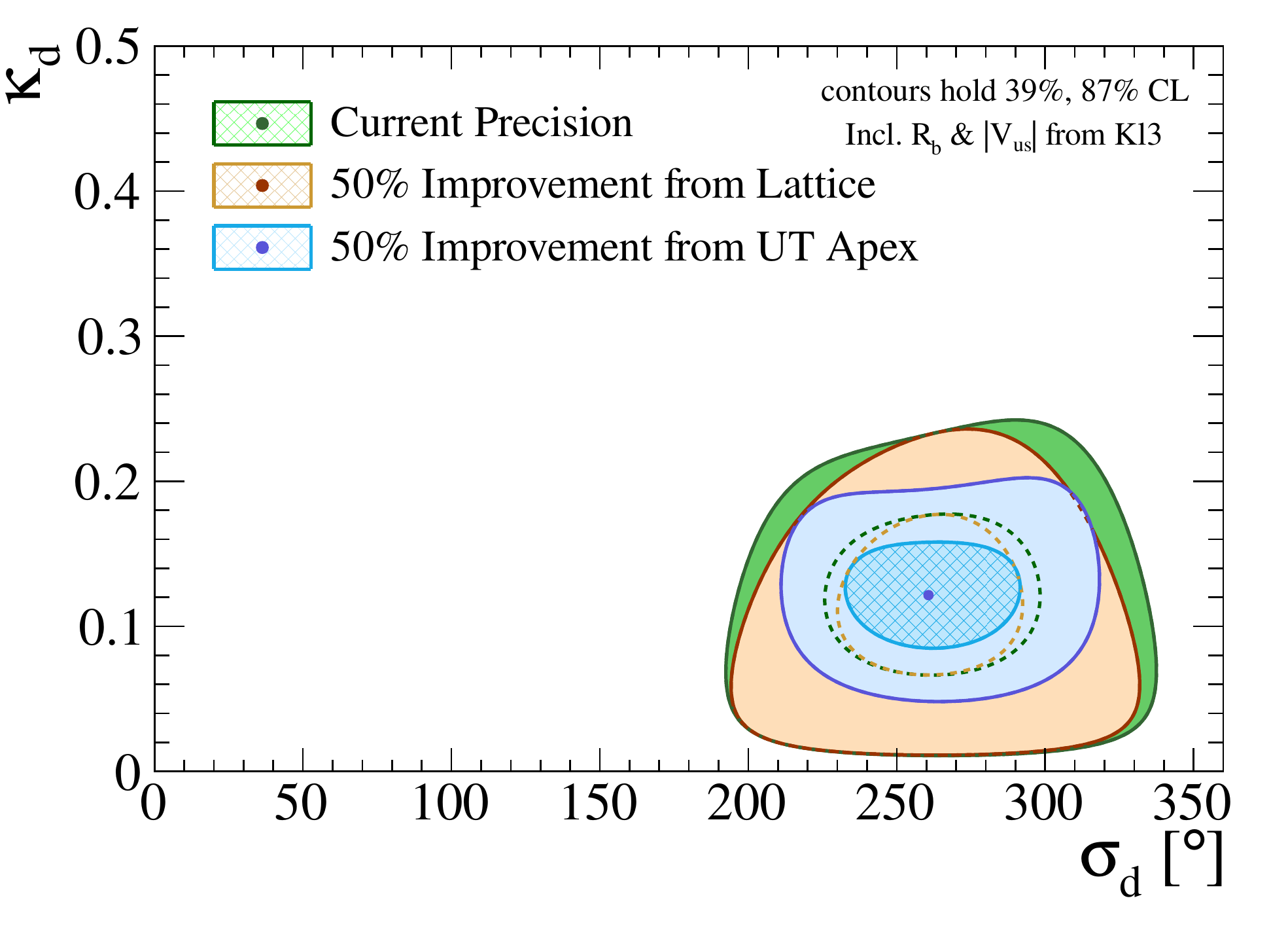}
    \hfill
    \includegraphics[width=0.49\textwidth]{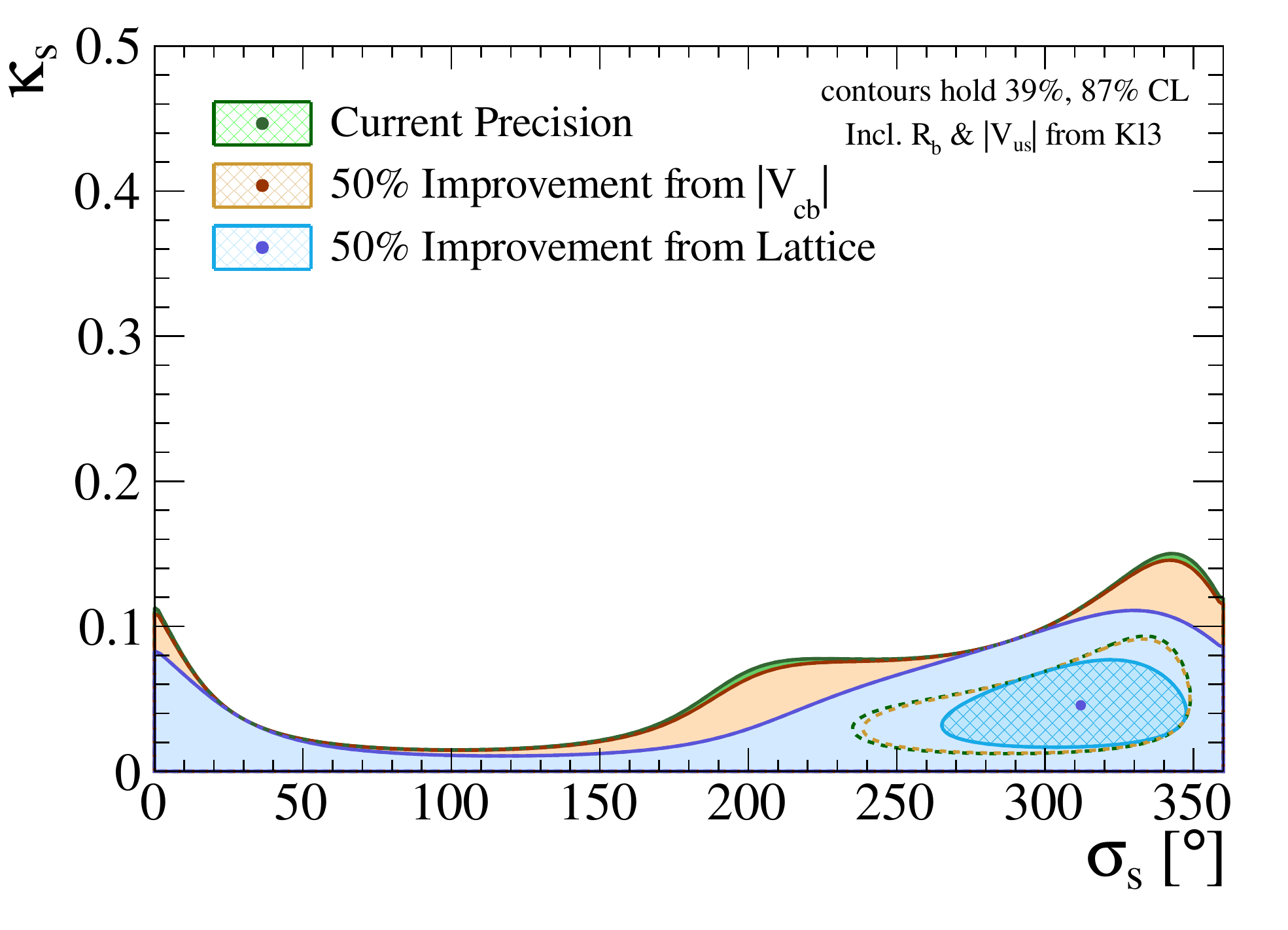}
    
    \includegraphics[width=0.49\textwidth]{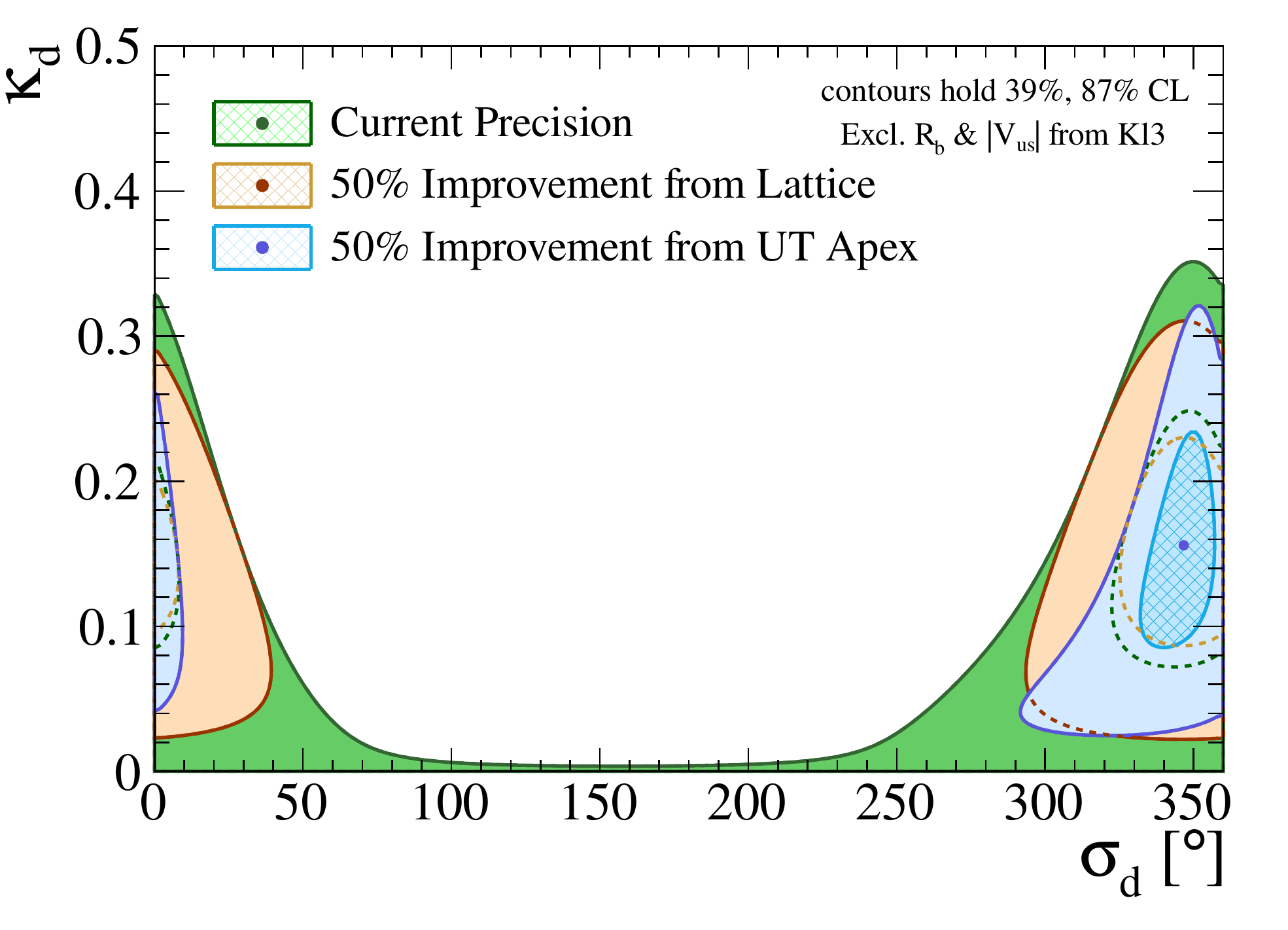}
    \hfill
    \includegraphics[width=0.49\textwidth]{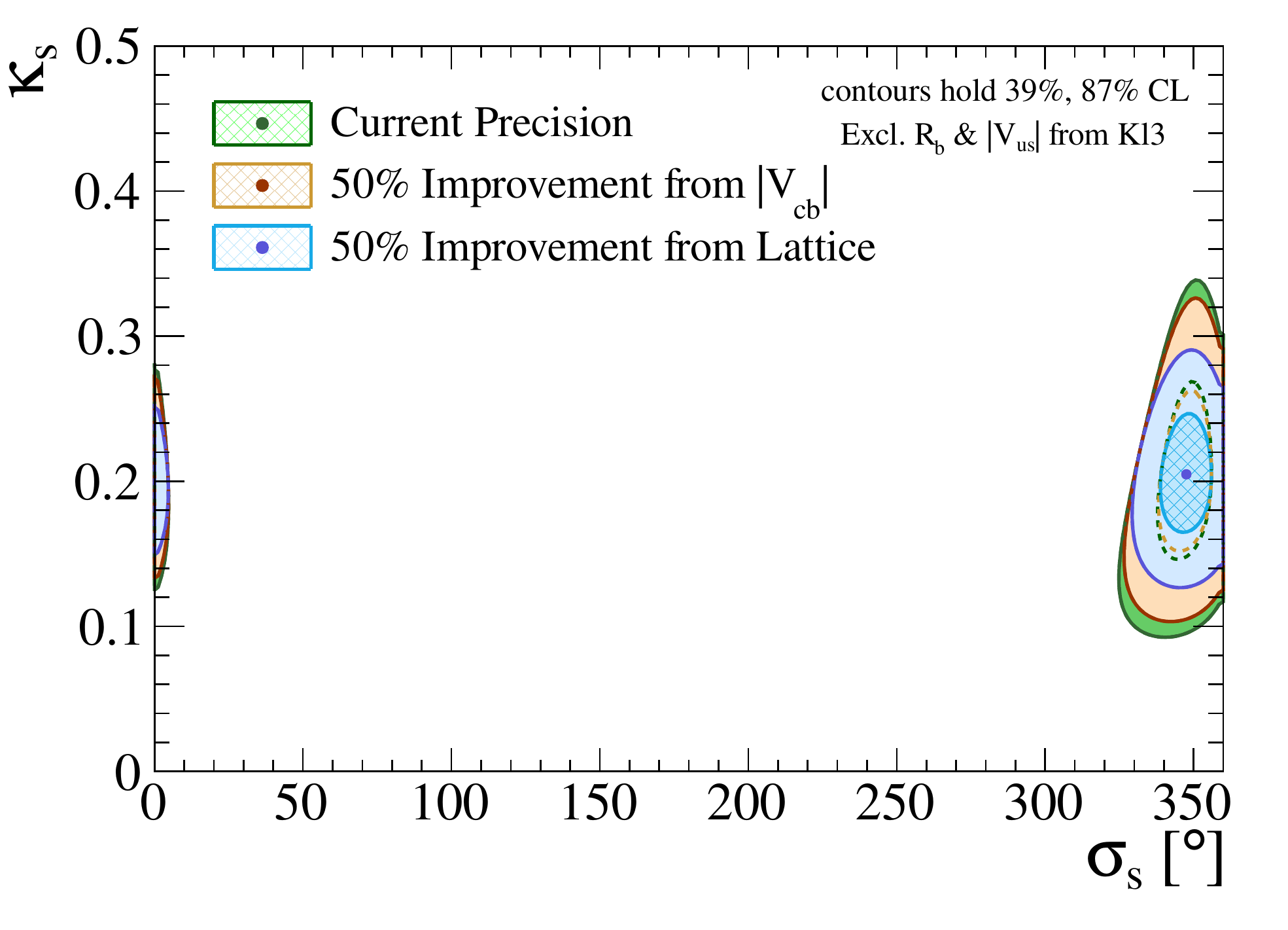}

    \includegraphics[width=0.49\textwidth]{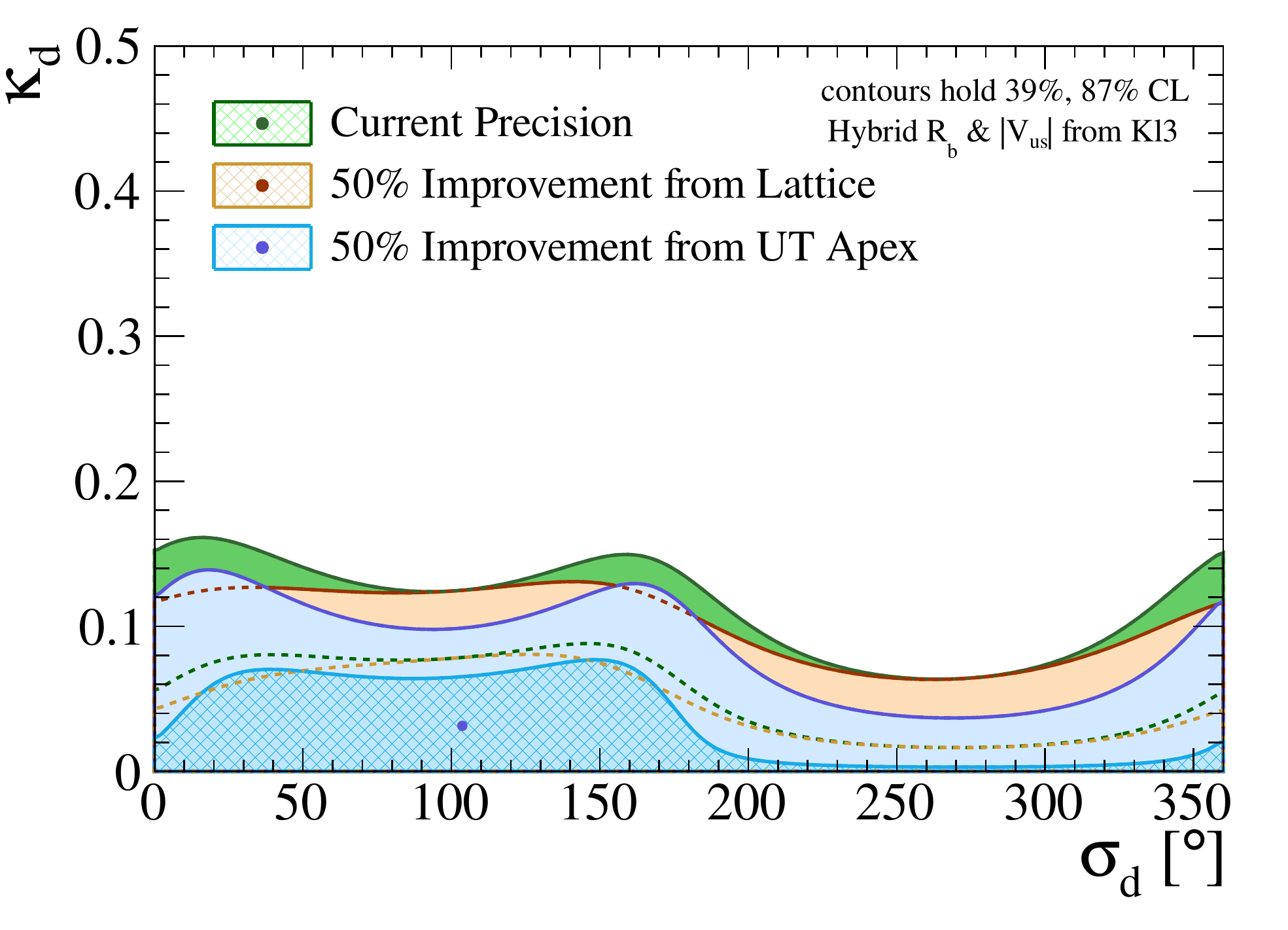}
    \hfill
    \includegraphics[width=0.49\textwidth]{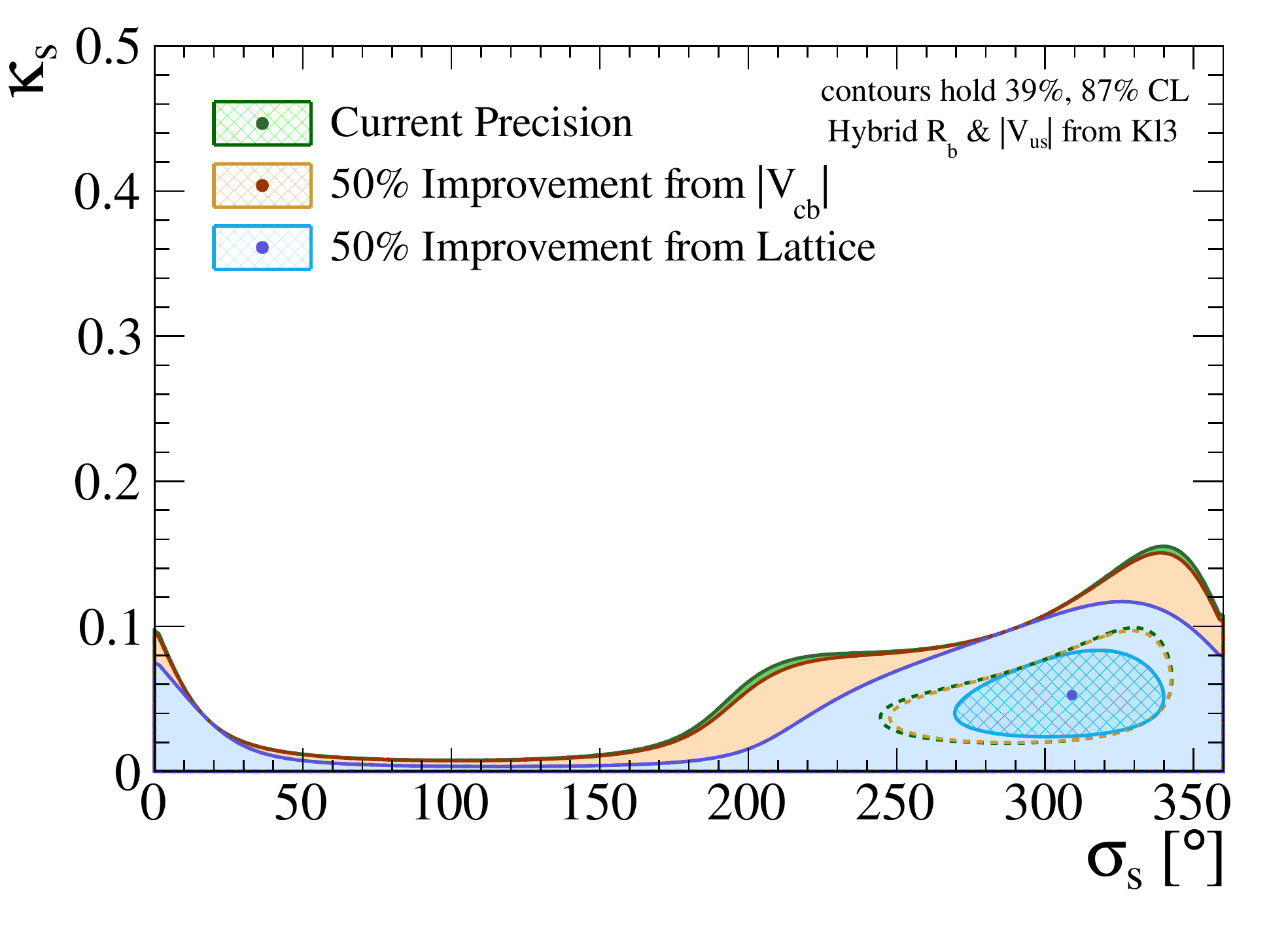}
    \caption{Comparison of the two-dimensional confidence regions for the NP parameters $\kappa_d$ and $\sigma_d$ (Left), and $\kappa_s$ and $\sigma_s$ (Right) assuming a hypothetical reduction of 50\% in the uncertainty on the CKM matrix element $|V_{cb}|$, the lattice calculations, or the UT apex.
    Top: Inclusive scenario. 
    Middle: Exclusive scenario.
    Bottom: Hybrid scenario.}
    \label{fig:Future}
\end{figure}

The NP searches in $B_q^0$--$\bar B_q^0$ mixing are limited by the SM predictions for $\phi_d$, $\Delta m_d$ and $\Delta m_s$.
Only the SM value for $\phi_s$ is still more precisely known than its experimental measurement.
The value of $\phi_d^{\text{SM}}$ is limited by our knowledge of the UT apex, and in particular $R_b$, while the largest uncertainty in the calculations of $\Delta m_d^{\text{SM}}$ and $\Delta m_s^{\text{SM}}$ are due to the non-perturbative parameters.
Improving these calculations has turned out to be difficult, and the associated time-scale unpredictable. 
We therefore refrain from making specific estimates for the future.
Instead we will only illustrate the impact that improvements on the UT apex, lattice calculations and the CKM matrix element $|V_{cb}|$ have on the NP parameters $\kappa_q$ and $\sigma_q$, and explore their relative importance.
Table \ref{tab:future} numerically compares the current precision on $\kappa_q$ and $\sigma_q$ with what is achievable assuming a hypothetical reduction of the uncertainty on each of these three inputs by 50\%.
For the two inputs that have the largest impact, the two-dimensional confidence level contours are compared in Fig.\ \ref{fig:Future}.

For the $B_s$-meson system, the precision on $\kappa_s$ and $\sigma_s$ is limited by the uncertainty on the lattice calculations.
Improvements to these calculations will reduce the allowed confidence region the most, and are therefore eagerly anticipated. 
In contrast, the impact from improvements in our knowledge of the UT apex is essentially negligible.
This is in particular the case for $\phi_s$.
In Ref.\ \cite{LHCb:2018roe}, discussing the prospects for LHCb Upgrade II, an experimental precision on the measurement of $\phi_s^{\text{eff}}$ from $B_s^0\to J/\psi\phi$ of 4 mrad was forecasted.
This is still a factor 2 bigger than the current precision on $\phi_s^{\text{SM}}$, given in Eqs.\ \eqref{eq:phisSM_I3}--\eqref{eq:phisSM_H3}.
The search for NP in $\phi_s$ will thus remain limited by the experimental data, irrespective of our knowledge on the UT apex.
This conclusion was already noted in Ref.\ \cite{Charles:2020dfl}, and can be easily understood:
Both $\phi_s^{\text{SM}}$ and $\Delta m_s^{\text{SM}}$ only depend on $\bar\rho$ and $\bar\eta$ at next-to-leading order, and their contributions are suppressed by a factor $\lambda^2$ compared to the leading order terms, while the lattice parameters appear quadratically in the leading order expression of $\Delta m_s^{\text{SM}}$.
Also $|V_{cb}|$ appears quadratically in $\Delta m_s^{\text{SM}}$, but the impact of an improved measurement remains small as long as the lattice results dominate the error budget.

The future prospects for finding NP in the $B_s$-meson system are highly dependent on the assumptions made.
From the considered scenarios shown in Fig.\ \ref{fig:Future}, the exclusive scenario assuming a 50\% improvement from lattice appears most exciting.
If this were to be realised, we could claim NP in $B_s^0$--$\bar B_s^0$ mixing with a significance of more than 5 standard deviations.
In the other scenarios, the situation will be more challenging.

For the $B_d$-meson system, the situation is very different.
Here, improvements in the determination of the UT apex and the reduction of the uncertainties from lattice calculations have an equally big impact on the allowed confidence regions for $\kappa_d$ and $\sigma_d$.
In particular, the limitation due to the UT apex is surprising, as was also pointed out in Ref.\ \cite{Barel:2020jvf} in the context of NP searches in $\phi_d$.
The uncertainties on the SM predictions for $\phi_d$ in Eqs.\ \eqref{eq:phidSM_I3}--\eqref{eq:phidSM_H3} are about 40\% larger than the uncertainty on the experimental measurement in Eq.\ \eqref{eq:phid_JpsiK}.
This situation will not significantly change in the near future, as we can illustrate using the forecasts for the LHCb Upgrade II \cite{LHCb:2018roe}.
LHCb expects to reduce the uncertainty on $R_b$ to 1\% and on $\gamma$ to $0.35^{\circ}$ by the end of the HL-LHC programme.
This corresponds to a precision on $\phi_d^{\text{SM}}$ of $0.48^{\circ}$.
By contrast, LHCb expects to be able to measure the mixing-induced CP asymmetry of the decay $B_d^0\to J/\psi K_{\text{S}}^0$ with an uncertainty of only 0.003.
This corresponds to a precision on $\phi_d^{\text{eff}}$ of $0.24^{\circ}$, and can be translated to a measurement of $\phi_d$ after taking into account the contributions from higher-order penguin topologies.
In Ref.\ \cite{Barel:2020jvf}, we argued that it is possible to control these penguin corrections with similar experimental precision, thus resulting in an estimated precision on $\phi_d$ of $0.34^{\circ}$.
Thus at the end of the HL-LHC programme, the uncertainty on the SM prediction will still be 40\% larger than the experimental measurement.
Measurements from the Belle II experiment will help to reduce this difference, but cannot completely bridge the gap in precision.
This makes it all the more important to critically analyse the UT fit and carefully select its inputs.
The determination of the SM values of $\bar\rho$ and $\bar\eta$ should remain a high priority as it impacts the vast majority of NP searches, ranging from $B_q^0$--$\bar B_q^0$ mixing to rare decays like $B_q^0\to\mu^+\mu^-$.

The future prospects for finding NP in the $B_d$-meson system are again highly dependent on the assumptions made.
From the considered scenarios shown in Fig.\ \ref{fig:Future}, the inclusive scenario assuming a 50\% improvement on the UT apex stands out.
Here, we could find hints for NP in $B_d^0$--$\bar B_d^0$ mixing with a significance of more than 3 standard deviations.
This is less promising than the $B_s$-meson system.
The difference is mainly due to the smaller value we find for $\kappa_d$ in the current data.

As a final remark, note that improvements on the non-perturbative parameters can also be achieved by combining the lattice calculations with LQSR results.
However, dedicated averages like those provided by FLAG \cite{Aoki:2021kgd} for the lattice calculations are still missing.
Nonetheless, such averages, if available, can have a significant impact on the study of $B$-meson decays, both on the NP searches in $B_q^0$--$\bar B_q^0$ mixing, and in the determination of $|V_{ub}|$ and $|V_{cb}|$, and thus the UT apex.

\begin{table}
    \centering
\begin{tabular}{||l|l|c|c|c|c||}
    \hline
    \hline
    Fit & Scenario & $\kappa_q$ & $\sigma(\kappa_q)$ & $\sigma_q$ & $\sigma(\sigma_q)$ \\
    \hline
    $B_d$ I3
    & Current                     & 0.121 & $-0.055/+0.056$ & $261^{\circ}$ & $-35^{\circ}/+37^{\circ}$ \\
    & 50\% improvement $|V_{cb}|$ & 0.121 & $-0.055/+0.056$ & $261^{\circ}$ & $-35^{\circ}/+37^{\circ}$ \\
    & 50\% improvement lattice    & 0.121 & $-0.055/+0.056$ & $261^{\circ}$ & $-31^{\circ}/+31^{\circ}$ \\
    & 50\% improvement UT apex    & 0.121 & $-0.036/+0.037$ & $261^{\circ}$ & $-28^{\circ}/+31^{\circ}$ \\
    \hline
    $B_d$ E3
    & Current                     & 0.156 & $-0.084/+0.093$ & $347^{\circ}$ & $-25^{\circ}/+21^{\circ}$ \\
    & 50\% improvement $|V_{cb}|$ & 0.156 & $-0.081/+0.089$ & $347^{\circ}$ & $-24^{\circ}/+21^{\circ}$ \\
    & 50\% improvement lattice    & 0.156 & $-0.069/+0.074$ & $347^{\circ}$ & $-22^{\circ}/+21^{\circ}$ \\
    & 50\% improvement UT apex    & 0.156 & $-0.071/+0.078$ & $347^{\circ}$ & $-15^{\circ}/+10^{\circ}$ \\
    \hline
    $B_d$ H3
    & Current                     & 0.031 & $-0.031/+0.057$ & $104^{\circ}$ & $-104^{\circ}/+256^{\circ}$ \\
    & 50\% improvement $|V_{cb}|$ & 0.031 & $-0.031/+0.056$ & $104^{\circ}$ & $-104^{\circ}/+256^{\circ}$ \\
    & 50\% improvement lattice    & 0.031 & $-0.031/+0.050$ & $104^{\circ}$ & $-104^{\circ}/+256^{\circ}$ \\
    & 50\% improvement UT apex    & 0.031 & $-0.031/+0.046$ & $104^{\circ}$ & $-104^{\circ}/+256^{\circ}$ \\
    \hline
    $B_s$ I3
    & Current                     & 0.045 & $-0.033/+0.048$ & $312^{\circ}$ & $-77^{\circ}/+37^{\circ}$ \\
    & 50\% improvement $|V_{cb}|$ & 0.045 & $-0.032/+0.046$ & $312^{\circ}$ & $-73^{\circ}/+37^{\circ}$ \\
    & 50\% improvement lattice    & 0.045 & $-0.028/+0.032$ & $312^{\circ}$ & $-47^{\circ}/+36^{\circ}$ \\
    & 50\% improvement UT apex    & 0.045 & $-0.033/+0.048$ & $312^{\circ}$ & $-77^{\circ}/+37^{\circ}$ \\
    \hline
    $B_s$ E3
    & Current                     & 0.205 & $-0.059/+0.064$ & $347.6^{\circ}$ & $-9.8^{\circ}/+8.5^{\circ}$ \\
    & 50\% improvement $|V_{cb}|$ & 0.205 & $-0.053/+0.058$ & $347.6^{\circ}$ & $-9.4^{\circ}/+8.5^{\circ}$ \\
    & 50\% improvement lattice    & 0.205 & $-0.040/+0.042$ & $347.6^{\circ}$ & $-8.8^{\circ}/+8.5^{\circ}$ \\
    & 50\% improvement UT apex    & 0.205 & $-0.059/+0.064$ & $347.6^{\circ}$ & $-9.7^{\circ}/+8.5^{\circ}$ \\
    \hline
    $B_s$ H3
    & Current                     & 0.053 & $-0.034/+0.046$ & $309^{\circ}$ & $-65^{\circ}/+34^{\circ}$ \\
    & 50\% improvement $|V_{cb}|$ & 0.053 & $-0.033/+0.044$ & $309^{\circ}$ & $-61^{\circ}/+33^{\circ}$ \\
    & 50\% improvement lattice    & 0.053 & $-0.029/+0.030$ & $309^{\circ}$ & $-39^{\circ}/+31^{\circ}$ \\
    & 50\% improvement UT apex    & 0.053 & $-0.034/+0.046$ & $309^{\circ}$ & $-64^{\circ}/+34^{\circ}$ \\
    \hline
    \hline
\end{tabular}
    \caption{Numerical comparison of the NP parameters $\kappa_q$ and $\sigma_q$ assuming a hypothetical reduction of 50\% in the uncertainty on the CKM matrix element $|V_{cb}|$, the lattice calculations, or the UT apex.
    The fit results are given separately for the inclusive (I3), exclusive (E3) and hybrid (H3) $R_b$ scenario with $|V_{us}|$ determined with the $K\ell3$ approach.}
    \label{tab:future}
\end{table}
%
%
%
\subsection[NP in gamma]{NP in $\gamma$}
As discussed in Section \ref{sec:gamma}, the two determinations of $\gamma$, based on the analysis of $B\to DK$ decays and based on the isospin analysis of the $B\to\pi\pi$, $\rho\pi$, $\rho\rho$ decays, are in good agreement with each another.
Given the very different origins of these measurements, this is a non-trivial result.
In the future, improved precision on the input measurements to these analyses could result in discrepancies between the two $\gamma$ determinations.
Averaging over both results, as we have done in Section \ref{sec:gamma}, would then no longer be justified, and the SM  determination of the UT apex would have to be revisited.
As it is a priori unclear which of the two measurements, if not both, is affected by NP contributions, independent information from additional observables will then be necessary to resolve the situation.
At the same time, this will also provide exciting new opportunities to search for NP, both in $\gamma$ itself and in $B_q^0$--$\bar B_q^0$ mixing, which is strongly correlated with the coordinates of the UT apex.

In Sections \ref{sec:Rt} and \ref{sec:Rt_RD}, we have explored alternative scenarios that do not rely on the measurements of $\gamma$.
However, all these scenarios have a common bottleneck: the SM determination of $R_t$.
Both the determination based on the measurements of $\Delta m_d$ and $\Delta m_s$, and the determination based on the branching fractions of the rare decays $B_q^0\to\mu^+\mu^-$ require us to make additional assumptions about FUNP, such that the ratios between the $B_d$ and $B_s$ observables remains SM-like.
Although this may be sufficient to find evidence of NP, it does not allow for a general model-independent approach and exploring multiple strategies simultaneously may be necessary to understand the origin of the NP contributions.
\subsection[Opportunities for B2mumu]{Opportunities for $ \bar{\mathcal{B}}(B_q\to\mu^+\mu^-)$}

\begin{figure}
    \centering
    \includegraphics[width=0.49\textwidth]{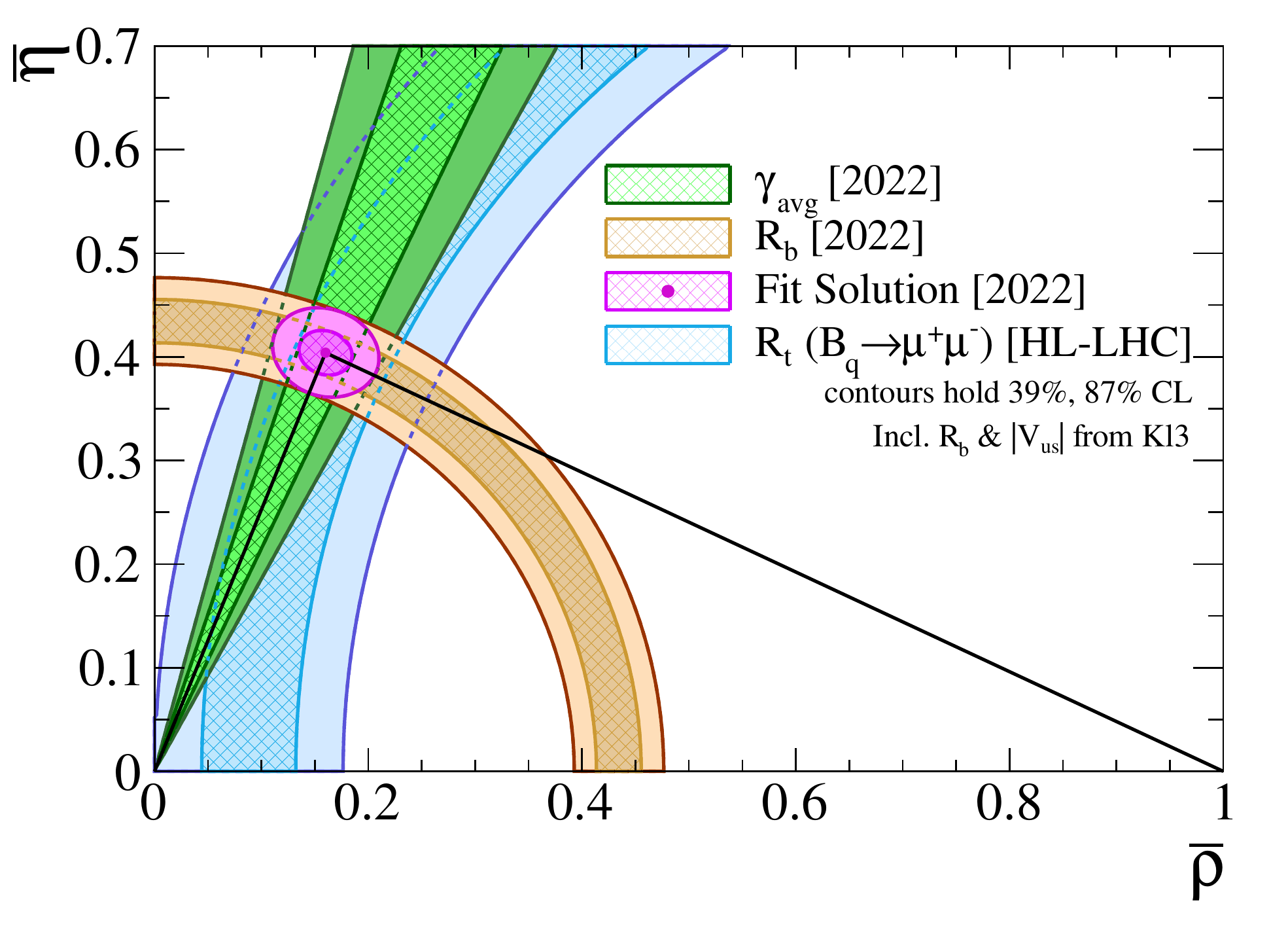}
    \includegraphics[width=0.49\textwidth]{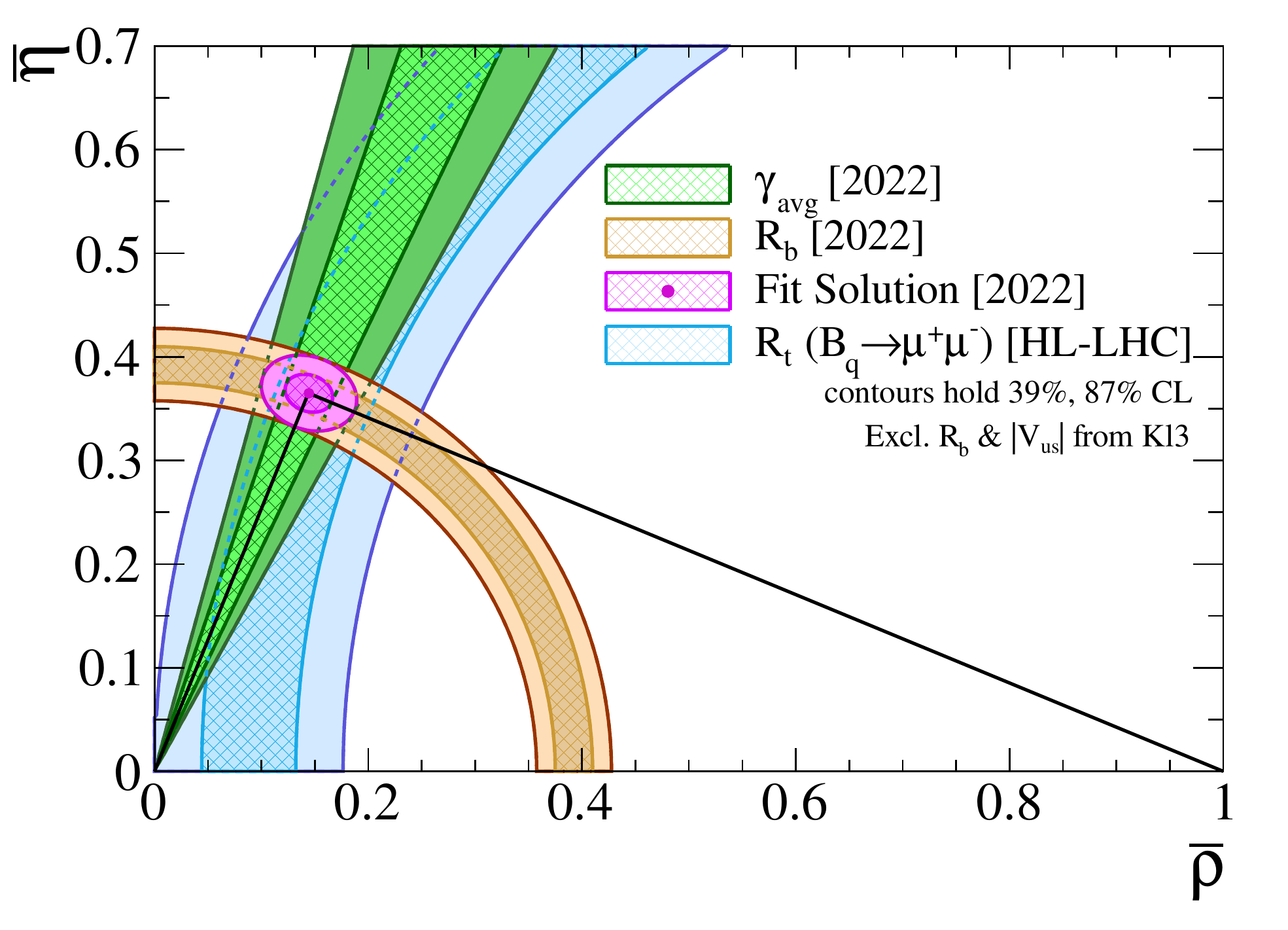}
    
    \includegraphics[width=0.49\textwidth]{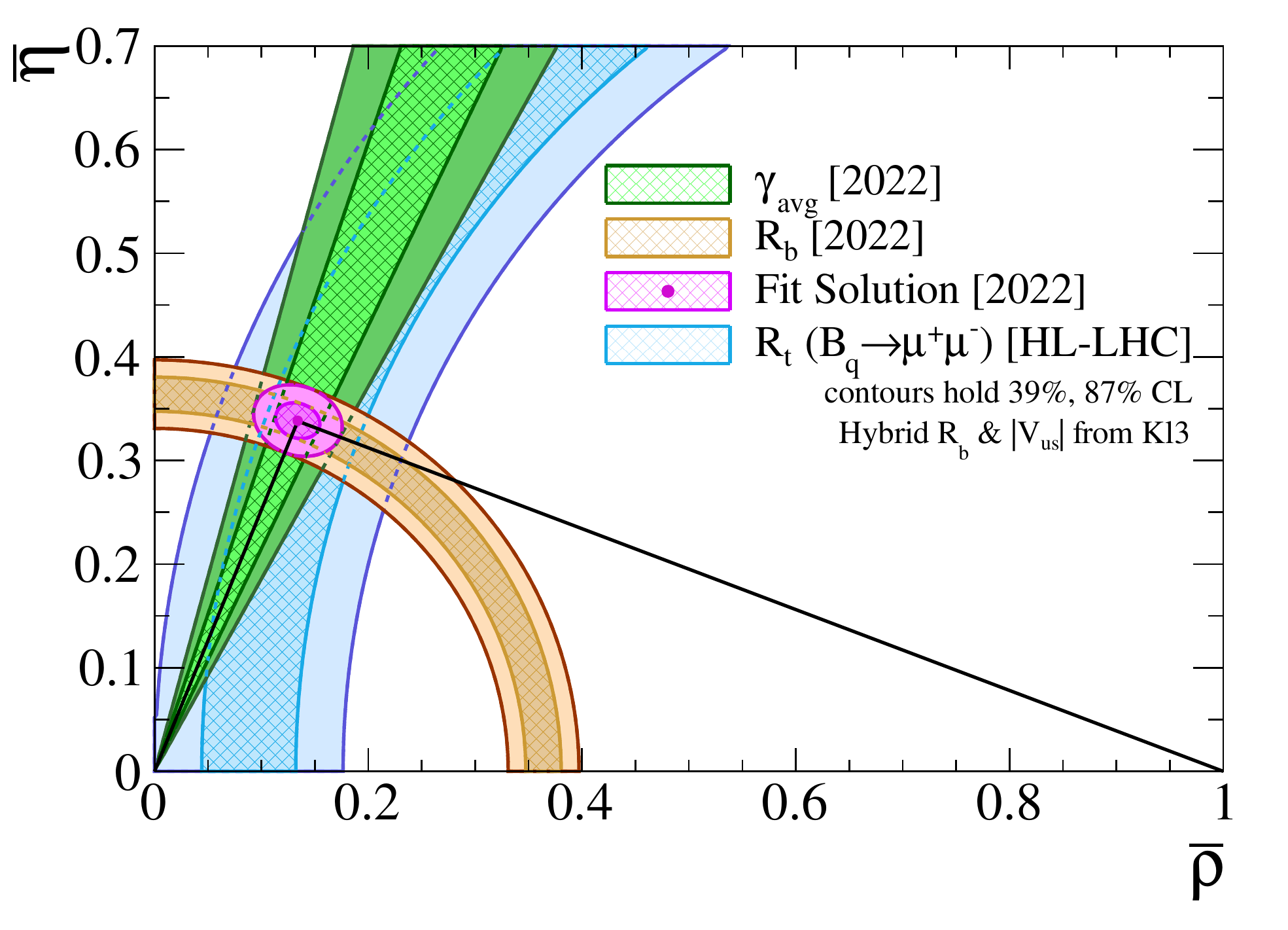}
    \caption{Comparison between the current SM determination of the UT apex and the expected constraint for the LHCb Upgrade II on $R_t$ from the ratio of branching fractions of $B_d^0\to\mu^+\mu^-$ and $B_s^0\to\mu^+\mu^-$.
    Left: Inclusive scenario. 
    Right: Exclusive scenario.
    Bottom: Hybrid scenario.}
    \label{fig:Future_Rt}
\end{figure}

The ratio of branching fractions between $B_d^0\to\mu^+\mu^-$ and $B_s^0\to\mu^+\mu^-$ provides an alternative opportunity to determine the UT side $R_t$.
However, as the precision on $R_t^2$ scales directly with the precision on the ratio of branching fractions, the current experimental uncertainties are still too large to already explore this option in detail.
But we can illustrate the potential for the LHC upgrade programme.
The LHCb collaboration expects to be able to reduce the uncertainty on the ratio of branching fractions to 34\% by 2025 and to 10\% by the end of the HL-LHC era \cite{LHCb:2018roe}.
Assuming the SM value from Eqs.\ \eqref{eq:Rt_classic_I3}--\eqref{eq:Rt_classic_H3}, this results in the prospects
\begin{equation}
    R_t = 0.93 \pm 0.16\:\text{(2025)}\:,\qquad
    R_t = 0.931 \pm 0.047\:\text{(Upgrade II)}\:,
\end{equation}
where the uncertainties are still fully dominated by the branching fraction measurements.
All other uncertainties combined only contribute at the 1\% level and can thus safely be ignored.
In comparison with the determinations in Eqs.\ \eqref{eq:Rt_Dms_I3}--\eqref{eq:Rt_classic_H3}, the LHCb Upgrade II prospects are still a factor 2 to 4 larger, as illustrated in Fig.\ \ref{fig:Future_Rt}.
We thus would require a precision on the ratio of branching fractions of 5\% or better to match the current determination based on $R_b$ and $\gamma$, and 2.5\% or better to match the current determination based on $R_b$, $\Delta m_d$ and $\Delta m_s$.

A second future application for the ratio of branching fractions between $B_d^0\to\mu^+\mu^-$ and $B_s^0\to\mu^+\mu^-$ is the quantity \cite{Fleischer:2017ltw}
\begin{align}
    U_{\mu\mu}^{ds} 
    & \equiv \sqrt{\frac{|P_{\mu\mu}^d|^2 + |S_{\mu\mu}^d|^2}{|P_{\mu\mu}^s|^2 + |S_{\mu\mu}^s|^2}}\:, \\
    & = \left[ \frac{\tau_{B_s}}{\tau_{B_d}} \frac{1-y_d^2}{1-y_s^2}
    \frac{1 + \mathcal{A}_{\Delta\Gamma}^d y_d}{1 + \mathcal{A}_{\Delta\Gamma}^s y_s}
    \frac{\sqrt{m_{B_s}^2 - 4 m_{\mu}^2}}{\sqrt{m_{B_d}^2 - 4 m_{\mu}^2}}
    \left(\frac{f_{B_s}}{f_{B_d}}\right)^2
    \left|\frac{V_{ts}}{V_{td}}\right|^2
    \frac{\bar{\mathcal{B}}(B_d\to\mu^+\mu^-)}{\bar{\mathcal{B}}(B_s\to\mu^+\mu^-)}
    \right]^{1/2}\:.
\end{align}
It has the advantage over the individual $B_d^0\to\mu^+\mu^-$ and $B_s^0\to\mu^+\mu^-$ branching fraction measurements that common parameters and uncertainties drop out in the ratio.
Therefore it provides a more powerful test of the SM, where $U_{\mu \mu}^{ds} = 1$.

\newpage
\section{Conclusions}\label{sec:conclusion}
In this paper, we have presented a comprehensive study of NP in $B_q^0$--$\bar B_q^0$ mixing. 
Our main goal is to determine the allowed parameter space for NP, having a critical look at analyses of the determination of the UT apex, which is needed as input for the corresponding SM predictions.
In particular, we explore the impact of the discrepancies between inclusive and exclusive measurements of the CKM matrix elements $|V_{ub}|$ and $|V_{cb}|$, which allow us to fix the side $R_b$ of the UT. 
Specifically, we perform separate analyses for the inclusive and exclusive determinations, respectively, and for a hybrid scenario, which combines the exclusive $|V_{ub}|$ with the inclusive $|V_{cb}|$ value.  
Generally, we find sizeable differences between these combinations, demonstrating that it would be very desirable and important to finally resolve the tensions between inclusive and exclusive determinations of the CKM parameters.

Combining the $R_b$ side with information on the angle $\gamma$, we can determine the apex of the UT. 
This angle is usually determined through $B\to DK$ decays, which proceed only via tree topologies. 
In our analysis, we utilise another avenue to obtain information on $\gamma$, which is provided by $B\to\pi\pi$, $\rho\pi$, $\rho\rho$ modes and isospin relations allowing us to control penguin uncertainties. 
Typically, these measurements are interpreted in terms of the UT angle $\alpha$. 
However, using information on the mixing phase $\phi_d$, they actually determine $\gamma$. Interestingly, we find full agreement between these two determinations of $\gamma$, which is non-trivial in view of the different dynamics governing the corresponding decays.
Within the SM, such an agreement is expected. 
Should a discrepancy arise in the future, it would indicate NP contributions to the corresponding decay amplitude, thereby requiring a more involved analysis. 
In view of the current agreement, we make an average between the two values and assume that this is the SM value of $\gamma$.

Using this information on $R_b$ and $\gamma$, we perform determinations of the UT apex for the various determinations of the CKM matrix elements $|V_{ub}|$  and $|V_{cb}|$.
Furthermore, within the SM, indirect CP violation in the neutral kaon system, described by the observable $|\varepsilon_K|$, allows us to obtain another constraint on the UT apex, given by a hyperbola in the $\bar\rho$--$\bar\eta$ plane. 
This curve depends strongly on $|V_{cb}|$, thereby suggesting a probe to distinguish between the inclusive and exclusive determinations of this CKM element. 
In our analysis, we find indeed this feature: the exclusive case gives sizeable tension, the inclusive case is more consistent, while the hybrid scenario results in the most compatible picture with the UT apex. 
Due to this strong dependence, it will be interesting to see how the interplay between the $|\varepsilon_K|$ hyperbola and the UT apex will evolve in the future.

The determination of the UT from $R_b$ and $\gamma$ plays a key role for the SM predictions of the $B_q^0$--$\bar{B}_q^0$ mixing parameters. 
It is important to stress that the corresponding UT analysis does not depend on any information from $B_q^0$--$\bar{B}_q^0$ mixing. 
We find that the choice of input parameters -- inclusive, exclusive or hybrid -- significantly impacts the picture emerging for the NP parameters in  $B^{0}_q$--$\bar{B}^{0}_q$ mixing. 

First, we constrain in a model-independent way the NP parameters separately for $B_s$ and $B_d$ systems. 
The strongest evidence for NP would arise in the $B_s$ system for the exclusive CKM matrix elements, corresponding to a NP contribution of about $20 \%$ with a significance of 3.5 standard deviations.

Next, we consider a flavour universal scenario with equal NP contributions to the $B_d$ and $B_s$ systems. 
In such a situation, the NP effects would cancel in the ratio $\Delta m_d/\Delta m_s$.
Consequently, we could use it to fix the side $R_t$ of the UT. Combining this side with $R_b$ allows us then to determine the UT apex without any information on $\gamma$. 
We find a picture very similar to the situation in the fit for the $B_s$ system, implying that this system dominates the analysis. 
While in the exclusive case the fit regions overlap well with those for the general NP analysis for the $B_s$ system, we find that the shapes of the contours for $B_d$ and $B_s$ systems are very different for the inclusive and hybrid scenarios, thereby indicating that the flavour universal scenario might not be realised in nature. 
However, in view of the current uncertainties, we cannot rule out such kind of NP. 

Finally, we go beyond the flavour universal scenario and relax the corresponding assumption when determining the NP parameters. 
While we still use the ratio $\Delta m_{d}/\Delta m_{s}$ to determine the UT apex together with the information on $R_b$, we extract $\kappa_q$, $\sigma_q$ separately using the individual measurements of $\Delta m_q$ and $\phi_q$.  
We find that the solutions for the $B_d$ and $B_s$ systems are statistically compatible with each other, but do not look the same. 
Consequently, the data favour a NP scenario which is not flavour universal in this analysis, although it cannot be excluded with the current precision. 
Let us finally emphasize that this analysis does not require information on $\gamma$. 
It will be interesting to see how the picture for the  NP parameters will evolve in the future. 

Our results for the NP parameters in the $B_s$ system can be applied to the analysis of the rare leptonic decay $B^0_s\to\mu^+\mu^-$. 
The corresponding SM branching ratio, using unitarity of the CKM matrix, depends on the CKM element $|V_{cb}|$. 
We have explored the impact of different determinations of this quantity, finding that it plays the key role for the uncertainty of the SM branching ratio. 
This dependence can be eliminated through the ratio of the $B^0_s\to\mu^+\mu^-$ branching ratio with the mass difference $\Delta m_s$. 
We generalise this approach by including possible NP contributions to $B_s^0$--$\bar B_s^0$ mixing through our analysis and constrain the space for NP in $B_s^0\to\mu^+\mu^-$ through the current measurement of the branching ratio. 
Following these lines, the CKM factors and the UT apex enter only through the analysis of NP in $B_s^0$--$\bar B_s^0$ mixing. 
Another interesting mode is the $B^0_d \to \mu^+ \mu^-$ decay which has not yet been observed. 
In the SM, the ratio of the branching ratios of $B^0_d \to \mu^+ \mu^-$ and $B^0_s \to \mu^+ \mu^-$ allows us to determine the $R_t$ side of UT. 
We have illustrated this interesting option for a future scenario, comparing it with the current picture of the UT.

We have extrapolated our analysis of the NP parameters of $B^{0}_q$--$\bar{B}^{0}_q$ mixing to the future high-precision era by considering various projections for the CKM matrix elements, the UT apex fit and lattice calculations.
For the $B_d$ system we find that the SM determination of the UT apex represents a key limitation for the NP searches.
On the other hand, the $B_s$ system has no such limitation and looks particularly interesting for revealing new sources of NP.
In the future, studies of $B_q^0$--$\bar B_q^0$ mixing will remain a central element of the analysis to further constrain NP. 
It will be exciting to obtain the full picture providing links to other probes such as anomalies in semileptonic rare $B$ decays.

\section*{Acknowledgements}
This research has been supported by the Netherlands Organisation for Scientific Research (NWO). 
PvV acknowledges support from the DFG through the Emmy Noether research project 400570283, and through the German-Israeli Project Cooperation (DIP).
%
%
%
\phantomsection 
\addcontentsline{toc}{section}{References}
\setboolean{inbibliography}{true}
\ifx\mcitethebibliography\mciteundefinedmacro
\PackageError{LHCb.bst}{mciteplus.sty has not been loaded}
{This bibstyle requires the use of the mciteplus package.}\fi
\providecommand{\href}[2]{#2}

\end{document}